\documentclass[fleqn,usenatbib]{mnras}

\usepackage[T1]{fontenc}

\DeclareRobustCommand{\VAN}[3]{#2}
\let\VANthebibliography\thebibliography
\def\thebibliography{\DeclareRobustCommand{\VAN}[3]{##3}\VANthebibliography}

\usepackage{graphicx}	
\usepackage{amsmath}	

\usepackage{newtxtext,newtxmath}

\usepackage{pdflscape}

\usepackage{xspace}

\newcommand{\persec}{s$^{-1}$\xspace}

\providecommand{\pic}[2][1]{\includegraphics[width=#1\linewidth]{#2}}

\title[W-FAST KBO occultation search]{A search for Kuiper Belt occultations using the Weizmann Fast Astronomical Survey Telescope}

\author[G. Nir et al.]{
Guy Nir,$^{1,3}$\thanks{E-mail: guynir42@gmail.com (GN)}
Eran O.~Ofek,$^{1}$
David Polishook$^{2}$
Barak Zackay$^{1}$
and Sagi Ben-Ami$^{1}$
\\
$^{1}$Department of Particle Physics and astrophysics, the Weizmann Institute of Science, Rehovot, Israel\\
$^{2}$Department of physics core facilities, Weizmann Institute of Science, 76100 Rehovot, Israel. \\
$^{3}$University of California, Berkeley, Department of Astronomy, Berkeley, CA 94720
}

\date{Accepted XXX. Received YYY; in original form ZZZ}

\pubyear{2023}

\begin{document}
\label{firstpage}
\pagerange{\pageref{firstpage}--\pageref{lastpage}}
\maketitle

\begin{abstract}
Measuring the size distribution of small (km-scale) KBOs
can help constrain models of Solar System formation 
and planetary migration. 
Such small, distant bodies are hard to detect with 
current or planned telescopes, 
but can be identified as sub-second occultations
of background stars. 
We present the analysis of data from the 
Weizmann Fast Astronomical Survey Telescope (W-FAST), 
consisting of fast photometry of $\sim10^{6}$ star-hours
at a frame rate of 10--25\,Hz. 
Our pipeline utilizes a matched-filter approach
with a large template bank, including red-noise treatment, 
and injection of simulated events for estimating the detection efficiency.
The KBO radius at which our survey is 10\% (50\%) efficient is 1.1 (2.0)\,km.
The data from 2020--2021 observing seasons were analyzed 
and no occultations were identified. 
We discuss a sample of sub-second false-positive events, 
both occultation-like and flare-like, 
which are still not fully understood 
but could be instructive for future surveys looking for short-duration events. 
We use our null-detection result to set 
limits on the km-scale KBO number density. 
Our individual radius bin limits are consistent with most previous works, 
with $N(r>1\text{km})\lessapprox 10^6$\,deg$^{-2}$ (95\% confidence limit).
Our integrated (all size) limits, 
assuming a power law normalized to large ($\approx 45$\,km) KBOs
gives a power law index $q<3.93$ (95\% confidence limit).
Finally, our results are in tension 
with a recently reported KBO detection from the ground, 
at the $p=4\times10^{-4}$ level. 
\end{abstract}




\section{Introduction}

Trans Neptunian Objects (TNOs) are a population of bodies found outside the orbit of Neptune. 
Some families of TNOs are thought to be remnants 
of the early stages of Solar System formation, 
and may be untouched by interactions with any of the planets
\citep{collisional_models_Dohnanyi_1969, KBO_sculpting_Morbidelli_Brown_Levison_2003, kuiper_belt_size_distribution_Kenyon_2004}. 
As such, they could hold important clues to the formation history
of our Solar System and to understanding the conditions present 
around young stars before their planets formed. 

A subset of TNOs are Kuiper Belt Objects (KBOs), 
which reside in a belt at 30--50\,AU from the Sun, 
and clustered mostly a few degrees from the ecliptic plane. 
While hundreds of KBOs larger than a few km have been identified, 
the smaller, but more abundant population remains hard to detect
\citep{kuiper_belt_survey_HST_Bernstein_2004, kuiper_belt_survey_Subaru_archival_Fuentes_2008, kuiper_belt_survey_Subaru_Fraser_2009, kuiper_belt_survey_Subaru_beam_Fuentes_2009, 
kuiper_belt_survey_Palomar_Schwamb_2010, kuiper_belt_structure_CFEPS_Petit_2011}.

Measuring the size and inclination distributions 
of km-sized KBOs holds clues 
to the formation and evolution of the Solar System \citep{KBO_sculpting_Morbidelli_Brown_Levison_2003, kuiper_belt_size_distribution_Kenyon_2004, KBO_initial_sizes_Schlichting_2013}, 
the origin of comets, \citep{scattered_disk_jupiter_family_Duncan_1997, kuiper_belt_jupiter_family_Levison_1997, scattered_disk_jupiter_family_Volk_2008, KBO_Jupiter_family_Wesley_2022}
and the material strength of these bodies \citep{KBOs_power_law_Pan_Sari_2004}.

Their small size and considerable distance from the Sun
make them too faint to be imaged directly by current or planned telescopes. 
However, it is possible to detect km-sized KBOs using serendipitous occultations
of background stars \citep{occultation_idea_Bailey_1976, KBO_occultation_diffraction_Roques_1987},  
even though the rate of these occultations is very low. 
At any given moment, $\mathcal{O}(10^{-9})$ stars are occulted by KBOs, 
and the duration of these occultations is typically a fraction of a second. 

A search for such occultations in the Hubble Space Telescope 
Fine Guidance Sensor (HST FGS) yielded two detections of sub-km KBO occulters
\citep{KBO_single_object_Schlichting_Ofek_2009, abundance_kuiper_objects_Schlichting_Ofek_2012}, 
while searches for more occultations using ground-based telescopes 
is ongoing \citep{TAOS_survey_Zhang_2013, OASES_survey_Arimatsu_2017, Colibri_survey_Pass_2018, TAOS_II_survey_Huang_2021, Colibri_survey_Mazur_2022}. 
One detection from the ground has been published so far \citep{KBO_detection_from_ground_Arimatsu_2019}, 
that could be evidence for an overabundance of KBOs larger than 1\,km, 
generally consistent with findings from crater sizes on the surface of Pluto
\citep{kuiper_belt_sizes_Pluto_craters_Morbidelli_2021}. 
We note that even the two relatively secure detections
in the HST FGS data have a false alarm probability of 2\% and 5\%, respectively. 

In this work we present the results of a dedicated survey for KBO occultations
carried out using the Weizmann Fast Astronomical Survey Telescope (W-FAST; \citealt{WFAST_system_overview_Nir_2021}), 
during the years 2020--2021. 
A custom data reduction pipeline,
described by \cite{wfast_kbo_pipeline_Nir_2023}
was developed to try to mitigate 
the different noise sources and false positives inherent to ground base photometry, 
as well as inject simulated events into the extracted lightcurves to 
measure the expected efficiency of the pipeline. 
Even though our analysis includes $\approx 750,000$ star-hours of usable data close to the ecliptic plane, 
the presence of correlated noise
(e.g., from the atmosphere; see \citealt{atmospheric_scintillation_Osborn_2015} 
and references therein)
is shown to reduce efficiency for the detection of small occulters, 
consistent with detecting zero occultations in two years' worth of data, 
using published estimates for the density of small KBOs. 
We present seven occultation candidates detected by our pipeline
that are most likely false positives, for various reasons. 
The different sources of false positives are discussed, 
and help demonstrate the difficulty of identifying true 
sub-second occultations in data from a single telescope. 

With zero detections in our dataset, we present upper limits
on the number density of KBOs near the ecliptic plane, 
and compare them to previously published results. 
Per individual radius bin, our limits are similar 
to the results of \cite{TAOS_survey_Zhang_2013}, 
that have accumulated a similar number of star-hours but
operated with three or four telescopes. 
We also combine the lack of detections in all size ranges
to compare our null result with the models given by 
previous KBO detections. 
The power law size distribution model given by HST FGS detections \citep{abundance_kuiper_objects_Schlichting_Ofek_2012} 
is consistent with our result, 
while the recent detection from the ground
\citep{KBO_detection_from_ground_Arimatsu_2019}
is inconsistent with our null result at $p=4\times 10^{-4}$ confidence level.

We present the observatory and the data analysis in \S\ref{sec:observatory and data}. 
We discuss the additional analysis performed on each of the occultation candidates in \S\ref{sec:analysis}. 
We summarize the coverage provided by the 2020--2021 dataset, 
as well as a summary of the candidate events, in \S\ref{sec:results}. 
We set limits on km-sized KBOs 
and discuss the challenges of detecting sub-km KBOs from the ground in 
\S\ref{sec:discussion}, 
and conclude in \S\ref{sec:conclusions}.

\section{The W-FAST occultation survey}\label{sec:observatory and data}

\subsection{Data acquisition and analysis}\label{sec:acquisition and analysis}

W-FAST is a 55\,cm, wide-field Schmidt telescope, 
equipped with a 7\,deg$^{2}$ field-of-view, fast-readout, low read-noise, sCMOS camera.
The system was designed to explore the sub-second transient sky 
and to search for trans-Neptunian objects using the occultation technique \citep{occultation_idea_Bailey_1976}.
The observatory is currently located in Neot Smadar, Israel (see site description in \citealt{LAST_system_overview_Ofek_2023}). 
During the 2020--2021 observing runs it had been located at Mizpe-Ramon, Israel\footnote{
The Wise Observatory in Mitzpe Ramon, with the coordinates: 30.59680689 N, 34.76205490 E, elevation 876\,m.
}. 
The W-FAST observatory is described in \cite{WFAST_system_overview_Nir_2021}, 
while the KBO analysis pipeline is discussed by \cite{wfast_kbo_pipeline_Nir_2023}. 

Given the high data rate produced by the camera (about 6\,Gbit\,s$^{-1}$),
we only saved cutouts around the 1000--5000 brightest stars in the field, 
rather than save full-frame images. 
The magnitude limit of stars depends on the exposure time and airmass, 
with stars of photometric S/N$=3\sigma$ having equivalent Gaia $B_p$ magnitude
of about 12-13.5. 
The spatial resolution, including atmospheric seeing and instrumental effects, 
moves between 5 and 10'', or 2 to 4.5 pixels with W-FAST's 2.33"/pix scale. 
The cutouts were chosen to have a square side length of 15 pixels, 
to allow for the width of the Point Spread Function (PSF) and also some room for mild tracking errors 
or vibrations of the telescope. 

We applied a custom photometry pipeline to these small images (see details in \citealt{WFAST_system_overview_Nir_2021,wfast_kbo_pipeline_Nir_2023}).
Dark and flat field corrections were conducted on the individual cutout images. 
Photometry was measured using a centring procedure starting with a weighted aperture, 
which gives better results in finding the average centroid offset of the stars in their cutouts, 
which are then used to position an un-weighted aperture on all stars. 
Some experimentation with PSF-weighted apertures was made, 
but the frame-to-frame flux measurements were less stable, 
presumably due to the time-varying PSF of images taken with short exposure times. 
In most cases apertures with a radius of 3 pixels were found to give the lowest noise. 
Background was measured on an annulus of 7.5 to 10 pixels, 
but was not subtracted from the flux measurements 
in order to reduce the total noise when looking for occultations. 
The background measurements were used to help disqualify bad events
or entire segments of data, e.g., that were taken during morning twilight. 
Once light-curves were produced, we searched them for occultations by applying a matched-filter \citep{matched_filter_Turin_1960}
with a template bank of 300--500 templates, 
spanning the parameter space of KBO occultations that we expect to be able to detect.
Parameters include the impact parameter, occulter radius, star radius, time offset and the transverse velocity between
the star and the occulter (e.g., \citealt{KBO_detectibility_Nihei_2007, oort_cloud_occultation_kepler_Ofek_2010}).
Since the light-curves include substantial \emph{red noise} (higher noise amplitude in lower frequencies)
we applied `whitening' to the filtering process (e.g., \citealt{gravitational_waves_non_gaussian_Zackay_2021, black_hole_binaries_GW_pipeline_Venumadhav_Zackay_2019}). 

Finally, in order to estimate the efficiency of our pipeline, 
we injected artificial occultations into the real light-curves 
and applied our pipeline to the modified data as well.
The resulting candidates are presented to a human scanner, 
without revealing any information that can bias his or her decision 
(e.g., the ecliptic latitude, or the Earth's projected velocity).
This step is crucial in order to estimate the real efficiency of the search 
and can not be replaced by analytical estimation.
A more detailed description of the pipeline is discussed by \cite{wfast_kbo_pipeline_Nir_2023}. 

\subsection{Observations}\label{sec:observations}

\begin{figure}
    \centering
    \pic[1]{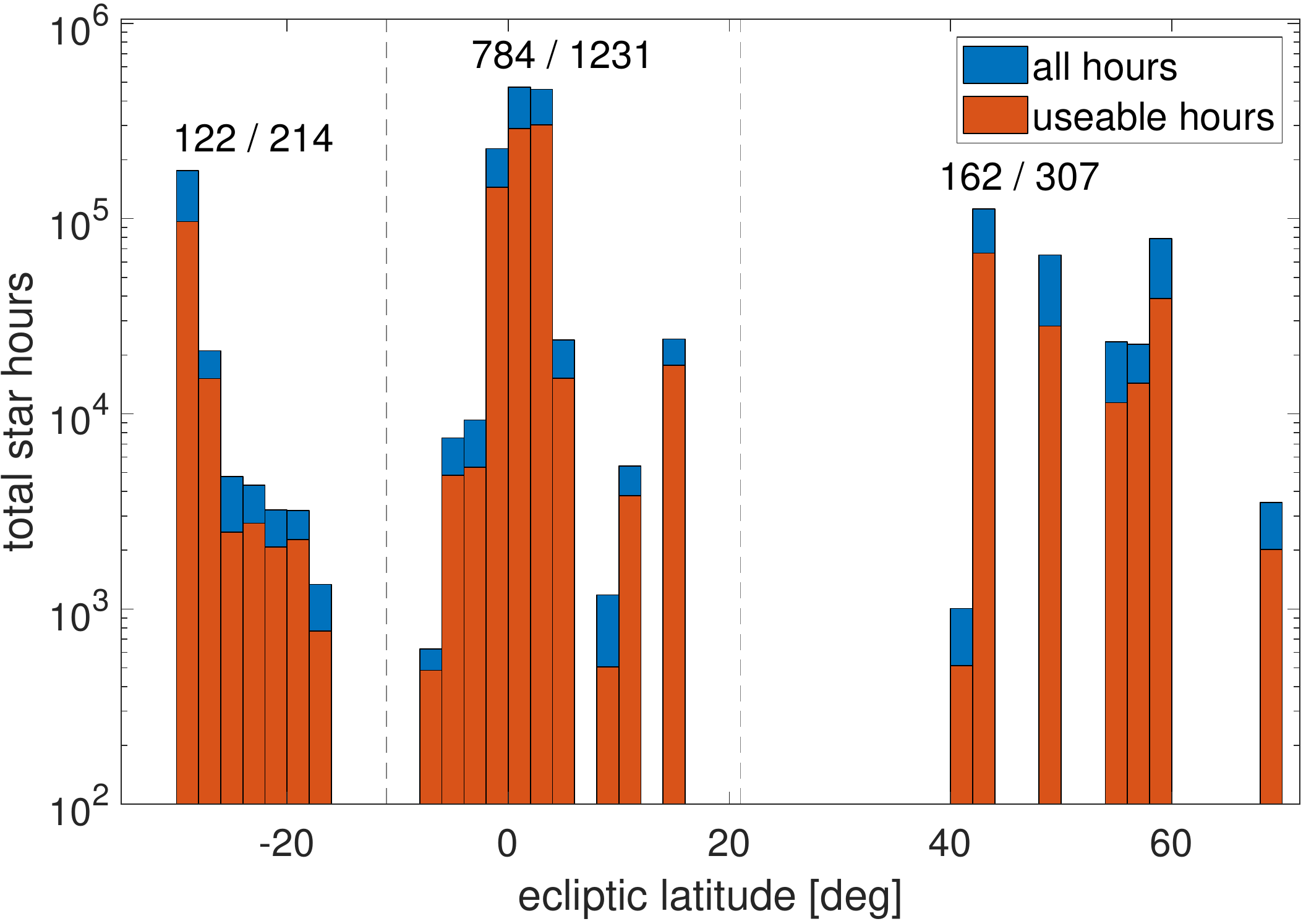}
    \caption{The number of star-hours in different ecliptic latitude bins. 
             The ecliptic latitude for all stars in a given field 
             is taken to be that of the center of the field 
             (with a width of $\approx 2.5$\,deg). 
             The bins in this figure are 2\,deg wide. 
             The red bars show the number of usable hours, 
             that were not lost to various quality cuts. 
             The blue bars represent the total star-hours observed. 
             In most bars the fraction of good data is about 60\% of the total. 
             We have clustered the observations into three sections:
             below ecliptic latitude $\beta<-11$\,deg, above $\beta>21$\,deg 
             and close to the ecliptic plane at $-11<\beta<21$\,deg. 
             The number of usable and total star-hours 
             is printed on the figure for each cluster (in thousands of hours).
             In the central cluster, 95\% of the star-hours are within $\pm 4$\,deg
             of the ecliptic plane. 
    }
    \label{fig:star-hours ecliptic latitude}
\end{figure}

During 2020, the telescope had been operating nightly from about May through September, 
with a few additional nights in October--November. 
During 2021, it had been operating continuously from April to November, 
with a few additional usable nights in December. 
The observations discussed in this work span approximately thirteen months of operations 
under generally good weather conditions. 
In total, the telescope had collected $\approx 840$ hours of usable, high-cadence data. 
The high cadence observations discussed in this paper were obtained using either 10 or 25\,Hz cadence. 
The 2020 data had all been taken at 25\,Hz, with 30\,ms exposure time (25\% dead-time). 
In 2021, following a software upgrade, the data were taken at 10\,Hz or 25\,Hz, 
with negligible dead-time. 
The total number of star-hours observed was $\approx 1.75\times 10^6$, 
of which about 60\% was usable, for a total of $\approx 10^6$ star-hours
(the remaining 40\% was lost to various quality cuts, as explained by \citealt{wfast_kbo_pipeline_Nir_2023}). 
About 40\% of the observations were lost due to various quality cuts
(see Table 1 of \citealt{wfast_kbo_pipeline_Nir_2023}). 
The distribution of star-hours across ecliptic latitudes is shown 
in Figure~\ref{fig:star-hours ecliptic latitude}. 
The star-hours for all stars in a field are recorded as though
the star's ecliptic latitude is given by the coordinates of the center 
of the field, not for the coordinates of each individual star. 
The bin widths in the plot are 2\,deg and the field of view of the camera is $\approx 2.5$\,deg. 
The star-hours in this figure were divided into clusters at ecliptic latitude
$\beta<-11$\,deg, $\beta>22$\,deg and $-11<\beta<22$\,deg. 
The usable and total star-hours in each cluster are printed on the plot.
In the central cluster, $\approx 95$\% of the star-hours
are 4\,deg from the ecliptic plane. 
In each bin, the number of usable star-hours is about 60\%
of the total star-hours. 
The observations taken off-ecliptic were used as a control group
for the KBO detection methods. 
There is a ratio of $\approx$1:3 between 
the control observations and on-ecliptic observations. 

\begin{figure}
    \centering
    \pic[1]{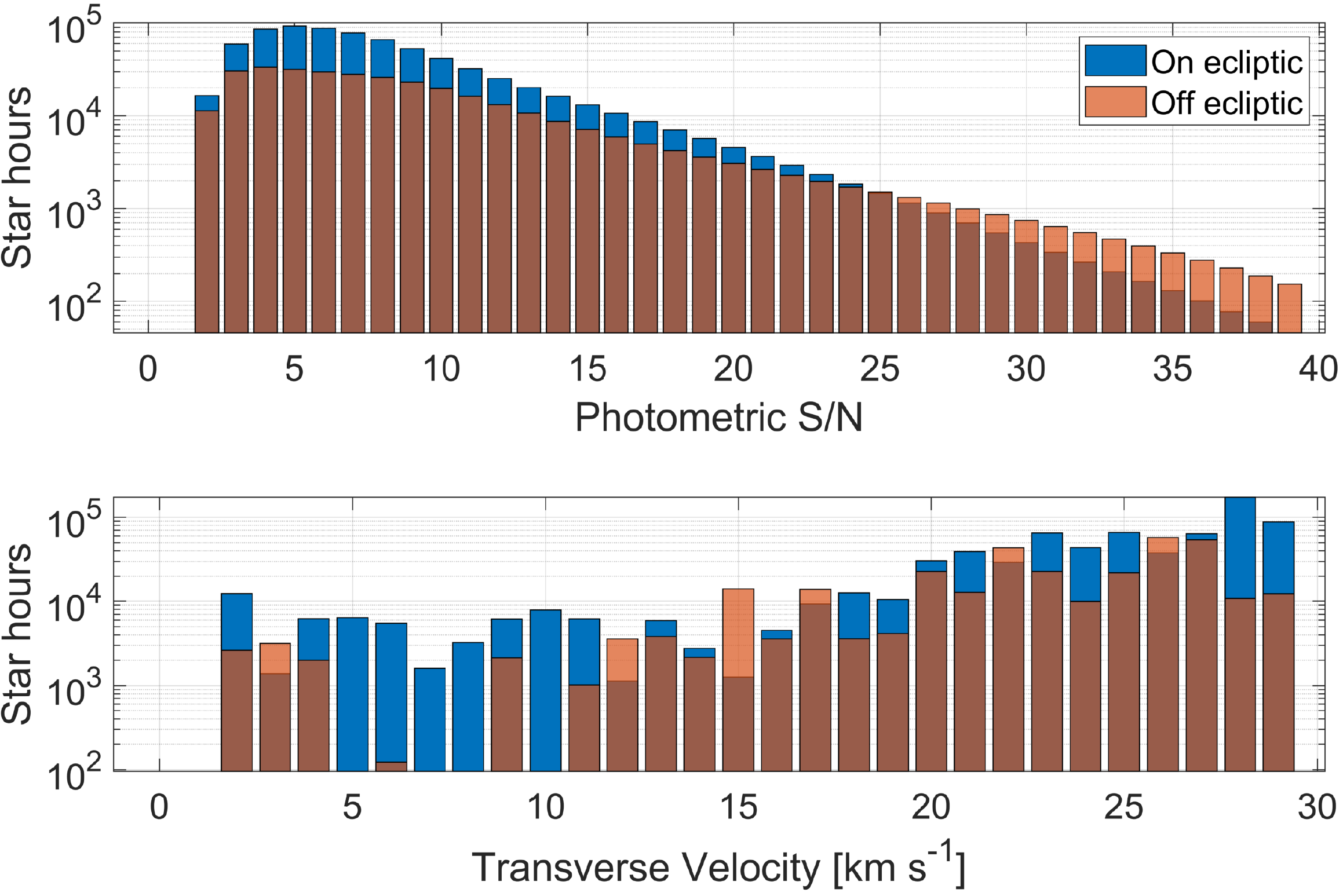}
    \caption{The number of (usable) star-hours as a function 
             of photometric S/N (top panel) 
             and transverse velocity (bottom panel). 
             The blue columns represent the star-hours taken 
             in fields close to the ecliptic, 
             while the red columns are for fields away from the ecliptic. 
             The ecliptic fields probably have lower S/N
             because they include many faint stars close to the 
             galactic center (which also coincides with the ecliptic)
             that contributed heavily to the total star-hours. 
             The transverse velocity shows that the survey was biased
             towards high-velocity fields, 
             with about ten times more star-hours on fields
             with transverse velocities above 20\,km\,s$^{-1}$.
             The limiting magnitude, corresponding to S/N$\approx 5$, 
             is 13.5 for 10\,Hz observations, and 12.5 for 25\,Hz observations. 
             While stars in that range are read-noise dominated, 
             stars with magnibrighter than $\approx 9$ are already dominated by
             scintillation and flat-field errors of 2--3\%, 
             such that the photometric noise never improves beyond S/N$\approx 50$.
    }
    \label{fig:star-hours snr vel}
\end{figure}

Since each star has different photometric properties, 
we also measured the distribution of star-hours 
across different photometric S/N values, 
which represents the noise in each individual measurement. 
The results are shown in the upper panel of Figure~\ref{fig:star-hours snr vel}. 
The histogram shows the distribution of star-hours for ecliptic and off-ecliptic fields. 
There is a correlation between the ecliptic fields and more numerous and fainter stars, 
since a large fraction of the star-hours were taken at the intersection of the ecliptic
plane with the galactic plane. These fields contain more stars, and quickly accumulate star-hours. 
Apparently, there are more faint stars in these fields, which makes for lower S/N 
compared to the off-ecliptic fields. 

Note that the S/N per measurement does not surpass $\approx 50$, 
despite some small fraction of stars being considerably bright. 
It seems that at high frame-rates the relative shot-noise decreases
as the photon count increases and at some point 
the error becomes a constant fraction of the flux, 
with a value of around 2\%. 
This irreducible photometric error could be due to atmospheric scintillations 
or differences in the gain of individual pixels that are not
treated completely by the flat field correction. 
Moving from 25\,Hz to 10\,Hz did not dramatically 
decrease this noise floor, so we will assume at least
part of the problem is due to flat-field errors. 
However, it should also be noted that some large
changes in the star's flux are sometimes seen, 
which appear to be atmospheric effects. 
We show some examples for such cases, where
the pixel brightness changes visibly even
on the same location on the sensor, 
when we discuss false-positives in \S\ref{sec:flare events}.

The observations also differ in their transverse velocity. 
This refers to the projection of the Earth's velocity at
the direction of the stars being observed. 
We assume all stars in the field have the same transverse velocity
as the center of the field at the time of observation. 
The total occultation velocity is the sum of Earth's projected velocity and the specific occulter velocity, 
$\approx 4.7$\,km\,\persec for circular, pro-grade orbits at 40\,AU. 
Since we do not know the specific KBO velocity \emph{a-priori}, 
and it is sub-dominant in most cases, we focus first on the Earth's
contribution to the overall velocity. 

The transverse velocity has two effects on the expected KBO yield:
(a) the rate of occultations grows linearly with the velocity, 
as more occulters traverse stars in a given time interval;
(b) the occultations become wider at low velocities, 
making them easier to detect as there could be more images 
taken while the flux is affected by the occulter
(see Ofek \& Nir, in prep.). 
These are two contradictory factors, and we attempted to
observe the ecliptic plane in a balanced range of velocities. 
Unfortunately, most of our observations are at high velocities, 
which has a big effect on the efficiency for detecting KBO occultations. 
The distribution of star-hours as a function of transverse velocity 
(taking into account only the Earth's projected motion) 
is shown in the bottom panel of Figure~\ref{fig:star-hours snr vel}. 
There are roughly ten times more star-hours at high velocities (20--30\,km\,\persec)
than at lower velocities. 
While this increases the expected number of occultations, 
it makes them hard to detect and discern them from data artefacts, 
as discussed in \S\ref{sec:false detections}. 
In the future, we intend to dedicate more time 
observing the ecliptic plane fields
far from opposition.

\section{Analysis methods}\label{sec:analysis}

We analyze each candidate event that has passed our 
detection threshold and quality cuts 
(see \citealt{wfast_kbo_pipeline_Nir_2023}). 
Briefly, we applied a matched-filter approach 
\citep{matched_filter_Turin_1960} using a template bank
made by simulating occultation lightcurves, 
including diffraction effects and finite star sizes. 
We corrected for the effects of correlated noise on 
the power spectral density using the data itself 
to estimate the noise in different frequencies. 
We checked each light curve for instances where
the flux is correlated between stars or correlated
to other indicators such as instantaneous point spread function width
or star centroid positions, among other quality cuts, 
and used those to disqualify data that was affected
by various systematics. 
Events that surpassed the detection threshold, 
i.e., time frames where the flux of any star, 
after being matched-filtered by any of the kernels, 
and normalized by the noise estimator, 
had a value higher than the threshold (nominally set to 7.5)
were inspected by one of the authors, 
and more false positives were ruled out. 
The remaining candidates are discussed in this work. 

For each candidate we used the parameter estimation presented by \cite{wfast_kbo_pipeline_Nir_2023}, 
and ran some additional tests that were added after the full analysis was done.
These are used to validate (and rule out) occultation events 
for various reasons (e.g., atmospheric scintillations, detector artefacts). 
We discuss these tests in the following subsections. 

\subsection{Parameter estimation}\label{sec:mcmc}

\begin{figure}
    \centering
    \pic[0.8]{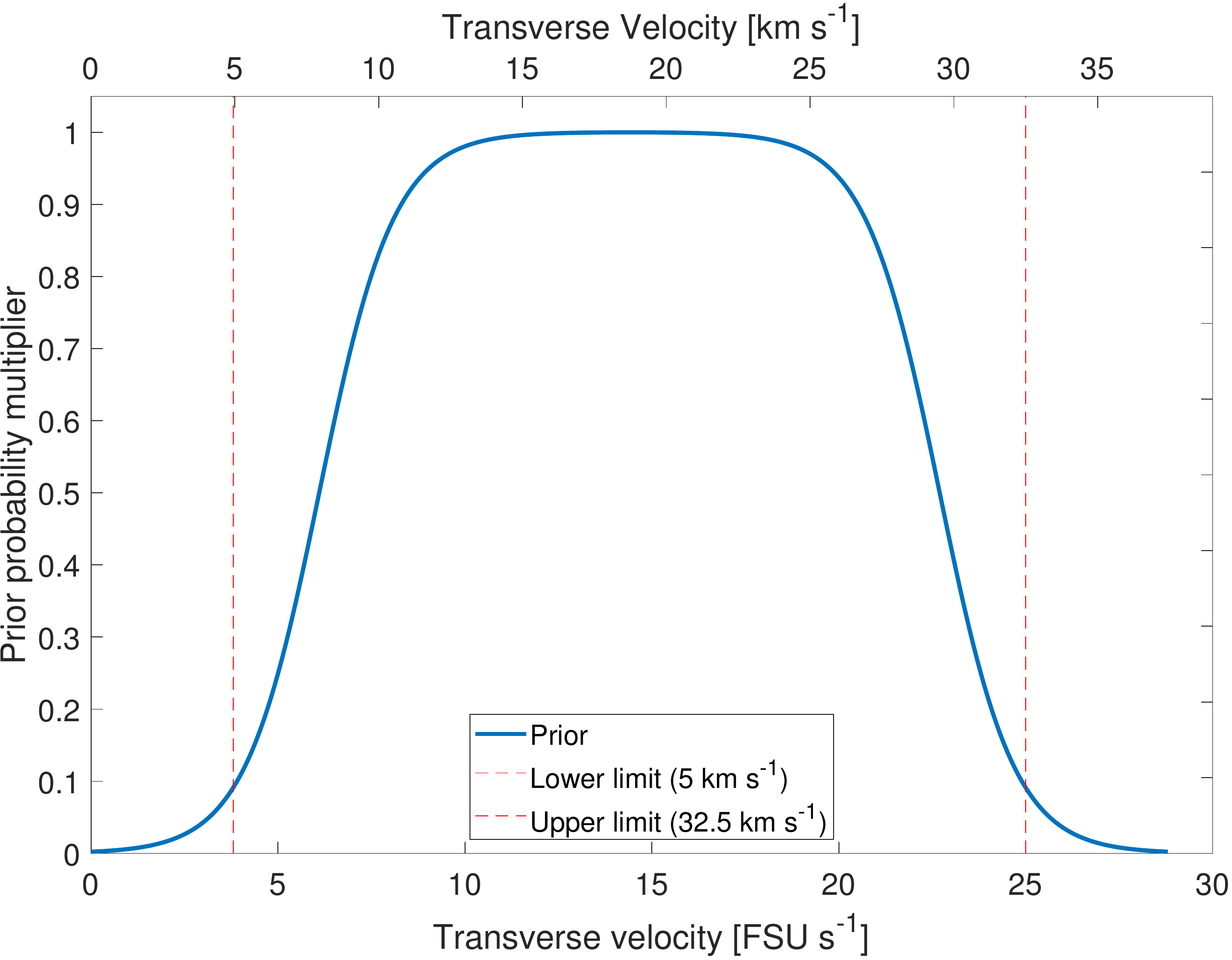}
    \caption{The prior used to limit the velocity parameter $v$
             given by Equation \ref{eq:velocity prior}.
             The exponential edges provide a soft stop
             for the MCMC sampler, without biasing the 
             results for velocities in the middle of the range. 
    }
    \label{fig:velocity prior}
\end{figure}

For each of the events we perform Markov Chain Monte Carlo (MCMC) sampling
to get an estimate of the distribution of occultation parameters posteriors, 
given the light-curve of each event. 
We follow the same analysis presented by \cite{wfast_kbo_pipeline_Nir_2023} with 
the following modifications: 
(a) we fit the time offset between the middle of the light-curve 
and the middle of the occultation template, 
allowing for some offset between the two, 
even though we begin the analysis by centering the light-curve
on the peak of the event (the point where the matched-filter result is maximal); 
(b) we set a non-uniform prior on the velocity 
(see Equation~\ref{eq:velocity prior} and Figure~\ref{fig:velocity prior}); 
(c) we run 20000 points in each chain, throwing away (`burning') half of each chain; 
(d) we use 50 chains with random starting positions, and use an additional rule to 
reject entire chains. 

\cite{wfast_kbo_pipeline_Nir_2023} had rejected chains where the $\chi^2$ per degree of freedom
was substantially larger than the median of all chains 
(three times the median absolute deviation of all chains). 
We adopt this but first disqualify chains where most of the templates
are so shallow they are essentially undetectable.
To quantify this rule, we measure the median 
signal of the templates in the chain.  
We define the signal of a specific template as
\begin{equation}
    S = \sqrt{\sum_i (f_i - \langle f \rangle)^2 \sigma_\text{phot}},
\end{equation}
where $f_i$ is the template flux in frame $i$ 
(normalized such that the flux outside the occultation is 1), 
$\langle f \rangle$ is the mean flux of the template, 
and $\sigma_\text{phot}$ is the measured noise of the actual event. 
$S$ is an estimate of the S/N that would be measured
for such an occultation, 
triggering by exactly the same template 
(an ideal matched-filter). 
We rule out any chain with a median $S<3.5$, 
which is much lower than our actual 
detection threshold of 7.5. 
The decision to rule out chains based on the 
different values of $\chi^2$ is made using only
chains that survived the signal rule. 

The goal of the priors used for the five parameters of the fit
is to limit the parameters to reasonable values
that can arise from real occultations, 
and reflect our previous knowledge of the event characteristics. 
We attempted not to bias the results and use either flat
or nearly flat priors where possible. 
For example, we chose to leave a uniform prior on the
occulter radius $r$, despite knowing that the size 
distribution is generally fast dropping with occulter radius. 
The range of allowed values is $0.1 \leq r \leq 3$\,FSU 
(Fresnel Scale Unit, which is the typical size scale for diffractive occultations
given by $\sqrt{\lambda D /2}=1.3$\,km for our observations, 
using the observed wavelength $\lambda=550$\,nm 
and the estimated distance to the Kuiper belt of $D=40$\,AU). 

The impact parameter, $b$, and the time offset $t$
are both nuisance parameters and have no physical reasons
to be close to any value. 
We used a flat prior on $b$, but because we are first centering 
the light-curve to fit the template, 
we assumed the time offset should be small, 
and applied a Gaussian prior, 
\begin{equation}
    \mathcal{P}(t) \propto \exp(-\frac{t^2}{2 \sigma_t^2}), 
\end{equation}
where $\sigma_t=100$\,ms. Note that the priors are not normalized because the Metropolis-Hastings algorithm only ever uses the probability ratios between points so the normalization of the priors generally cancels out.
The prior on the stellar size, $R$, 
is also given as a Gaussian, reflecting 
our belief in the successful fit of the star's temperature
based on the measured Gaia colors:
\begin{equation}
    \mathcal{P}(R) \propto \exp(-\frac{(R-R_0)^2}{2 \sigma_R^2}), 
\end{equation}
where $R_0$ is the fit result and $\sigma_R=R_0/10$, all in Fresnel units. 
The colors are based on cross-matching the star's position with the Gaia data release 2 \citep{GAIA_data_release_two_2018}.
We used a velocity prior that was not used by \cite{wfast_kbo_pipeline_Nir_2023}.
The shape of the prior is set to a two-sided exponential function 
with a flat plateau in the middle:
\begin{equation}\label{eq:velocity prior}
    \mathcal{P}(v) \propto P_0 + \exp(\frac{v - v_h}{\sigma_v}) + \exp(\frac{v_\ell - v}{\sigma_v}), 
\end{equation}
where $v_\ell=3.8$\,FSU\,s$^{-1}$ ($\approx 5$ km s$^{-1}$) 
and $v_h=25$\,FSU\,s$^{-1}$ ($\approx 32.5$ km s$^{-1}$) 
are the edges of the range of likely values 
of transverse velocities of the Earth-KBO system, 
and $P_0=0.1$ is a constant that tunes the size of the flat
region in between the edges. 
A plot of the velocity prior values is shown in 
Figure \ref{fig:velocity prior}. 
This prior allows for limiting the sampler from 
going into un-physical values without placing a hard stop on the chains, 
but at the same time does not heavily bias the resulting values
towards high or low velocities inside the allowed range. 

Finally, we are mostly interested in the posterior results
for the physical parameters of the occultation, $r$ and $v$, 
so we present the posteriors after marginalizing over the
remaining $t$, $b$ and $R$. 
For the single parameter marginal posteriors we find 
the shortest interval containing 68\% of the distribution
and use that as confidence intervals, as discussed in \cite{wfast_kbo_pipeline_Nir_2023}. 
We use the parameter estimates to check if the light-curve
is consistent with the transverse velocity of the Earth, 
for the field in which the event was detected. 
E.g., the MCMC analysis results for the first candidate event
are shown in Figure~\ref{fig:mcmc 2020-07-01}.

\subsection{Nearest neighbor analysis}\label{sec:nearest}

During regular analysis we disqualify events
that occur at a time when other stars in the field
show similar features to the occultation candidate
(See \citealt{wfast_kbo_pipeline_Nir_2023}). 
For the short list of candidates we have left 
we also check if nearby stars show flux variability 
at slightly earlier or later times. 
This may happen if any atmospheric event passes across
multiple stars but not exactly at the same time
(e.g., a cloud or focusing/defocusing bubble of air). 

The typical angular distance between bright stars in our field
is 1--6 arc-minutes. 
Assuming the typical velocity is 15 arc-seconds\,s$^{-1}$, 
it is also reasonable that stars separated
by $\sim 3$ arc-minutes will be affected 
by a localized atmospheric event
within a time interval of $\sim 10$\,s. 
This is the time-scale of a double batch
of 200 frames (8 or 20\,s, depending on the cadence). 

We choose the closest stars to the occulted star, 
that have a photometric $S/N>5$, 
and plot the flux for that batch of 200 frames
when the candidate event occurred. 
This qualitative method gives us another chance 
to find localized atmospheric events that would
otherwise not trigger our data quality cuts
because they do not occur simultaneously on different stars. 
We show the nearest neighbor analysis for one of the 
likely-atmospheric events in Figure~\ref{fig:neighbors 2020-07-01}. 

\subsection{Outlier analysis}\label{sec:outliers}

Another possible false positive could be caused by
instrumental effects (sensor defects, telescope reflections, etc.). 
To check if a candidate event's flux is stable over longer 
periods outside the 200 frames where it was detected, 
we loaded the long term flux values for all stars 
in a region of 2000 frame before and after the event. 
This translates to 80 or 200 seconds of data in either direction, 
for 25 or 10\,Hz cadences, respectively. 

For each star we subtract a smoothed flux, 
that we get using a Savitzky-Golay filter 
with a 3rd order polynomial and a 101 frame window size. 
The values of this mean-subtracted flux are iteratively fit to a Gaussian distribution, 
while removing any outliers of more than three times the noise RMS. 
We count the number of outliers removed by this \emph{sigma clipping} procedure
and check if the target star has an unusually large number of outliers ($>50$)
compared to other stars during that time period 
(typically 20--30 outliers in the time range we check). 

Stars with many outliers can be disqualified as having 
unstable flux over time (e.g., due to a nearby bad pixel).
We also display the distribution of stars with many outliers
on the image plane, to see if any clustering of unstable stars
is evident on the camera's sensor. 
An example for such an outlier analysis is shown for one of 
the likely-atmospheric events in Figure~\ref{fig:outliers 2020-07-01}.

\subsection{A note about main-belt and near Earth occulters}

Some care should be taken when deducing the physical parameters of an occulting object
based on the estimated parameters of an occultation. 
Throughout this work we assume the occulter is in the Kuiper belt, 
with a nominal distance of 40\,AU. This simplifies the parameter estimation process 
but a more careful analysis should be done on any real occultations, 
including allowing the distance (and thus, the Fresnel scale) to vary as well. 

A considerably smaller distance could occur when considering other occulter populations. 
An occultation caused by a main belt asteroid could resemble a KBO occultation
in terms of the width and depth of the event. 
Taking as an example an object with a circular orbit with a radius of 2.5\,AU, 
the Fresnel scale would be $\approx 250$\,m (as seen from Earth, at opposition). 
The relative velocity of such objects would be about 44 FSU\,\persec, 
which can be confused with high velocity events in fields with high transverse velocity. 
However, the number of main belt asteroids is estimated to be about $10^7$ objects above 250\,m
\citep{asteroids_population_Davis_2002, asteroids_population_Subaru_Maeda_2021}. 
Spread out over thousands of square degrees of the main belt, 
the estimated density would be $10^3$--$10^4$ objects per degree squared. 
So the number density per square degree of main belt occulters 
is at least an order of magnitude smaller than the estimate of $\approx 10^5$ 
km-scale KBOs per square degree given by \cite{abundance_kuiper_objects_Schlichting_Ofek_2012}. 

Near Earth objects would have an even smaller Fresnel scale, 
and thus, in most cases, occultations that are substantially shorter than the W-FAST exposure time. 
Their population is also considerably smaller and the chance for occultations is
expected to be lower than that of the main belt population.

\section{Results}\label{sec:results}

We present the results of the analysis pipeline over the data collected in 2020--2021. 
We summarize the status of a few occultation-like false positives
which are ruled out for various reasons. 
We then discuss the efficiency for detecting occulters of various sizes and 
estimate the expected number of detections based on existing size distribution power law models. 

\subsection{Occultation candidates}\label{sec:candidates}

Our search resulted in eight occultation candidates passing
both the quality cuts and the human inspection, 
as described in \cite{wfast_kbo_pipeline_Nir_2023}. 
All these events were eventually rejected 
based on deeper inspection of the data. 

One of these events, upon second inspection, 
was shown to be close to a warm pixel, 
which produced slightly higher values 
but was not flagged by the data pipeline. 
We summarize the properties of all remaining seven candidates
in Table~\ref{tab:candidate summary}, 
where it is clear that none of the candidates
is fully consistent with what we expect from a KBO occultation. 
Some are clearly false positives, 
such as events with many outliers in the long term light-curves
(this includes the events of 2021-04-11, 2021-04-12 and 2021-04-16). 
Note that the candidates' star in these cases is subject to an 
unusually high number of flux outliers. 
Most other stars in the same dataset do not show many outliers
(more than 90\% of the stars, as seen in the top panels of the relevant figures, 
e.g., Figure~\ref{fig:outliers 2021-04-11}). 
Thus we are confident that the flux outliers are the cause of these
candidates, and not merely a background phenomena that coincidentally
accompanied these events. 

Other candidates seem likely to be atmospheric, such as the events
of 2020-07-01 or 2021-09-14 that show both a decrease and an increase
of flux, appearing ``flare-like'', which makes them suspiciously similar
to a sample of 30 flare events, with sub-second time-scales, 
that are most likely to be non-astrophysical, but a mixture
of atmospheric events and satellite glints \citep{satellite_glints_Nir_2021}. 
These flares are discussed further in \S\ref{sec:flare events}. 

A single event was detected in an off-ecliptic field (the event at 2020-07-01)
making it unlikely to be a KBO occultation. 
It is also the only event detected in the 2020 data. 
The remaining six events were taken in 2021 on ecliptic fields. 
Five of the events occurred in the first half of April 2021, 
raising further suspicion that they are not real occultations. 
Note that the flare events were found in both the 2020 and 2021 data sets. 

\newcommand{\tableaddedspace}{\rule{0pt}{2em}}
\begin{landscape}
\begin{table}
    \caption{Summary of candidate occultations. 
             The ecliptic latitude ($\beta$) is given for the center 
             of the field of view of the observations. 
             The event S/N is the integrated matched-filter result 
             for the entire event, while the photometric S/N is the 
             single-frame flux vs.~noise measured over the previous 2000 frames. 
             The stellar size ($R_\star$) is given in Fresnel Scale Units (FSUs), 
             and was calculated using a fit to Gaia DR2 colors. 
             The FWHM (Full Width at Half Maximum) is given in arc-seconds, where the combined
             atmospheric and instrumental seeing is commonly in the 4.5--$5.0''$ range. 
             The airmass and transverse Earth velocity ($V_\text{Earth}$) 
             are calculated for the center of the field at the time of the event. 
             The parameter ranges for occulter radius $r$ and occultation velocity $v$
             are from the MCMC fit (see \S\ref{sec:mcmc}). 
             In cases where the posterior is bi-modal we show two options
             for the ranges as (i) and (ii). 
             Events that are marked as having ``neigh.~stars'' are those
             where there is some variability detected by eye in the 
             light-curves of the neighboring stars. 
             Events that are marked as ``flare-like'' display strong 
             flux increase in addition to dimming. 
             The number of outliers is calculated in the 4000 frames
             around the event time, where most stars will 
             produce 20--30 outliers just due to statistical fluctuations.              
    }
    \label{tab:candidate summary}

        \begin{tabular}{ccccccccccccccccc}
    
        \hline
        Event time (UTC) & coordinates & $\beta$ & event & $M_G$ & $T_\text{eff}$ & $R_\star$ & phot. & $f$  & FWHM     & air- & $V_\text{Earth}$ &  $r$ range & $v$ range     & neigh. & flare- &  num.     \\ 
                         &             & (deg)   &  S/N  &       & ($^\circ$K)    & (FSU)     &  S/N  & (Hz) & (arcsec) & mass & (km\persec)      &  (km)      & (km\,\persec) & stars  &  like & outliers \\ \hline

        \tableaddedspace 2020-07-01 20:58:01 & 19:32:42.6+21:25:23.4 &  42.8 &  7.63 &  9.72 & 5342 &  1.1 & 22.41 &   25 &  5.34 &  1.09 &  27.2 &                                                       1.9--3.4 & \begin{tabular}{@{}l@{}} (i) 8--10 \\ (ii) 13--28\end{tabular} &  No & Yes & 25 \\ 
        \tableaddedspace 2021-04-01 17:29:08 & 06:31:34.0+20:33:54.8 &  -1.7 &  7.54 & 11.41 & 6640 &  0.4 & 18.64 &   10 &  5.65 &  1.13 &   4.0 &                                                       0.5--1.5 &  19 -- 28  & Yes &  No & 21 \\ 
        \tableaddedspace 2021-04-03 17:00:36 & 06:31:57.7+20:21:56.8 &  -1.7 &  8.42 & 11.89 & 6464 &  0.5 & 19.24 &   10 &  5.18 &  1.09 &   5.0 &                                                                                                               1.4--3.3 & 22--29 &  No &  No & 19 \\ 
        \tableaddedspace 2021-04-11 18:36:45 & 10:20:42.6+09:00:32.8 &  -1.7 &  7.65 & 12.92 & 6488 &  0.3 & 10.64 &   10 &  4.84 &  1.08 &   9.5 &                                                                                                               0.6--0.9 & 15--21 & Yes &  No & 79 \\ 
        \tableaddedspace 2021-04-12 20:22:19 & 10:22:38.2+07:27:50.7 &  -1.7 &  7.54 & 12.80 & 5388 &  0.4 &  8.49 &   10 &  5.26 &  1.19 &   9.3 & \begin{tabular}{@{}l@{}}(i) 1.0--2.5 \\ (ii) 0.6--1.0\end{tabular} & \begin{tabular}{@{}l@{}} (i) 17--30 \\ (ii) 6--8.5\end{tabular} & Yes &  No & 62 \\ 
        \tableaddedspace 2021-04-16 20:45:56 & 13:32:52.1-09:48:57.8 &   0.0 &  8.07 & 11.94 & 5886 &  0.4 &  8.31 &   25 &  4.58 &  1.37 &  21.2 &                                                                                                               0.6--2.1 & 17--25 & Yes &  No & 119 \\ 
        \tableaddedspace 2021-09-14 17:58:01 & 22:00:23.2-11:34:15.3 &   1.3 &  7.53 & 11.84 & 5961 &  0.6 & 11.53 &   10 &  7.00 &  1.61 &  20.3 &  0.55 -- 0.9 & 7 -- 8.5                                      & No & Yes & 21\\ \hline

    \end{tabular}

\end{table}
\end{landscape}

Two events are single-frame events where there is insufficient information 
to confirm or rule out the events, but the low frame rate and 
low projected Earth velocity make them unlikely to be KBO occultations. 
For several events, there is some level of inconsistency between 
the field's projected Earth velocity 
and the best fit velocity of the putative occultation. 
Simply put, the dip in the light-curve was either too narrow or too wide
for the expected event duration.
The expected Earth's projected velocity can be compared to 
the MCMC fit to each event in Table~\ref{tab:candidate summary}. 
Note that the Earth's projected velocity does not include
the unknown occulter velocity 
(for a pro-grade, circular orbit at 40\,AU, 
the occultation velocity will be smaller than the Earth's projected velocity
by $\approx 4.7$\,km\,\persec). 
Because we see several events that are inconsistent with the Earth's projected velocity, 
we cannot assume that the events that are consistent are true occultations, 
as they can, just as likely, be artefacts that just happened to occur on a field
with the ``correct'' projected velocity. 
In any case, in Table~\ref{tab:candidate summary} we show that even the events with a consistent 
velocity are ruled out by other tests, 
such as the number of outliers or variability in nearby stars' flux. 

We discuss each occultation event separately in \S\ref{sec:individual candidates}, 
and rule out each of them as being a true occultation. 
In \S\ref{sec:flare events} we discuss a sample of short-duration flux-increases (``flares''). 
Assuming some of the flares in that sample are atmospheric, 
further strengthens the hypothesis that some of our occultation candidates
are also caused by atmospheric events.

\subsection{Efficiency for detecting occultations}\label{sec:efficiency}

A rough estimate of the number of expected detections for each star
can be calculated using 
the average transverse velocity, the maximum impact parameter, 
and the surface density of occulters. 
The transverse velocity can be taken to be roughly 20\,km\,\persec, 
The impact parameter is the angular
distance at closest approach between 
the center of a star and the center of an occulter. 
We can estimate the maximum impact parameter at which the event
is still detectable to be about a Fresnel-scale, $\approx 1$\,km 
(projected at the distance of the occulter). 
Thus each star subtends a strip of width 2\,km and length 20\,km every second. 
At a distance of 40\,AU, this covers an angular size 
of $\approx 1.3\times 10^{-11}$\,deg$^2$ every hour. 

The measured angular density of KBOs, given by \cite{abundance_kuiper_objects_Schlichting_Ofek_2012}, 
is $N(r>250\text{m}) = 1.1 \times 10^7$\,deg$^{-2}$, 
with a differential power law index $q=-3.8$,
so the cumulative distribution is $N(>r)\propto r^{-2.8}$. 
If we assume we can detect occulters with a radius of 0.5\,km, 
the number density becomes $\approx 1.6\times 10^6$\,deg$^{-2}$. 
Thus, we will require $\approx 50,000$ star-hours to detect a single occultation. 
If we can only detect occulters of radius 1.5\,km or larger, 
the surface density becomes $7.3\times 10^4$\,deg$^{-2}$, 
and the number of star-hours per detection becomes $\approx 10^6$. 
Using these rough estimates it seems that with 750,000 star-hours 
the survey presented in this work should be able to detect a few 
occultations with an occulter radius of $\sim 1$\,km. 

\begin{figure}
    \centering
    \pic[1]{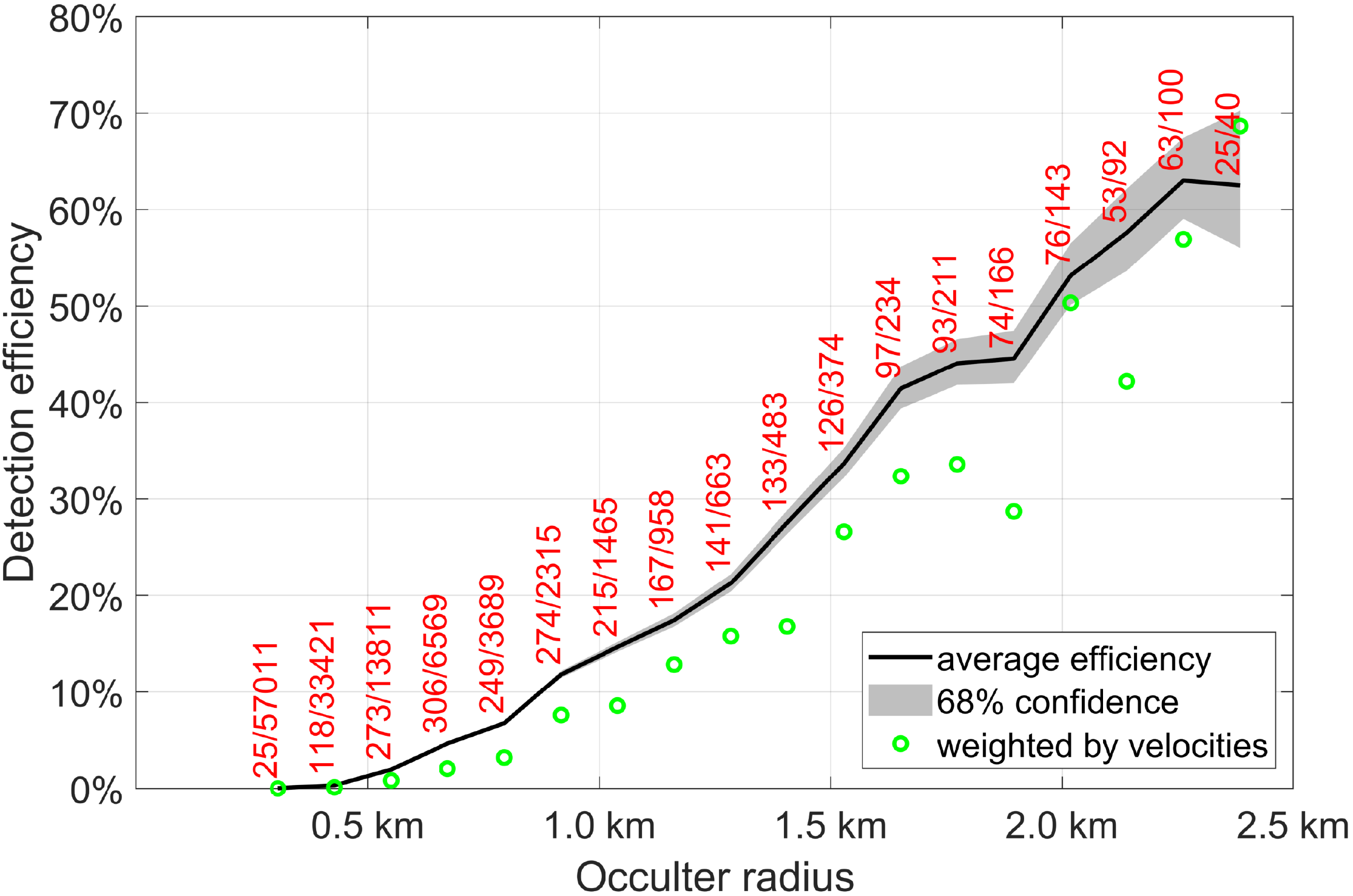}
    \caption{The detection efficiency for occulters of various sizes
             over the data collected during the 2020--2021 observing seasons. 
             The black line represents the overall efficiency, assuming 
             a uniform distribution of transverse velocities of observations. 
             The grey areas represent the uncertainty due to Poisson (counting) errors. 
             The red numbers show the number of simulated, injected events in each size category
             (detected and total number of events). 
             The events were generated using a flat distribution of velocity and impact parameter. 
             The stellar S/N and angular size were taken from 
             the actual star's light-curve and Gaia data, respectively.    
             The occulter radius was drawn from a steep power law distribution. 
             The distribution of impact parameters for real occultations is expected to be uniform, 
             but the distribution of velocities in our observations is weighted towards higher velocities
             (see Figure~\ref{fig:star-hours snr vel}). 
             A weighted average efficiency is shown as green circles, 
             showing that the real efficiency in our dataset is lower than
             suggested by the raw detection fractions across all events. 
             In any case, it is evident that the detection rate at small radii ($<1$\,km)
             is only a few percent at best. 
    }
    \label{fig:efficiency}
\end{figure}

A more careful analysis needs to also take into account
the efficiency for detecting occulters of different radii. 
During the analysis, we have systematically injected simulated
occultations with different parameters into the raw light-curves, 
and tested what fraction of the events were detected. 
Using $\approx 10^5$ simulated events we calculate the detection efficiency
at each occulter radius (with bins of size 0.1\,km). 
See \cite{wfast_kbo_pipeline_Nir_2023} for more details on these simulations. 
The results are shown in Figure~\ref{fig:efficiency}. 
The black line represents the average efficiency for that
occulter radius bin, with grey areas representing the Poisson
(statistical) uncertainties associated with the number
of detected events. 
The detected and total number of simulated events in each bin 
are shown as red numbers above the line. 

Each simulated event has parameters drawn from random distributions. 
The transverse velocity and impact parameter are chosen from a uniform distribution, 
while the radius of the occulter is chosen from a steep power law 
that is both more realistic and helps generate enough detections 
even at the smaller sizes where events are hard to detect. 
The photometric S/N is taken directly from the light-curve, 
and the stellar angular size is calculated using the star's 
cross-match to the Gaia DR2 database \citep{GAIA_data_release_two_2018}.
Since stars are chosen randomly from the stars tracked in each field, 
the distributions of stellar size and S/N 
are representative of the sample as a whole. 
The uniform velocity distribution is not a good representation of our data, 
since most of the data has been collected at high transverse velocity ($>20$\,km\,\persec). 
Using the distribution of star-hours in different velocity bins, 
we re-weigh the efficiency to calculate more realistic (and lower) 
efficiency values and show them as green circles on the plot. 

The efficiency for detecting small occulters, 
with radius $<1$\,km, is only a few percent, 
either when averaging the detections of all velocities, 
or when re-weighing based on the time spent in each velocity. 
It is thus more obvious why the simple estimates
made at the beginning of this section are highly-optimistic. 
In fact, such low efficiency means more than an order of magnitude
more star-hours are required to make a single detection. 

We also note that 
the injected events are only put into regions of the light-curve 
where there are no quality flags or other event candidates. 
So, the efficiency fraction does not include the fraction of time that is flagged as ``bad data''. This is accounted for when calculating the coverage by only including the ``good data'' when estimating the total coverage and number of detections. 
Furthermore, each simulated event has been vetted by a human ``scanner'', 
that also may reduce the number of detected events. 
We noticed that more than 90\% of simulated events that had triggered
were also correctly identified by the vetting process, 
so we ignore the contribution of the human in estimating the efficiency. 

To estimate the total number of expected occultations, 
we calculate the efficiency separately for each value of the relevant parameters, 
and multiply that by the number of star-hours in that parameter bin. 
The coverage is the number of square degrees that were monitored, 
in which some occulter radius range would have been detected, 
weighed by the efficiency relevant for that selection of stars, 
transverse velocity and impact parameters:
\begin{equation}\label{eq:coverage}
    \Omega(r) = \int_{-b_\text{max}}^{b_\text{max}}\int_{\text{obs}} \varepsilon(r, v(t)) v(t) dt db, 
\end{equation}
where $v(t)$ is the projected velocity in each observation, 
$\varepsilon$ is the detection efficiency for each occulter radius $r$
(calculated from injection simulations) 
and $b_\text{max}=2$\,FSU is the maximum impact parameter 
where we injected simulated events 
(most events with $b>b_\text{max}$ will not be detectable). 
The integral over $b$ is simply $2b_\text{max}$, 
and the integral over time is simply done by summing 
the star-hours at different velocities. 
For the coverage presented in this work we used only
star-hours collected at a distance of $\pm 5$\,deg from the ecliptic plane, 
and assume the density of KBOs in that region is uniform. 

\begin{figure}
    \centering
    \pic[1]{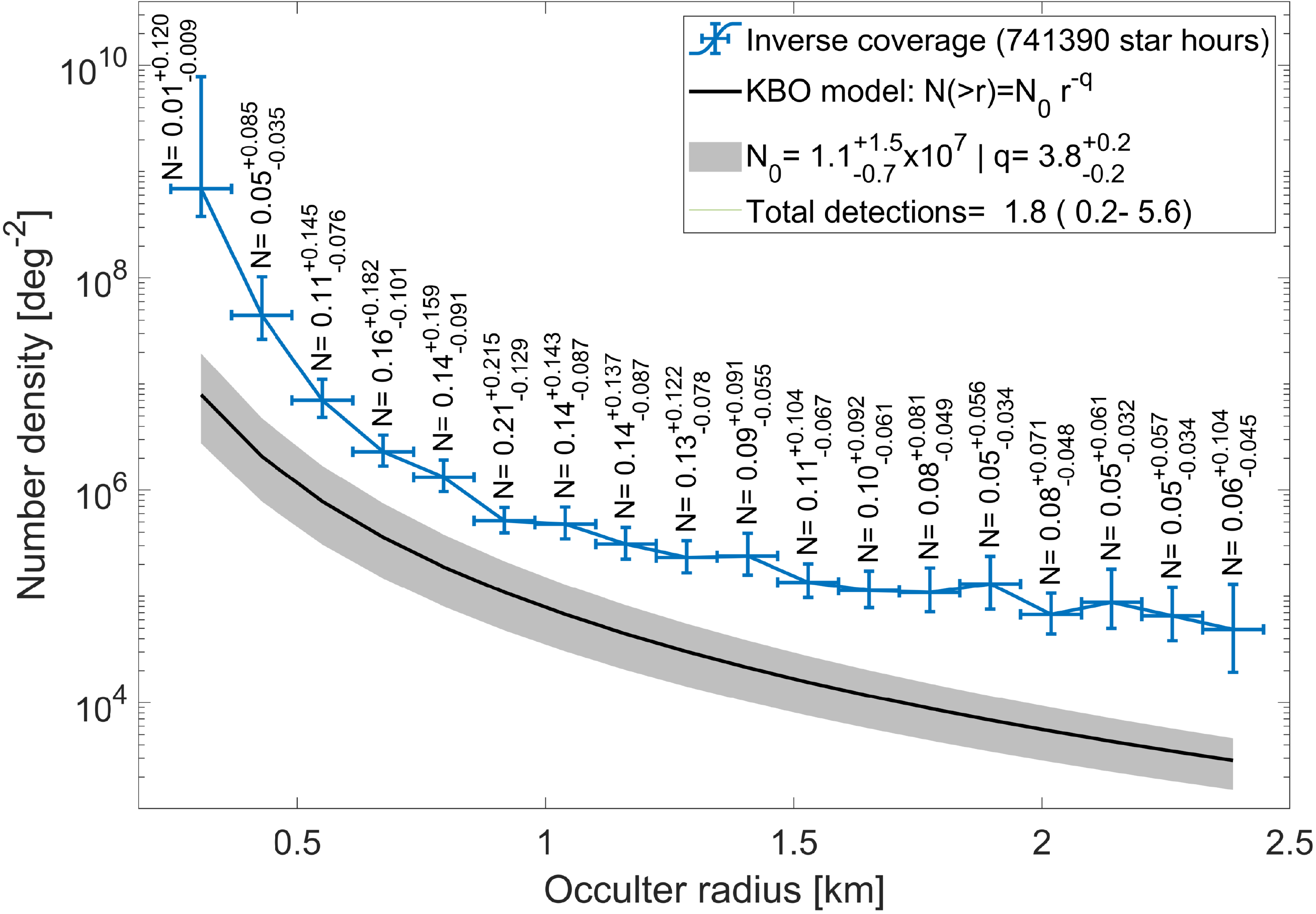}
    \caption{Inverse coverage as a function of occulter radius, 
             compared to the expected KBO number density (per radius bin of size 0.12\,km). 
             The blue crosses represent the inverse number of deg$^2$
             covered by the current dataset, including the efficiency for detection, 
             at each given occulter radius bin, 
             considering the star-hours taken at different observational parameters. 
             The horizontal errors simply span the width of each size bin, 
             while the vertical errors include the statistical errors due
             to the counting (Poisson) errors on the number of detected
             simulated events that were used to estimate the efficiency in each bin. 
             The black line and grey area represent the best model for the
             number density of KBOs close to the ecliptic given by
             \protect\cite{abundance_kuiper_objects_Schlichting_Ofek_2012}, 
             along with the uncertainties on the total number and the power law index. 
             The numbers above the crosses are the expected number of detections
             in each size bin, including errors from both the efficiency and 
             the KBO density model. 
             The total number of expected detections integrated over all sizes 
             in this plot is $1.8_{-1.6}^{+3.8}$. 
    }
    \label{fig:coverage}
\end{figure}

To compare the coverage to the expected number density
of occulters, we plot the inverse of the coverage in 
Figure~\ref{fig:coverage}. 
Counting only observations close to the ecliptic, 
we get a total of $\approx 740,000$ star-hours. 
The blue crosses represent the inverse coverage, 
with horizontal errors spanning the radius interval of each bin, 
and the vertical errors representing the statistical uncertainty 
derived from the efficiency calculation, which themselves come
from the counting statistics of injected detections. 
The black line and grey area represents the KBO model
given by \cite{abundance_kuiper_objects_Schlichting_Ofek_2012}, 
including the uncertainties of both the overall number of objects
and the power law index. 
The numbers above each radius bin represent the 
expected number of detections for that bin alone, 
with uncertainties calculated as the quadratic sum 
of statistical uncertainties from the efficiency and
the KBO number density. 
It is evident that, using the KBO density model, 
we expect to get less than one event in each bin. 
At low radii, there are many more occulters, but the low
efficiency reduces the coverage dramatically, 
as evident from the rise of the blue crosses on the left. 
For higher radii, the efficiency approaches 100\% but
the number of occulters is expected to drop quickly. 
The most sensitive part of the radius distribution is around 1\,km, 
but not by a large factor. 
The total number of expected detections, 
including KBOs of all sizes in this range, 
is $1.8_{-1.6}^{+3.8}$ events. 
The combined uncertainties of both the efficiency and the density model, 
make this number consistent with either five events on the optimistic side, 
and with zero events on the pessimistic side 
(with 68\% confidence intervals). 
If events were detected, they are not substantially 
more likely to be detected with a certain occulter radius, 
but with a slightly higher probability to be in the range of $0.5<r<1.5$\,km. 
It is also evident that if the number density had a dramatic increase above 1\,km, 
e.g., a break to a shallower power law that increases by a factor of a few
the number of occulters in the $1<r<2$\,km range, 
it would be improbable that we would not find any occultations in our data. 

Finally, we show that after treating the light-curves
for red noise, removing segments of bad data, 
and setting a high threshold for detection 
required to reduce the number of false-positives, 
we are left with a low efficiency, 
that can dramatically increase the number of 
star-hours required for even a handful of detections. 
The naive expectation that a few tens of thousands of star-hours
are sufficient to detect a KBO occultation underestimates
the difficulty of this kind of search from the ground, 
especially using a single telescope.\footnote{Even with multiple telescopes, the threshold
for detection would not necessarily drop substantially, 
although the additional telescopes would make it easier to 
identify which events are real (See \S\ref{sec:false detections}).} 
With more than a million star-hours we are still left 
without a single confirmed detection. 
We discuss the implication of our results in \S\ref{sec:conclusions}.

\section{Discussion}\label{sec:discussion}

\subsection{Limits on the km-scale KBO number density}\label{sec:limits}

Assuming none of the candidates in our sample is a real KBO occultation, 
we can place upper limits on the underlying density, 
using the efficiency calculated from the fraction of detected simulated events. 
The model we have used in Figure~\ref{fig:coverage}, 
adapted from the results of \cite{abundance_kuiper_objects_Schlichting_Ofek_2012},
already represents a good upper limit estimate, 
as the integrated number of events we expect from 
that model and our efficiency estimate is $1.8_{-1.6}^{+3.8}$ events. 
This is still consistent with zero detections, 
but a model with substantially larger density, 
particularly at $r>1$\,km, would already be 
in tension with our findings. 

A more quantitative calculation allows us 
to draw a contour of the 95\% confidence upper bound, 
based on the coverage in each occulter size bin. 
The coverage calculated in Equation~\ref{eq:coverage}, 
shown in Figure~\ref{fig:coverage}, 
can be used to calculate the expected number of KBO occultations
in either each radius bin, or integrated over all sizes 
where our survey is sensitive:
\begin{equation}
    N_\text{expected} = \int_{r_\text{min}}^{r_\text{max}} \Omega(r) n(r)dr,
\end{equation}
where the differential number of KBOs per solid angle is given by $n(r)\propto r^{-q}$.
We probe the range $0.25<r<2.5$\,km, which is 
the range that we used in the injection simulations, 
and also contains most of the detection potential. 
The individual estimated number of detections for each 
radius bin is shown above each point in Figure~\ref{fig:coverage}, 
assuming the model from \cite{abundance_kuiper_objects_Schlichting_Ofek_2012}. 

The lack of detections allows us to calculate the upper limits 
in various scenarios. 
In the following calculations,
we calculate 95\% upper limits by 
setting $N_\text{expected}=3$, 
which is the Poisson expectation value that gives a 5\% probability 
to measure zero occultations. 

\begin{figure}
    \centering
    \pic[1]{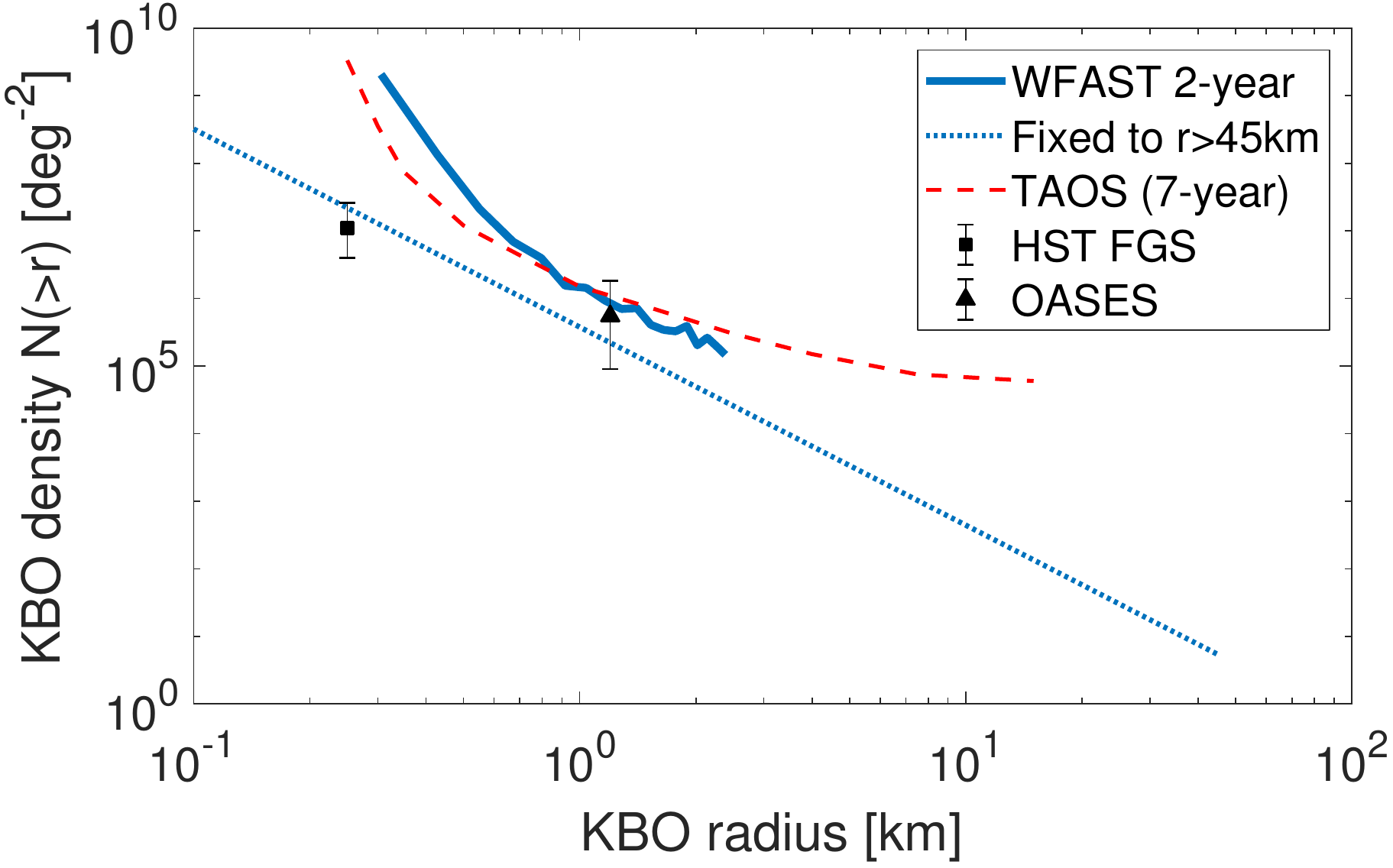}
    \caption{Measurements and upper limits on the number density of KBOs
             that are larger than some radius, $N(>r)$, 
             for various occultation surveys. 
             The solid, blue line is the upper limit by this work, 
             calculated individually for each radius bin. 
             The dashed red line is a similar upper limit 
             given by TAOS I \protect\citep{TAOS_survey_Zhang_2013}, 
             which is fairly consistent with our results, 
             including a similar number of star-hours 
             but with three or four telescopes. 
             The black square is the measurement given by
             \protect\cite{abundance_kuiper_objects_Schlichting_Ofek_2012}, 
             using the HST fine guidance sensor. 
             The black triangle is the measurement given by
             \protect\cite{KBO_detection_from_ground_Arimatsu_2019}, 
             using OASES (a ground based observatory with two telescopes). 
             The dashed, blue line is the steepest power law model that
             is normalized to the measured number density of KBOs at $r=45$\,km, 
             that is also consistent with our null detection over all radius bins. 
             }
    \label{fig:all limits}

\end{figure}

One way to set the upper limits is to calculate the 
number density limit individually for each bin. 
This result is plotted as the solid, blue line in Figure~\ref{fig:all limits}. 
The limits are fairly consistent with the results of TAOS \citep{TAOS_survey_Zhang_2013}, 
that also had accumulated $\approx 10^6$ star-hours over seven years. 

The dashed, blue line in Figure~\ref{fig:all limits} shows the upper limit on 
a power law model for KBOs which is normalized to the number density at 45\,km, 
given by \cite{kuiper_belt_survey_Subaru_beam_Fuentes_2009}, 
with a power law index $q<3.93$. 
Any steeper (more negative value) power law would generate 
an expected number of detections above 3, and would be ruled out 
at a 95\% confidence level by combining the lack of detection on all size bins.

\subsection{Comparison to previous results}\label{sec:previous results}

\begin{figure}
    \centering
    \pic[1]{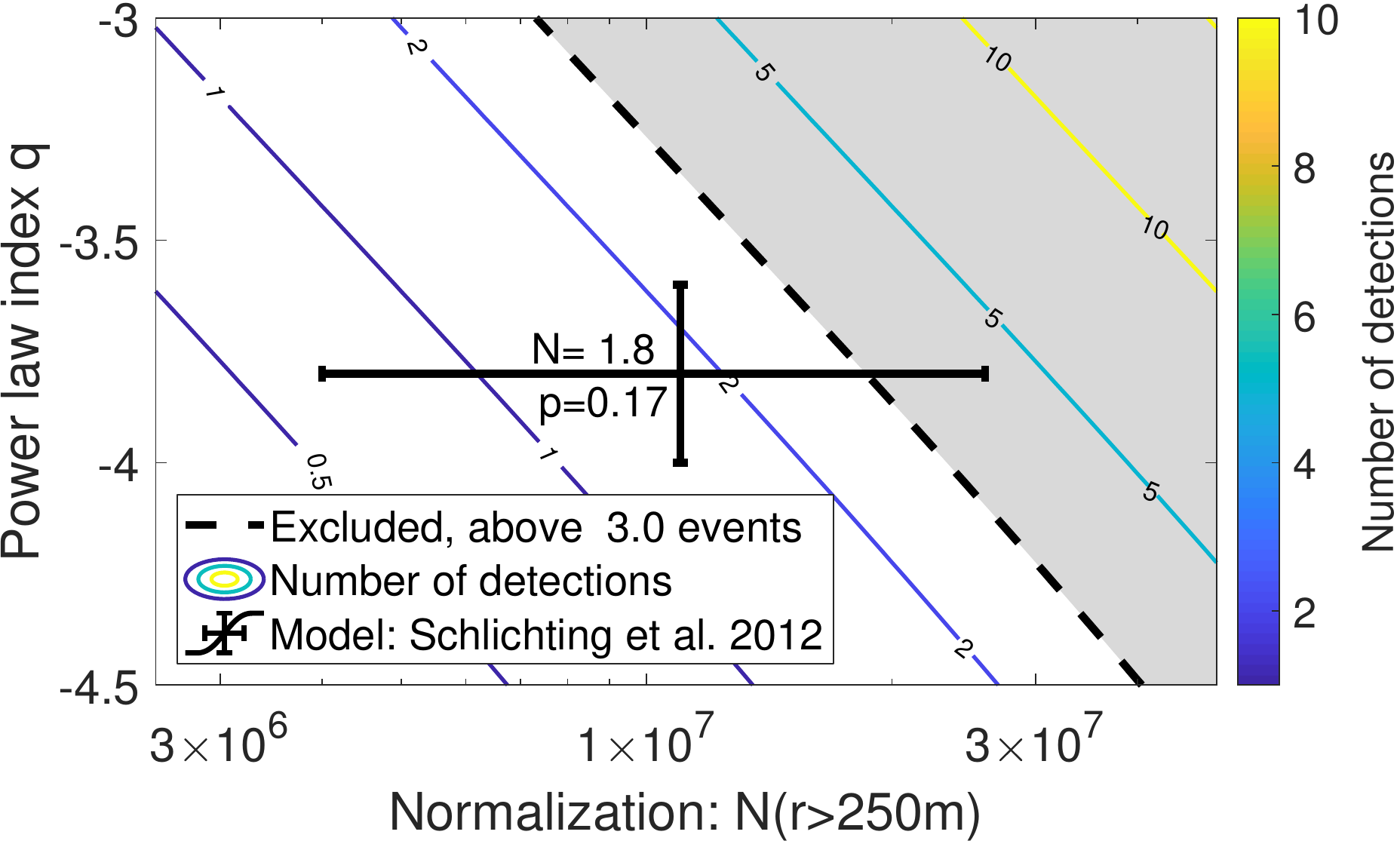}
    \caption{The parameter space of power law models for KBOs, 
             for models similar to the result of \protect\cite{abundance_kuiper_objects_Schlichting_Ofek_2012}. 
             The contours show the expected number of detections, 
             given our coverage, for a pair of model parameters
             defined by the normalization at 250\,m and the power law index. 
             The shaded region is excluded at the 95\% confidence level. 
             The fiducial value and uncertainties given by 
             \protect\cite{abundance_kuiper_objects_Schlichting_Ofek_2012} 
             are shown as black error bars. 
             The values printed next to the fiducial value
             show the expected number of detections in our survey ($N=1.8$), 
             and the Poisson probability to measure zero detections given 
             the stated expectation value ($p= 0.17$). 
             This model is in some tension with our results, 
             but is not ruled out with more than 83\% confidence. 
    }
    \label{fig:hst kbo limit}
\end{figure}

Previous detections of KBOs from the Hubble Space Telescope Fine Guidance Sensor (HST FGS)
include two occultation events, and a third candidate at high ecliptic latitude
which is deemed to be a false positive 
\citep{KBO_single_object_Schlichting_Ofek_2009, abundance_kuiper_objects_Schlichting_Ofek_2012}. 
From those measurements, and the number density calculated for large KBOs ($>45$\,km)
they propose a power law model for smaller KBOs, 
with a single power law index of $q=3.8\pm 0.2$. 
This result, shown as a black square in Figure~\ref{fig:all limits}, 
is consistent with the power law index limit 
from our survey ($q=3.93$, dotted, blue line). 
We can check the parameter space of models that
are defined by $N(>250\text{m})$ and $q$, 
with values close to those of the model given by \cite{abundance_kuiper_objects_Schlichting_Ofek_2012}. 
For each pair of parameter values we calculate $N_\text{expected}$, 
considering the coverage we have over all occulter size bins. 
The results are shown in Figure~\ref{fig:hst kbo limit}. 

\begin{figure}
    \centering
    \pic[1]{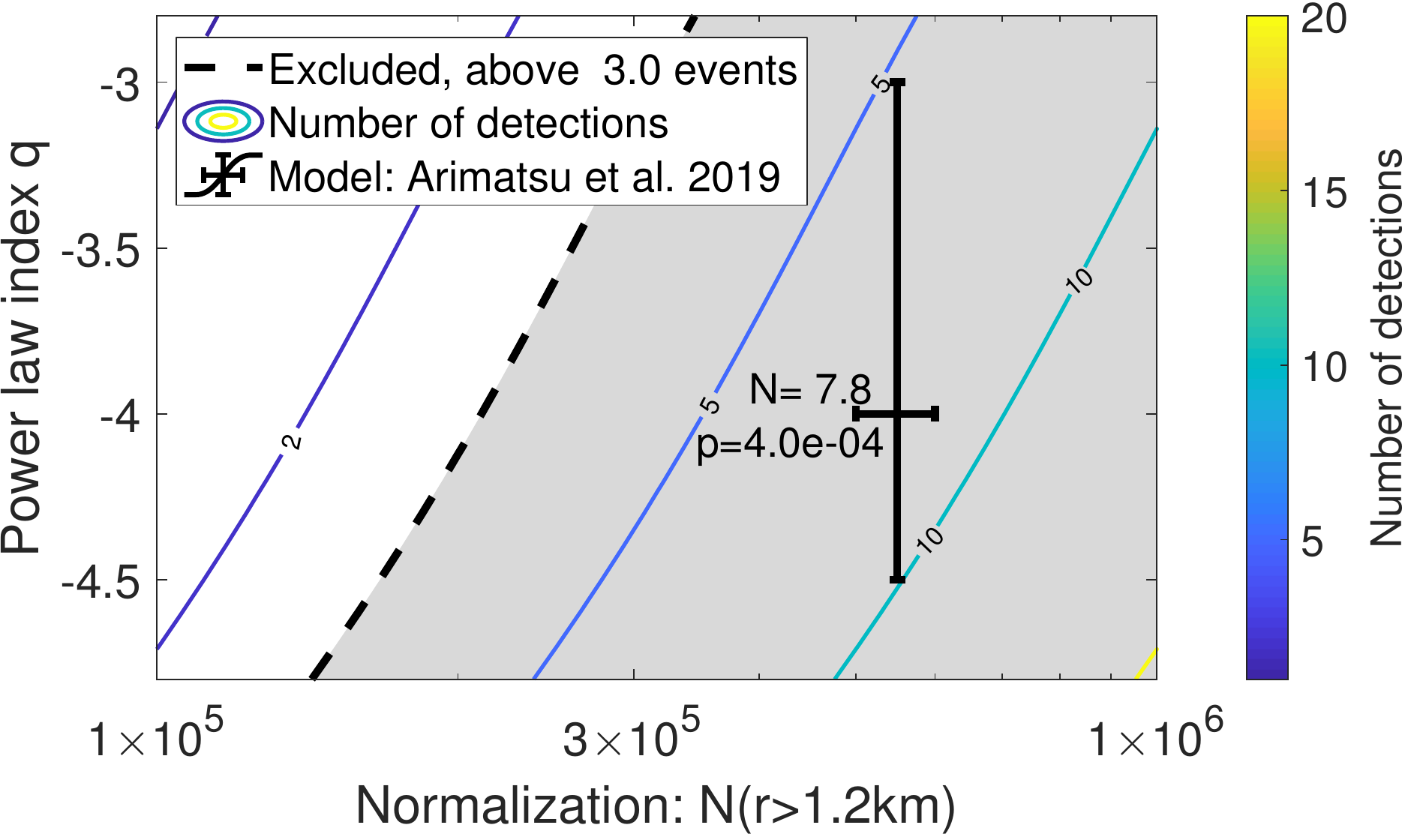}
    \caption{The parameter space of power law models for KBOs, 
             for models similar to the result of 
             \protect\cite{KBO_detection_from_ground_Arimatsu_2019}. 
             The contours show the expected number of detections, 
             given our coverage, for a pair of model parameters
             defined by the normalization at 1.2\,km and the power law index. 
             The shaded region is excluded at the 95\% confidence level. 
             The fiducial value and uncertainties given by 
             \protect\cite{KBO_detection_from_ground_Arimatsu_2019} 
             are shown as black error bars. 
             The values printed next to the fiducial value
             show the expected number of detections in our survey ($N=7.8$), 
             and the Poisson probability to measure zero detections given 
             the stated expectation value ($p=4\times 10^{-4}$). 
             Thus, the measured abundance reported by 
             \protect\cite{KBO_detection_from_ground_Arimatsu_2019}
             is inconsistent with our results. 
        }
    \label{fig:ground kbo limit}
\end{figure}

The most recent detection of a small KBO occultation comes from a ground survey, 
using a system of two telescopes with only a small sample of a few $10^4$ star-hours
\citep{KBO_detection_from_ground_Arimatsu_2019}.
This detection, if real, indicates an uptake in the size distribution
at a radius of 1--2\,km. 
This result is consistent with theoretical models, e.g., 
\cite{KBO_initial_sizes_Schlichting_2013}.
\cite{kuiper_belt_sizes_Pluto_craters_Morbidelli_2021}
also indicated an increase in the number of KBOs, 
based on analysis of the crater size distribution on Pluto. 
It is somewhat surprising, 
that with an order of magnitude more star-hours, 
no such occultations were seen in our survey. 
The power law upper limit model shown as a dashed, blue line
in Figure~\ref{fig:all limits} is already below the results
from \cite{KBO_detection_from_ground_Arimatsu_2019}. 
To quantify this, we test models similar to the parameters presented 
by \cite{KBO_detection_from_ground_Arimatsu_2019} 
and check which values are consistent with our null detection across all radius bins. 
The results are shown in Figure~\ref{fig:ground kbo limit}. 
Their density model would yield $N_\text{expected}=7.8$, 
which is inconsistent with our null detection
at the $p=4\times 10^{-4}$ level. 

There are some possible explanations for this tension:
(a) the sensitivity of the two telescopes used by \cite{KBO_detection_from_ground_Arimatsu_2019}
is substantially higher, due to the lower detection threshold 
allowed by cross-matching the light-curves of the two telescopes. 
The signal-to-noise-ratio of each individual detection 
they present is $\approx 5$, compared to the threshold of $7.5$ 
used in our survey. 
However, their estimate of the occulter size is 1.3\,km, 
which is large enough to be detected by our pipeline 
in at least 15\% of the parameter space (see \S\ref{sec:efficiency}).
While the sensitivity may affect the number of star-hours required to make a detection, 
the KBO density reported in previous works already takes the into account the sensitivity of those surveys. 
The density reported in \cite{KBO_detection_from_ground_Arimatsu_2019}
is inconsistent with our estimates, regardless of the number of star-hours, the number of telescopes, and the efficiency of that survey. 
Furthermore, the fact that the TAOS I survey 
\citep{TAOS_survey_Zhang_2013} found no occultations, 
even with three or four telescopes and $\approx 10^6$ star-hours, 
makes this explanation insufficient. 
(b) The detection reported by \cite{KBO_detection_from_ground_Arimatsu_2019}
is actually an unlucky coincidence of two 
atmospheric artefacts at the same time. 
This is also unlikely, since the two light-curves appear consistent in shape and timing. 
On the other hand, even looking just at the two seconds before and after
the candidate event, the OASES-01 light-curve (their Figure~2, blue line) 
shows at least one instance 
where the flux increases in amplitude similarly to the dip 
that is part of the occultation event. 
If even one telescope is frequently subject to random correlated fluctuations
of similar amplitude to the one constituting the occultation candidate
(such as those presented in \S\ref{sec:flare events}), 
it becomes considerably more likely that the event they reported
is caused by coincident fluctuations and not an occultation. 
(c) this single detection is a ``lucky accident'' with a probability of $4\times 10^{-4}$.

The analysis presented in the methods section 
of \cite{KBO_detection_from_ground_Arimatsu_2019}
shows that there is no overabundance of fluctuations outside the 
occultation candidate event, in the 20 minutes surrounding it, 
and no large atmospheric events coincident with it affecting other stars. 
Even so, in some of the event candidates discussed in Section~\ref{sec:candidates}
there is no evidence for a large number of 
outliers or any flux changes in nearby stars. 
A dedicated search for false positives over a larger 
part of their data, inspecting localized fluctuations, 
is required to gain more insight into
the nature of this event. 
Further detections (or lack thereof) of new occultations
in data collected after this event was identified in 2016, 
would also help shed light on the subject. 

It is interesting to note that even the baseline model 
of KBO density we used here is not beyond suspicion. 
Those results, based on two detections from space, 
reported in \cite{KBO_single_object_Schlichting_Ofek_2009, abundance_kuiper_objects_Schlichting_Ofek_2012}, 
would not be subject to the false-positive events caused by atmospheric fluctuations. 
Based on their own bootstrap simulations, 
\cite{abundance_kuiper_objects_Schlichting_Ofek_2012}
estimate the false alarm rate for their event to be 5\%. 
\cite{abundance_kuiper_objects_Schlichting_Ofek_2012}
also report a false detection on an off-ecliptic field. 
While they rule out that event because of its incompatible ecliptic latitude, 
the fact such false-positives exist
brings into question the validity of the on-ecliptic detection. 
The two events are not very different in the shape of their light-curves, 
and it is not clear why one should be more believable than the other, 
besides the position on the sky of the occulted star. 
If even images taken from space 
can be affected by such spurious detections, 
it is not surprising if later, ground based surveys 
are hard-pressed to identify real occultations
from various false-positives. 

Some of the uncertainty surrounding these past detections
could be cleared up once additional surveys, 
e.g., OASES, TAOS II and Colibri \citep{OASES_survey_Arimatsu_2017, TAOS_II_survey_Huang_2021, Colibri_survey_Mazur_2022}, 
would publish their results. 
Characterization of various systematics such as discussed in this work, 
especially from surveys with multiple telescopes, 
could help explain the plausibility that such
artefacts could appear in two telescopes at the same time.

\subsection{False detections}\label{sec:false detections}

When looking for short duration, rare events with only a single telescope,
care must be taken to avoid rare occurrences of different types of foregrounds. 
It is particularly difficult to know if an event is real without confirmation 
from another observatory, preferably one located a few tens of meters away
such that it observes through different instances of the atmosphere. 
Even with two telescopes, relatively common foreground events 
(such as scintillation or wind-driven tracking errors) 
can occur in both images simultaneously 
when large amounts of data are accumulated. 
In this work we have applied multiple layers of 
foreground rejection methods 
(automated quality cuts, human vetting, and additional 
\emph{ad hoc} tests for found detections discussed in \S\ref{sec:analysis}). 

Clearly there are some hard-to-reduce foreground events
to consider when searching for occultations by sub-km KBOs. 
The candidates presented in this work, though surpassing the 
detection threshold and not triggering any of the data 
quality cuts we applied, are all most likely to be 
instrumental or atmospheric artefacts. 
Some show many outliers, some have suspicious
variations in the flux of nearby stars, 
and most of them do not show the expected
width compared to the Earth's projected velocity. 
Five out of seven events occurred in the first half of 
April 2021, further hinting that the source
of those events is instrumental or environmental. 
In addition, we detected 30 flare events (see \S\ref{sec:flare events}), 
of which some also show dimming beside the brightening, 
making it clear that false detections are 
much more common than real detections, 
and indistinguishable from them. 

The large number of false detections we see in our data
also suggests that if not carefully controlled, 
such events could also occur coincidentally
in two telescopes, particularly under bad atmospheric conditions. 
One hint for such bad conditions comes from the clustering 
of candidates in April 2021, where five out of seven events are detected. 

A survey with two telescope might set a lower threshold
that allows for a higher false alarm fraction. 
A single telescope observing $10^6$ star-hours at 25\,Hz
would need a false-alarm rate of $\approx 10^{-11}$ to get one
false detection in the entire survey. 
A naive approach to adding another telescope would be 
to increase the single-telescope false-alarm rate by a factor of $10^{11}$ 
(e.g., by dropping the threshold from $7.5\sigma$ to $3\sigma$).
If atmospheric events such as presented in this work 
are instead concentrated in a two week period 
(instead of uniformly across two years)
then the coincidence rate of would be 50 times higher. 
If atmospheric instability has shorter typical durations, 
it may be that for short periods of times, 
two-telescope events become even more likely. 

If the distribution of foreground events is much flatter than a normal distribution
(i.e., events are in the ``long tails'' of the distribution), 
the rate could be less affected by the decrease of threshold. 
It should be noted, however, that five out of seven events have a 
total score (matched-filter S/N) of 7.5--7.7, indicating that
the distribution is not completely flat. 
This suggests that before setting a threshold for 
an experiment with more than one telescope, 
a study of the distribution of low-threshold events, 
as a function of time and S/N, 
should be made on each telescope, 
to make sure the combined false-alarm rate is well controlled.

\section{Conclusions}\label{sec:conclusions}

We presented the results of a dedicated occultation survey 
for km-scale KBOs using the W-FAST observatory, 
over the two year period of 2020--2021. 
Using a single telescope, 
we collected $\approx 740,000$ star-hours on low ecliptic latitudes, 
and combined with data quality cuts, human vetting and injected simulations, 
we establish a competitive upper limit on KBOs in the size range of 0.5--2.5\,km. 

We detect seven possible occultation candidates but rule them all out 
based on inspection of the long term light-curve, the expected event velocity, 
and the light-curves of neighboring stars. 
We caution that such false positives do not always show clear distinction
from real events when tested against only one of these trials, 
and that care must be taken in interpreting data from non-repeating events like stellar occultations. 
Since we do not fully understand the causes and the occurrence rates
of such false events, it is possible that short periods of, 
e.g., atmospheric instability, could cause the probability 
of false positives to increase dramatically, even when combining two telescopes. 

Our results are in agreement with the detection of KBO 
occultations from space \citep{KBO_single_object_Schlichting_Ofek_2009, abundance_kuiper_objects_Schlichting_Ofek_2012}. 
The expected number of detections based on their model would
yield 1.8 detections in our survey, 
which is still consistent with zero detections, with $p=0.15$. 
Our results are in tension with the single detection 
from the ground made by \cite{KBO_detection_from_ground_Arimatsu_2019}, 
where their model yields an expected 7.8 detections in our survey, 
which is ruled out with $p=4\times 10^{-4}$. 
The existence of an uptake of the KBO density is favoured by 
some theoretical models (e.g., \citealt{KBO_initial_sizes_Schlichting_2013}), 
but appears to be ruled out by our survey. 

The null detection presented in this work can also be used to put 
a general upper limit which is consistent with that presented 
by \cite{TAOS_survey_Zhang_2013}, using three and four telescopes of 
the TAOS I observatory, with a similar number of star-hours. 
This shows that careful data analysis and vetting can produce 
a detection efficiency similar to that of multiple telescopes. 
The main disadvantage of working with a single telescope is that
even a true occultation would be hard to believe in a group of 
false positives that look so similar to real events. 
For example, one of the events presented in Table~\ref{tab:candidate summary}
has consistent velocity to that of the Earth's projected velocity 
(the event of 2021 April 16), but is ruled out for other reasons. 
It is to be expected that false positive events would be consistent
with different velocities, and some of them would match the observations. 
Thus, it would be difficult to confirm even true events, 
that could just happen to pass all the tests. 
A second (or third) telescope would be useful in vetting these events, 
besides the advantage it could give in reducing the detection threshold. 

\section*{Acknowledgements}

E.O.O.~is grateful for the support of
grants from the 
Willner Family Leadership Institute,
André Deloro Institute,
Paul and Tina Gardner,
The Norman E Alexander Family M Foundation ULTRASAT Data Center Fund,
Israel Science Foundation,
Israeli Ministry of Science,
Minerva,
BSF, BSF-transformative, NSF-BSF,
Israel Council for Higher Education (VATAT),
Sagol Weizmann-MIT,
Yeda-Sela,
Weizmann-UK,
Benozyio center,
and the Helen-Kimmel center.
B.Z.~was supported by a research grant from the Willner Family Leadership Institute for the Weizmann Institute of Science.
S.B.A~is supported by the Peter and Patricia Gruber Award; 
the Azrieli Foundation; 
the André Deloro Institute for Advanced Research in Space and Optics; 
and the Willner Family Leadership Institute for the Weizmann Institute of Science.
S.B.A~is the incumbent of the Aryeh and Ido Dissentshik Career Development Chair.
We are grateful to the Wise Observatory staff for their support in taking the observations used in this work. 

\section*{Data Availability}

The full data set used in this work, including raw images, calibration files, and the higher level products such as lightcurves and event summaries are available upon reasonable request from the authors.


\bibliographystyle{mnras}
\bibliography{references.bib}

\appendix

\section{Individual occultation candidates}\label{sec:individual candidates}

The following section describes in detail each of the events 
passing all our quality cuts and human vetting, 
but before the velocity vetting. 
Additional analysis methods 
(velocity, outliers, and neighbors analyses, discussed in \S\ref{sec:analysis})
leads us to conclude that none of these events is a real occultation. 

\subsection{Occultation candidate 2020-07-01}\label{sec:candidate 2020-07-01}

This event is unusual in many respects. 
It is the only occultation event detected in 2020
(although six flare events were seen during that year). 
The event is also the only occultation identified in off-ecliptic fields. 
This event shows a dip and an increase of flux, 
making it half-way between an occultation and a flare
(see \S\ref{sec:flare events}). 

\begin{figure*}
    \centering
    \pic[0.9]{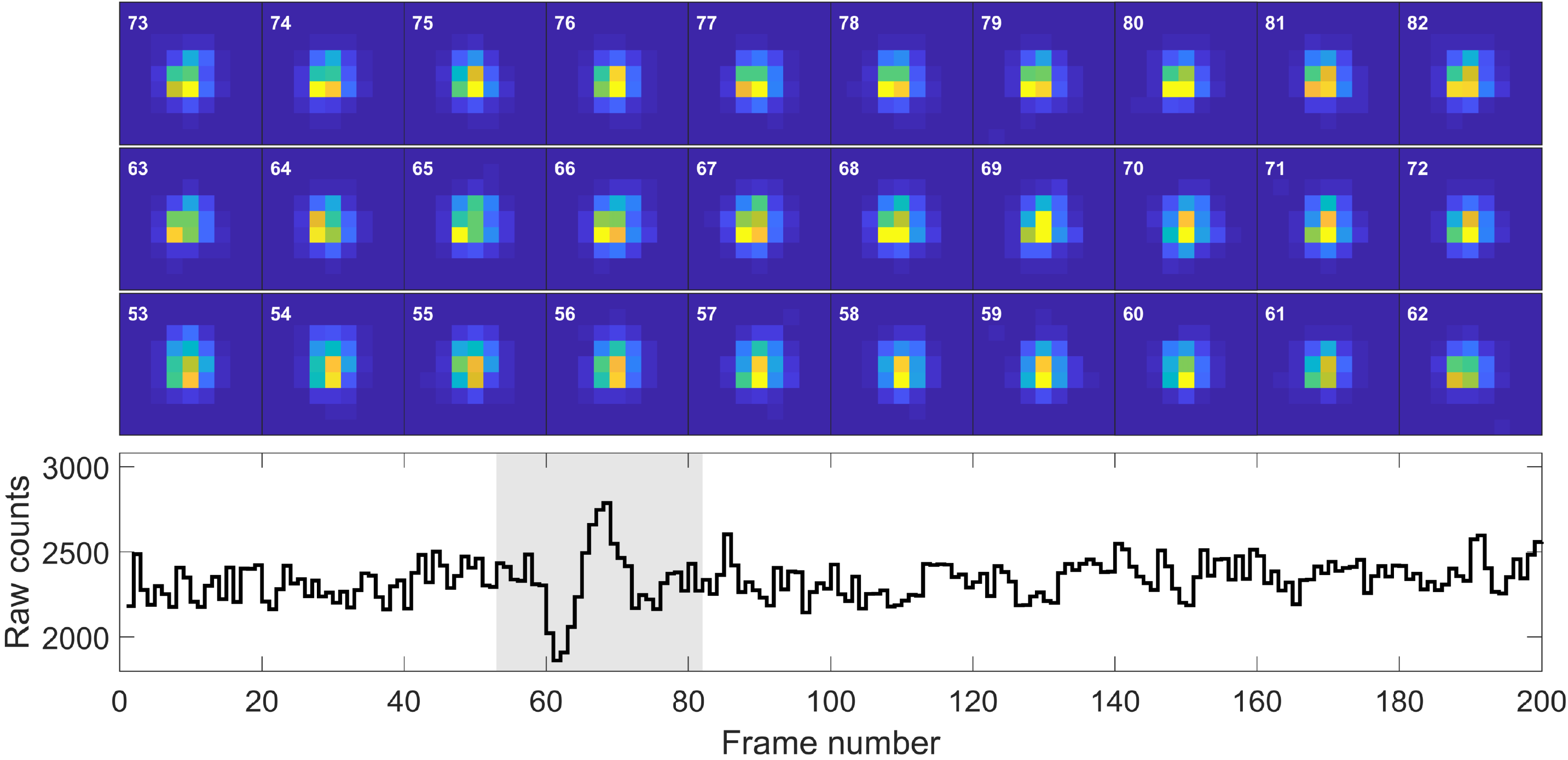}
    \caption{Raw counts and cutouts for the occultation candidate of 2020 July 01. 
             The cutouts are set to the same dynamic range, 
             showing that the dimming and brightening is not an artefact
             of the photometric reduction, but a feature of the underlying pixel measurements. 
             }
    \label{fig:flux cutouts 2020-07-01}
\end{figure*}

The shape of the light-curve, as well as the cutout images around 
the time of the event are shown in Figure~\ref{fig:flux cutouts 2020-07-01}. 
The shape of the light-curve is unusual: 
the intensity of light decreases as though in occultation, 
then increases dramatically. 
This may be due to diffraction fringes 
causing alternating decreases and increases of the measured flux. 
Such strong, asymmetric brightening can occur in cases of binary or elongated KBOs, 
as discussed in \cite{wfast_kbo_pipeline_Nir_2023}. 
However, the similarity to ``flares'' in our sample 
makes it more likely that this is an atmospheric event, 
that could very well have been classified as a flare. 
Indeed, some flares in our sample show brightening and dimming
similar to the shape of this event
(see \S\ref{sec:flare events}). 

\begin{figure*}
    \centering
    \pic[0.9]{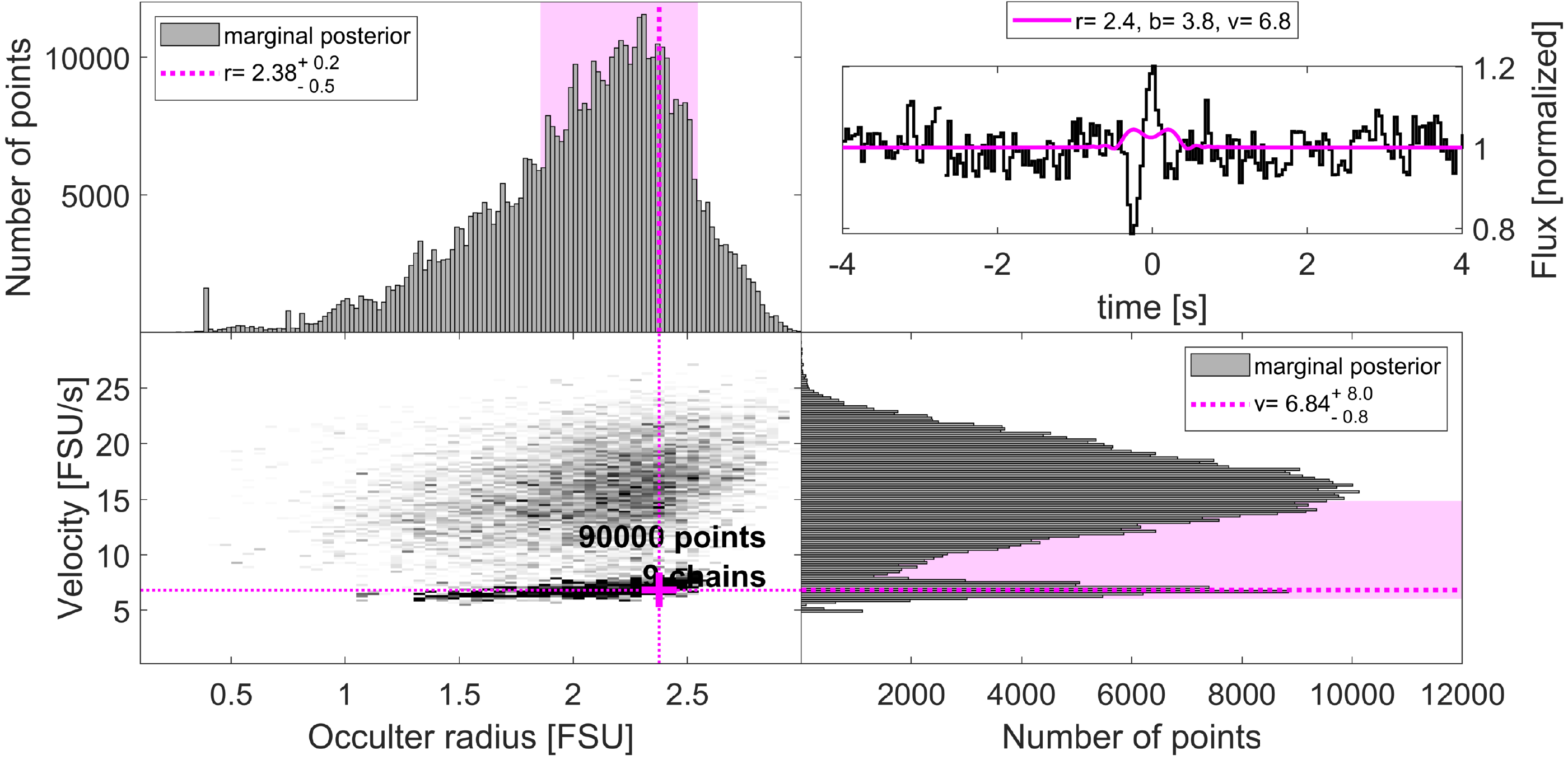}
    \caption{Results of an MCMC analysis (see \S\ref{sec:mcmc}) for the occultation candidate of 2020 July 01.
             The lower-left panel shows the posterior, 
             marginalized over the impact parameter, time offset and stellar size, 
             while the top-left and bottom-right panels show the marginalized
             posteriors leaving only occulter radius and velocity, respectively. 
             The pink, dotted lines and cross show the most likely (best-fit) value in the two parameters, 
             although the range of likely values includes many other combinations of radius and velocity, 
             as shown by the pink shaded areas, that highlight 
             the smallest interval containing 68\% of the posterior. 
             The top-right panel shows the raw light-curve and the best template (in pink),
             using the best-fit parameters described above. 
    }
    \label{fig:mcmc 2020-07-01}
\end{figure*}

The MCMC fit to the light-curve yielded poor fits, 
probably due to the asymmetric and undulating shape of the light-curve (see Figure~\ref{fig:mcmc 2020-07-01}). 
It would perhaps be more useful to fit this event with models of a binary occulter, 
but such modeling is more complicated and computationally demanding. 
Also, due to the similarity between this event 
and other false positives we decided to keep the analysis simple. 
We ran 50 chains with $10^4$ steps (following $10^4$ steps as burn-in), 
as discussed in \S\ref{sec:mcmc}. 
Nine chains converged (were not rejected),
with a clearly bi-modal distribution. 

The lower cluster of chains at low velocities 
($v\approx 6.8$\,FSU\,\persec $\approx 8.8$\,km\,\persec)
is inconsistent with the observation's transverse velocity of 27.2\,km\,\persec. 
The top cluster has higher velocities, but is also less dense. 
Both distributions have most of their mass at 
values of $r>2$\,FSU $\approx2.5$\,km. 
The large radii reflects the strong changes in the flux 
over many frames, which is hard to achieve with a small occulter. 
This, along with the high ecliptic latitude of the event, 
strengthens our conclusion that this event is a false-positive, 
caused either by the atmosphere or by instrumental effects. 

\begin{figure*}
    \centering
    \pic[0.9]{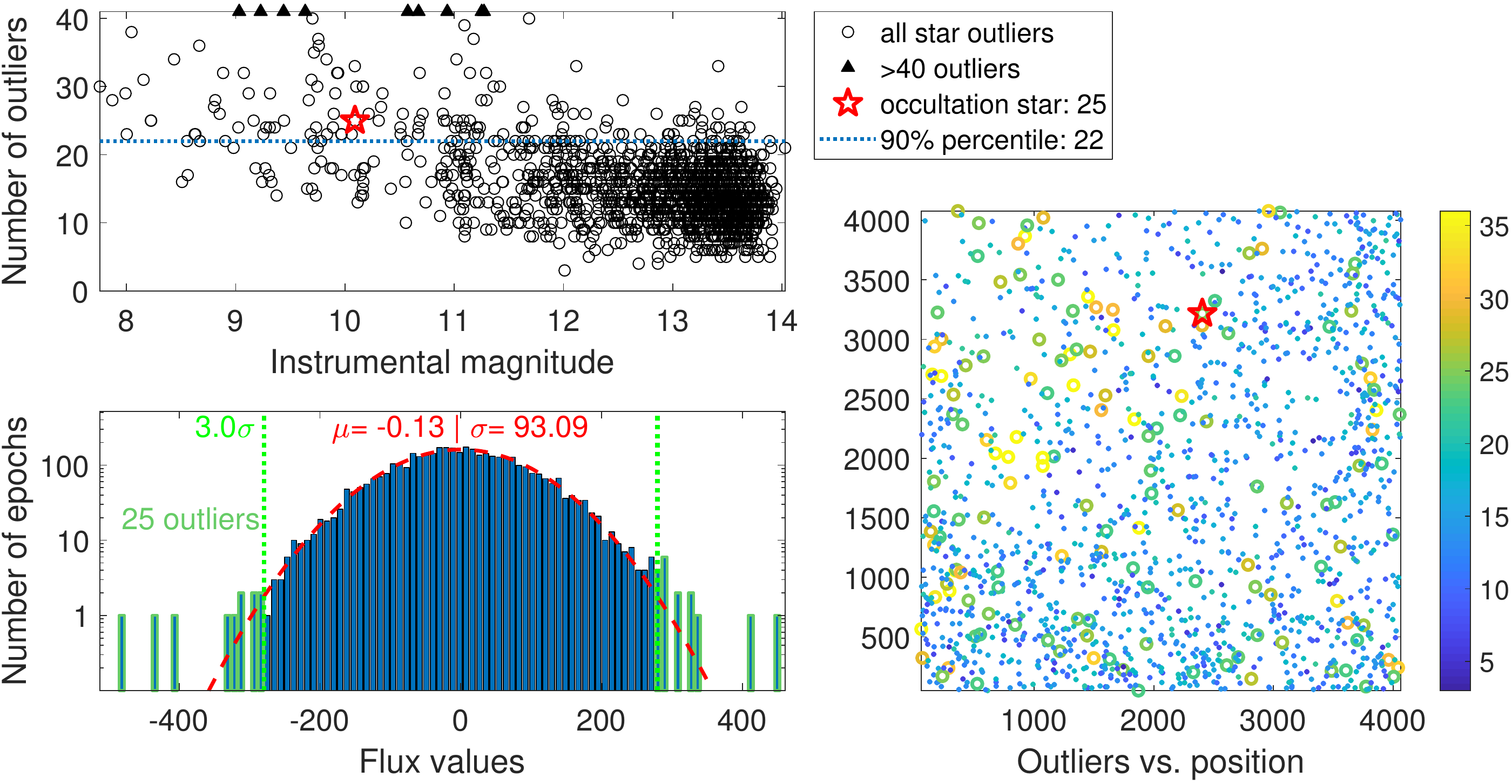}
    \caption{Outlier analysis (see \S\ref{sec:outliers}) 
             for the occultation candidate of 2020 July 01.
             The top panel shows the number of outliers for each star as black circles 
             as a function of each stars' instrumental magnitude
             (with an arbitrary zero point). 
             A few stars with outliers above the maximum value 
             in this plot were replaced with black triangles, 
             which helps focus the plot on the occultation star's range of outliers. 
             The red pentagon shows the position of the candidate star. 
             The blue, horizontal, dashed line shows the 90th percentile, equal to 22 outliers. 
             The bottom-left panel shows the histogram of the flux values as blue bars, 
             the 3$\sigma$ limits used to find outliers as green, dotted, vertical lines, 
             and the fit to a Gaussian as a red, dashed line. 
             The occultation star has 25 outliers, which is above
             the 90th percentile but does not seem to be an extreme value 
             compared to other bright stars on the left side of the top panel. 
             The bottom-right panel shows the number of outliers for stars
             as a function of their pixel coordinates on the focal plane. 
             Large circles represent stars with more outliers than the 90th percentile. 
             The candidate star is marked with a red pentagon. 
             There is no clustering of `bad stars' close to the candidate star. 
    }
    \label{fig:outliers 2020-07-01}
\end{figure*}

One way to test if an event is due to spurious atmospheric noise 
or bad pixels is to look at the number of flux values that
are dramatically different from the mean 
(flux outliers, see \S\ref{sec:outliers}). 
We show the number of outliers found for each star
in the top panel of Figure~\ref{fig:outliers 2020-07-01}. 
We see that 90\% of the stars have less than 22 outliers 
in that time range. 
The candidate star has 25 outliers 
which puts it in the top 90\% of the stars
in terms of number of outliers, 
but is not dramatically different from stars
of similar brightness. 
This indicates the cause of the event is not likely
an instrumental effect. 
We also show the histogram of the flux values for the 
candidate star, along with a fit to a Gaussian, 
in the bottom-left panel, highlighting the outlier distribution. 

In the bottom-right panel of Figure~\ref{fig:outliers 2020-07-01}
we plot the position of each star on the focal plane, 
color coding the marker for each star based on the number of outliers it has. 
This allows us to visualize any clustering of bad light-curves, 
perhaps indicating a local phenomena. 
In this case, there is some cluster of bad fluxes around $x,y = 1000,2000$
but this area is not close to the candidate star, marked with a red pentagon. 
This again indicates it is less likely that this event is due to local instrumental effects. 

\begin{figure*}
    \centering
    \pic[0.9]{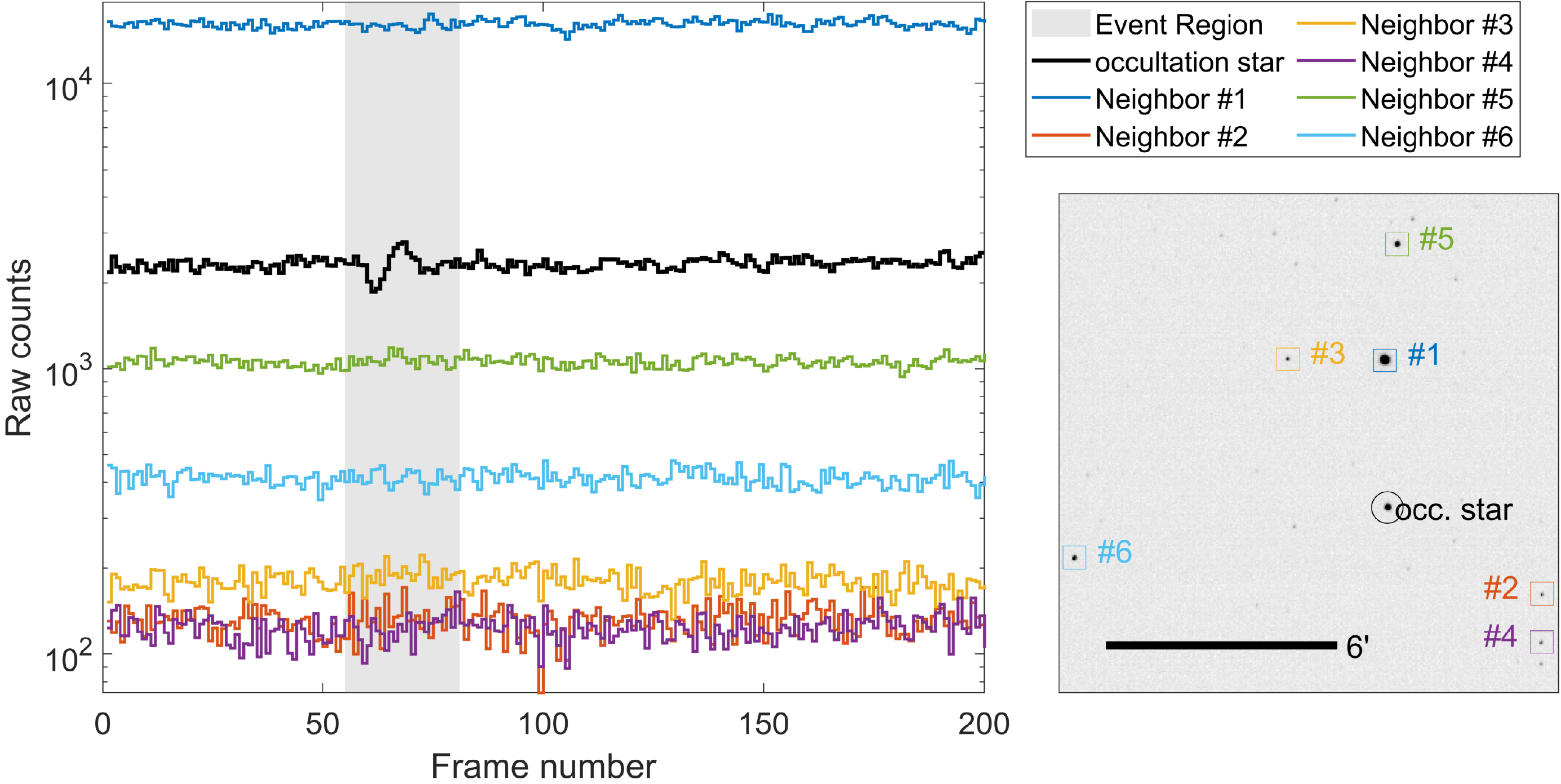}
    \caption{Flux values for nearby stars around the time of the event of 2020 July 01. 
             There does not seem to be any evidence of flux changes in any of the nearby stars
             during the 8\,s period surrounding the event.              
    }
    \label{fig:neighbors 2020-07-01}
\end{figure*}

Finally, in Figure~\ref{fig:neighbors 2020-07-01}
we show the fluxes of the nearest neighboring stars,
out of stars with photometric $S/N>5$
(see \S\ref{sec:nearest}). 
The light-curves do not seem to show similar 
flux variations as the occultation star, 
which weakens the hypothesis that this event is atmospheric. 

\subsection{Occultation candidate 2021-04-01}\label{sec:candidate 2021-04-01}

\begin{figure*}

    \centering
    \pic[0.9]{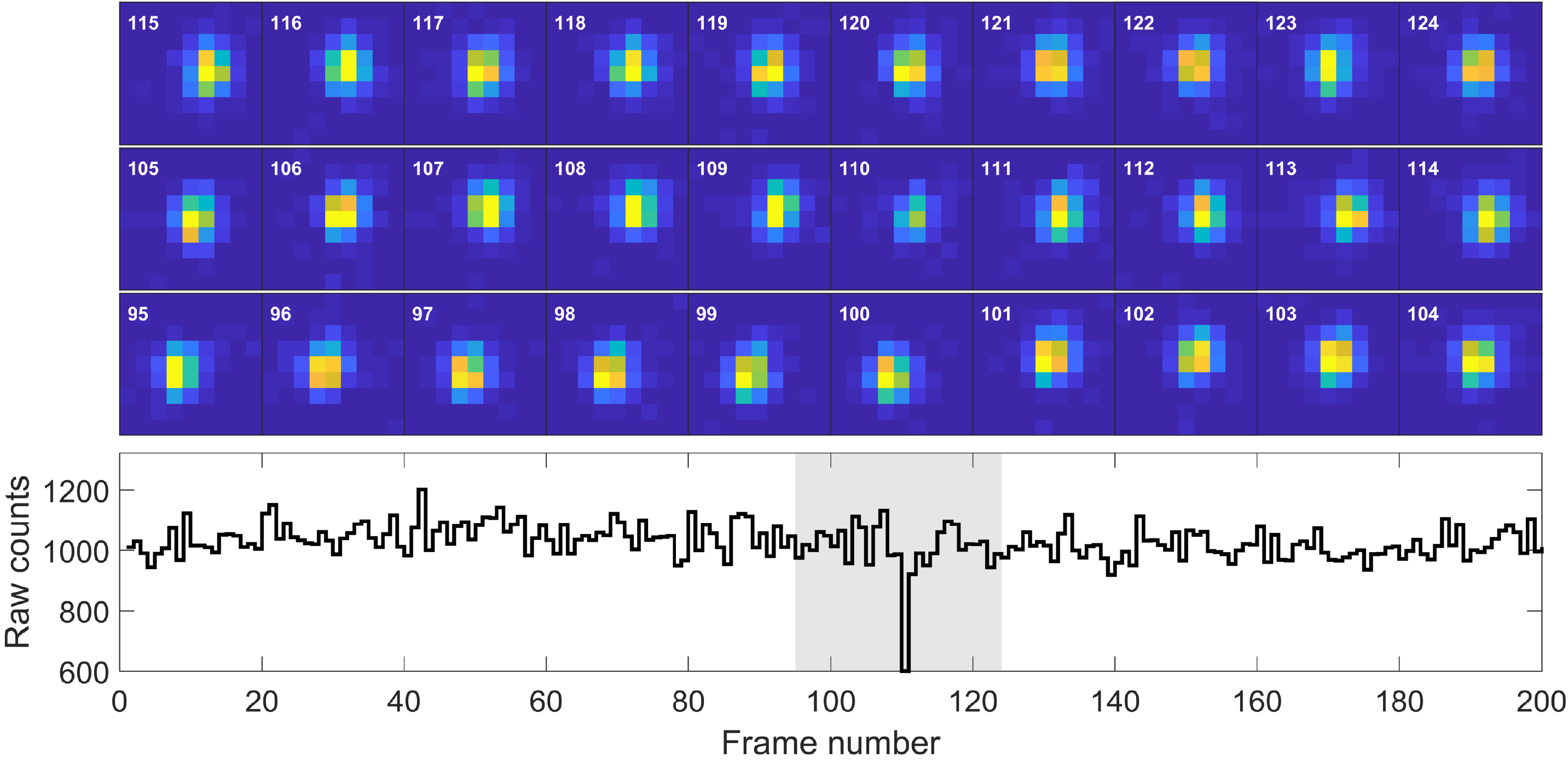}
    \caption{
         Raw counts and cutouts for the occultation candidate of 2021 April 01. 
         The cutouts are set to the same dynamic range, 
         showing that the dimming is not an artefact
         of the photometric reduction, but a feature of the underlying pixel measurements.
    }
    \label{fig:flux cutouts 2021-04-01}
\end{figure*}

This event is a single-frame occultation candidate. 
The shape of the light-curve, as well as the cutout images around 
the time of the event are shown in Figure~\ref{fig:flux cutouts 2021-04-01}. 

\begin{figure*}
    \centering
    \pic[0.9]{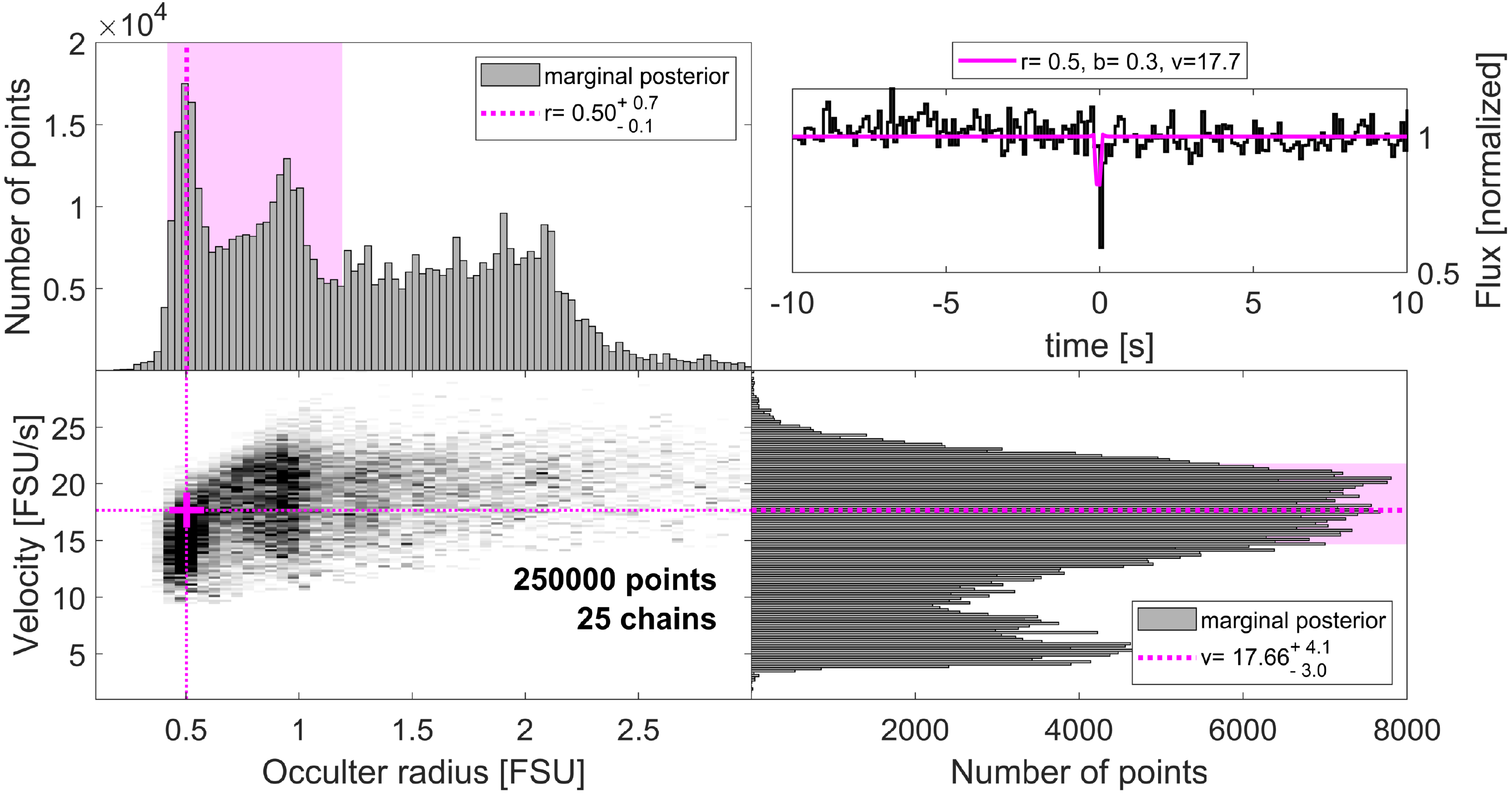}
    \caption{
        Results of an MCMC analysis (see \S\ref{sec:mcmc}) for the occultation candidate of 2021 April 01.
        The panels are the same as in Figure~\ref{fig:mcmc 2020-07-01}. 
        The posterior distribution is spread out over a large part of the 
        parameter space, but does not have much overlap with low velocities
        (the Earth's projected velocity for this event is $4.0$\,km\,\persec). 
        The best-fit lightcurve does not reproduce the depth of the single-frame dip, 
        and requires a velocity of 17.66\,FSU\,\persec$\approx 23$\,km\,\persec. 
    }
    \label{fig:mcmc 2021-04-01}

\end{figure*}

The MCMC fit to this event follows the same methods as used for all events
(see \S\ref{sec:mcmc} and \S\ref{sec:candidate 2020-07-01}). 
The posterior distribution, shown in Figure~\ref{fig:mcmc 2021-04-01}, 
does not converge on a single peak but traces a wide swath of the parameter range. 
Most striking is the low probability given to low velocities ($v<5$\,FSU s$^{-1}$). 
At very low transverse velocities, the events are expected to be wide, 
such that even at 10\,Hz sampling they should affect multiple frames. 
If the KBO occulter is not moving in retrograde, the total transverse velocity
should be even lower, making it more unlikely to appear as a single frame event. 
If the KBO is moving in retrograde,  
the total transverse velocity could instead be in the range 10--15\,FSU\,s$^{-1}$, 
which is consistent with the event light-curve. 

An outlier analysis of the star's long term light-curve, 
as described in \S\ref{sec:outliers} reveals that 
this star does not display a large number of outliers, 
with 21 outliers in the time range of 4000 frames around the event. 
Compared to other stars this is not above average. 

\begin{figure*}
    \centering
    \pic[0.9]{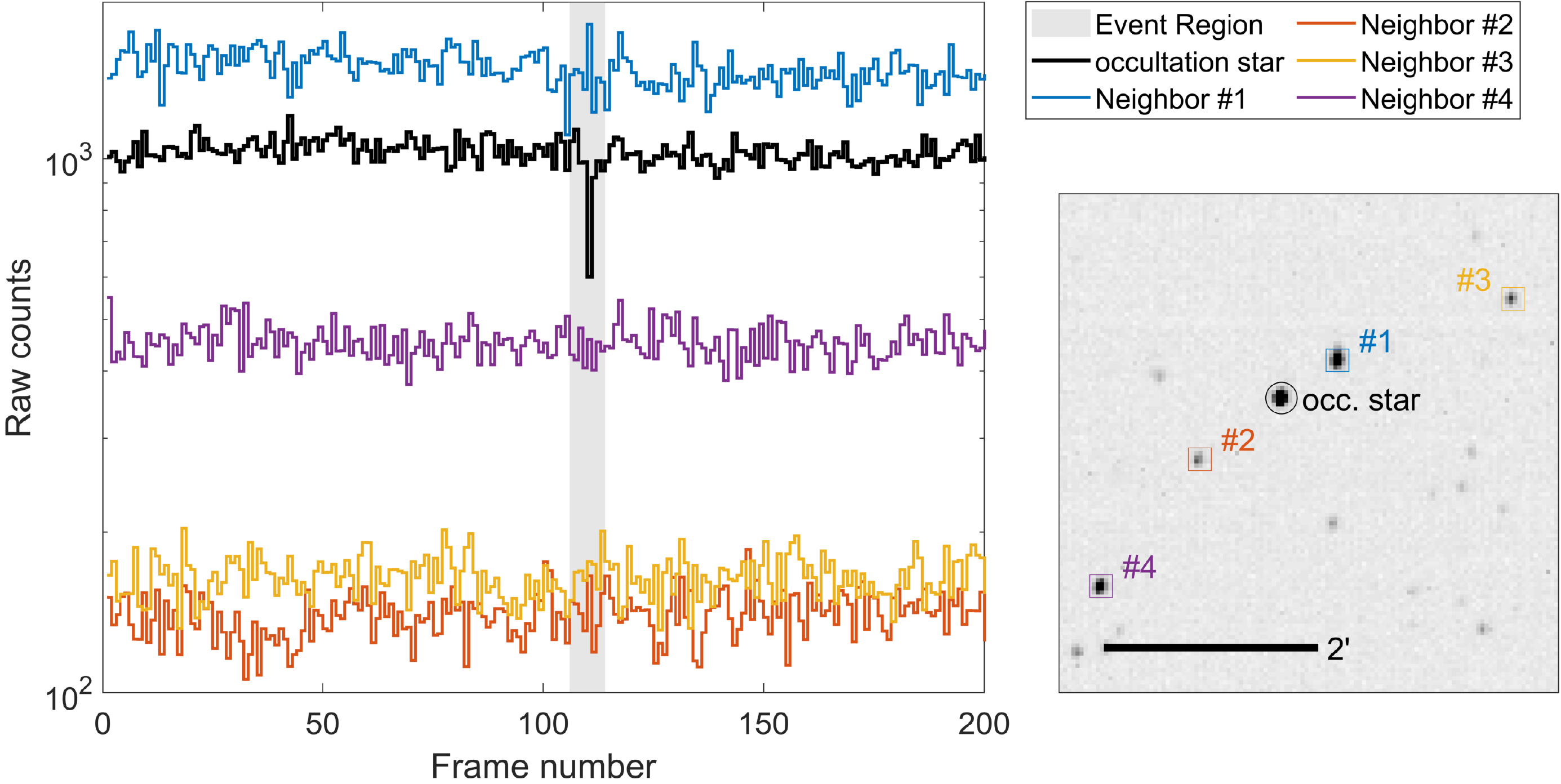}
    \caption{
        Flux values for nearby stars around the time of the event of 2021 April 01. 
        There is a small decrease in the flux of the most nearby star (\#1, blue line), 
        but no other strong dimming events is seen in other stars 
        during the 20\,s period surrounding the event.
    }
    \label{fig:neighbors 2021-04-01}
\end{figure*}

We also check if other stars nearby the event star 
are affected by any transient dimming. 
The results for four nearby stars with $S/N>5$ are shown in 
Figure~\ref{fig:neighbors 2021-04-01}. 
We see a suspicious single frame dimming in star \#1, 
but no other obvious transient events on any of the other stars. 

This event is hard to classify due to the dimming being confined to a single frame. 
The low velocity of the observations make it unlikely to be a true occultation
event with such a short duration. 
However, looking at the number of outliers and nearby stars we see no 
evidence for instrumental or atmospheric effects. 
This event, along with the next candidate observed only two days later, 
remain questionable, single-frame events.

\subsection{Occultation candidate 2021-04-03}\label{sec:candidate 2021-04-03}

\begin{figure*}
    \centering
    \pic[0.9]{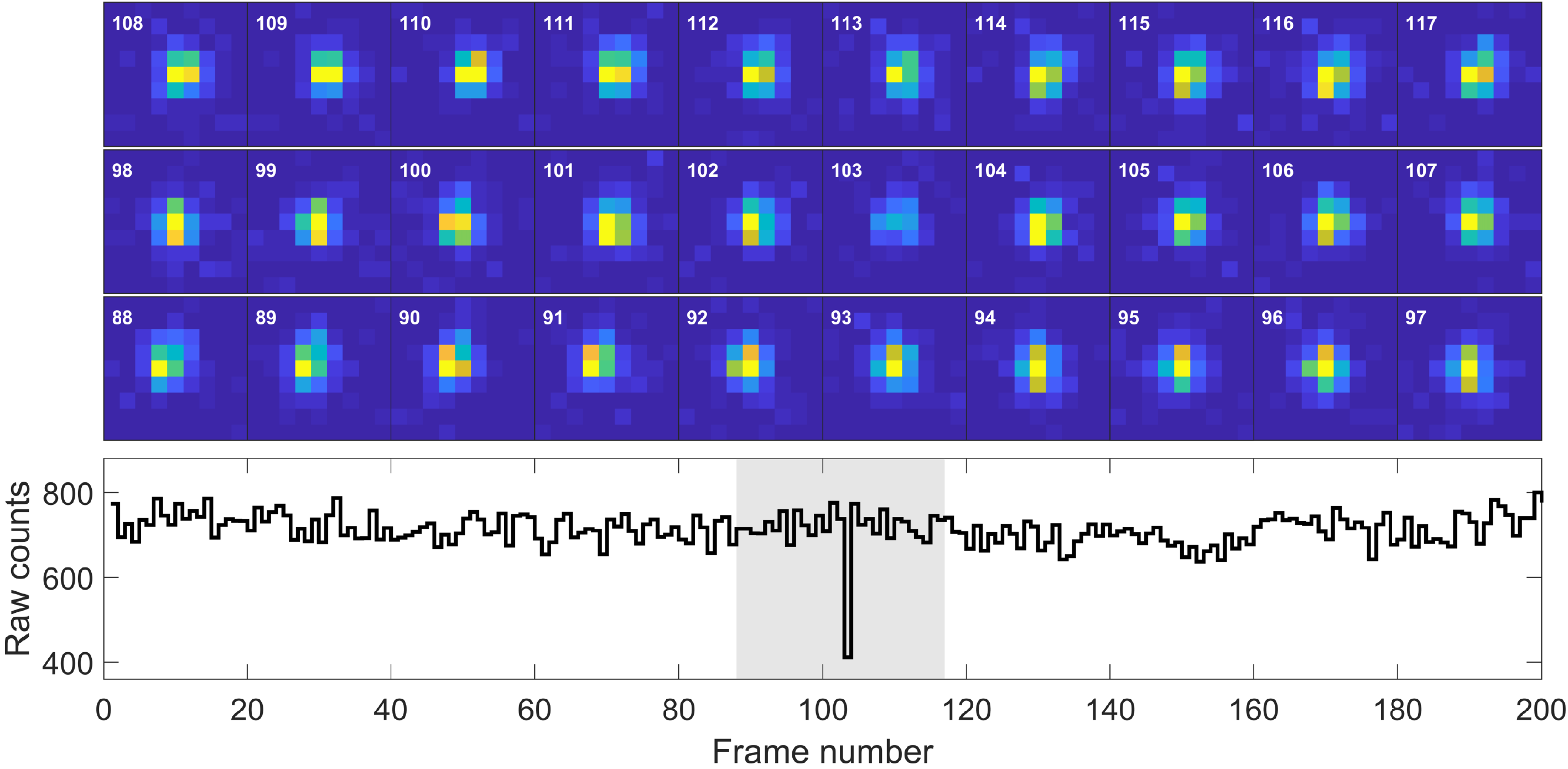}
    \caption{
         Raw counts and cutouts for the occultation candidate of 2021 April 03. 
         The cutouts are set to the same dynamic range, 
         showing that the dimming is not an artefact
         of the photometric reduction, but a feature of the underlying pixel measurements.
    }
    \label{fig:flux cutouts 2021-04-03}

\end{figure*}

Similar to the event of 2021 April 01, 
this event also shows a dimming on a single frame. 
The cutout images show a clear dimming which is not 
due to the photometric reduction, 
as shown in Figure~\ref{fig:flux cutouts 2021-04-03}. 

The MCMC posterior is consistent with a wide range of values 
for the occulter radius, $r$, but is fairly limited to 
high velocities $17<v<22.5$\,FSU\,s${^-1}$ ($22<v<29$\,km\,s$^{-1}$). 
This seems wholly inconsistent with the observed transverse velocity
of these observations, even for a retrograde, eccentric occulter. 

\begin{figure*}
    \centering
    \pic[0.9]{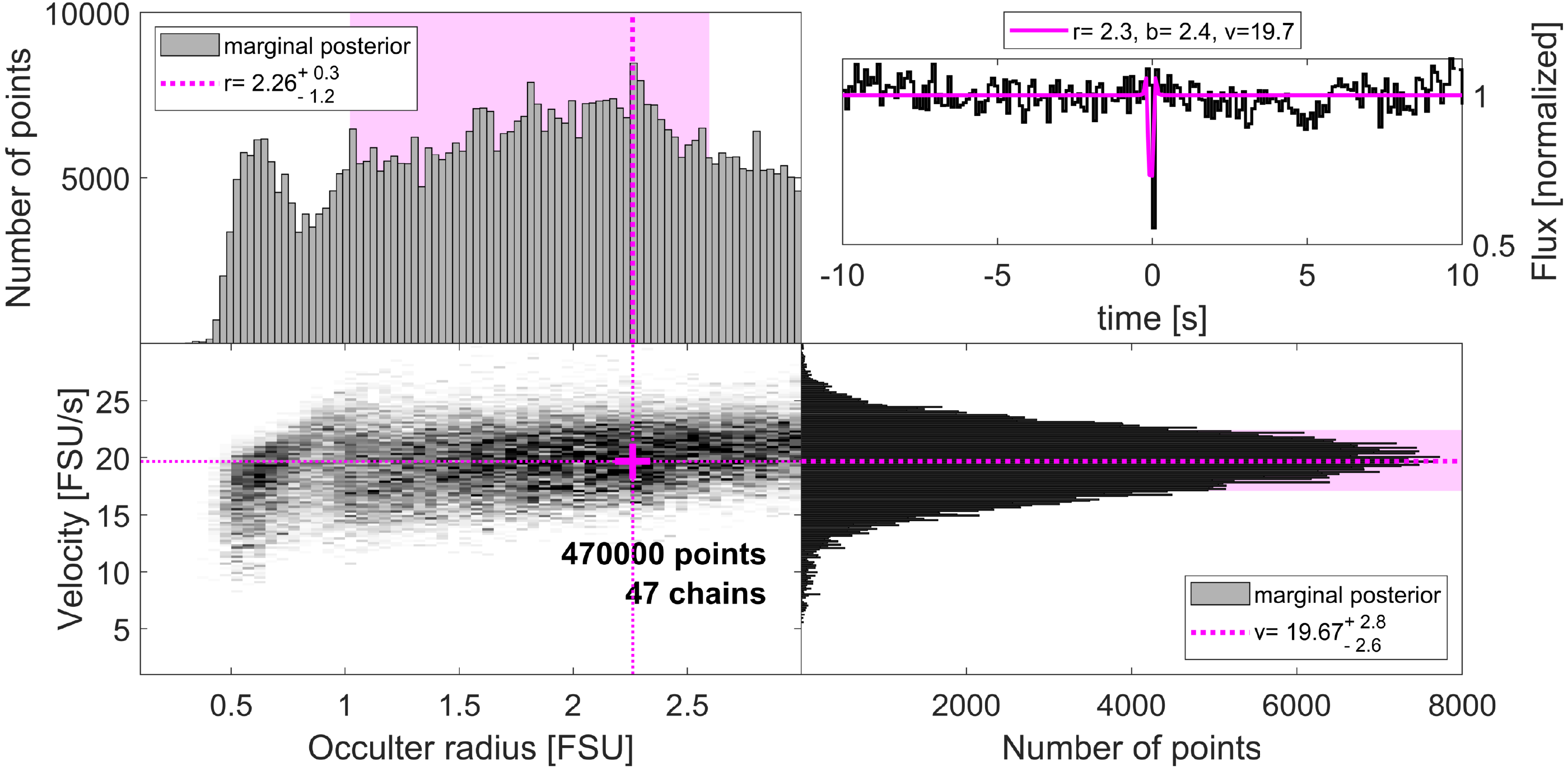}
    \caption{
        Results of an MCMC analysis (see \S\ref{sec:mcmc}) for the occultation candidate of 2021 April 03.
        The panels are the same as in Figure~\ref{fig:mcmc 2020-07-01}. 
        The posterior distribution is spread out over a large range in $r$, 
        but a rather limited range of high velocities ($\approx 25.5\pm 3.5$\,km\,\persec).
        By comparison, the Earth's projected velocity is $5.0$\,km\,\persec.
    }
    \label{fig:mcmc 2021-04-03}

\end{figure*}

Similarly to the previous event, there is not a substantial number of outliers
for the light-curve of this star in the 4000 frames around the event. 
The neighbor stars' light-curves also do not show any substantial transients
in the flux of any stars in the few arc-minutes around the event star. 
While we do not see any direct evidence for an instrumental or atmospheric
disturbance that could explain the dimming, 
the fact that both this and the event two days before show 
a single-frame dimming at a field with low expected transverse velocities
make it improbable that these two events are due to real occultations.

\subsection{Occultation candidate 2021-04-11}\label{sec:candidate 2021-04-11}

\begin{figure*}

    \centering
    \pic[0.9]{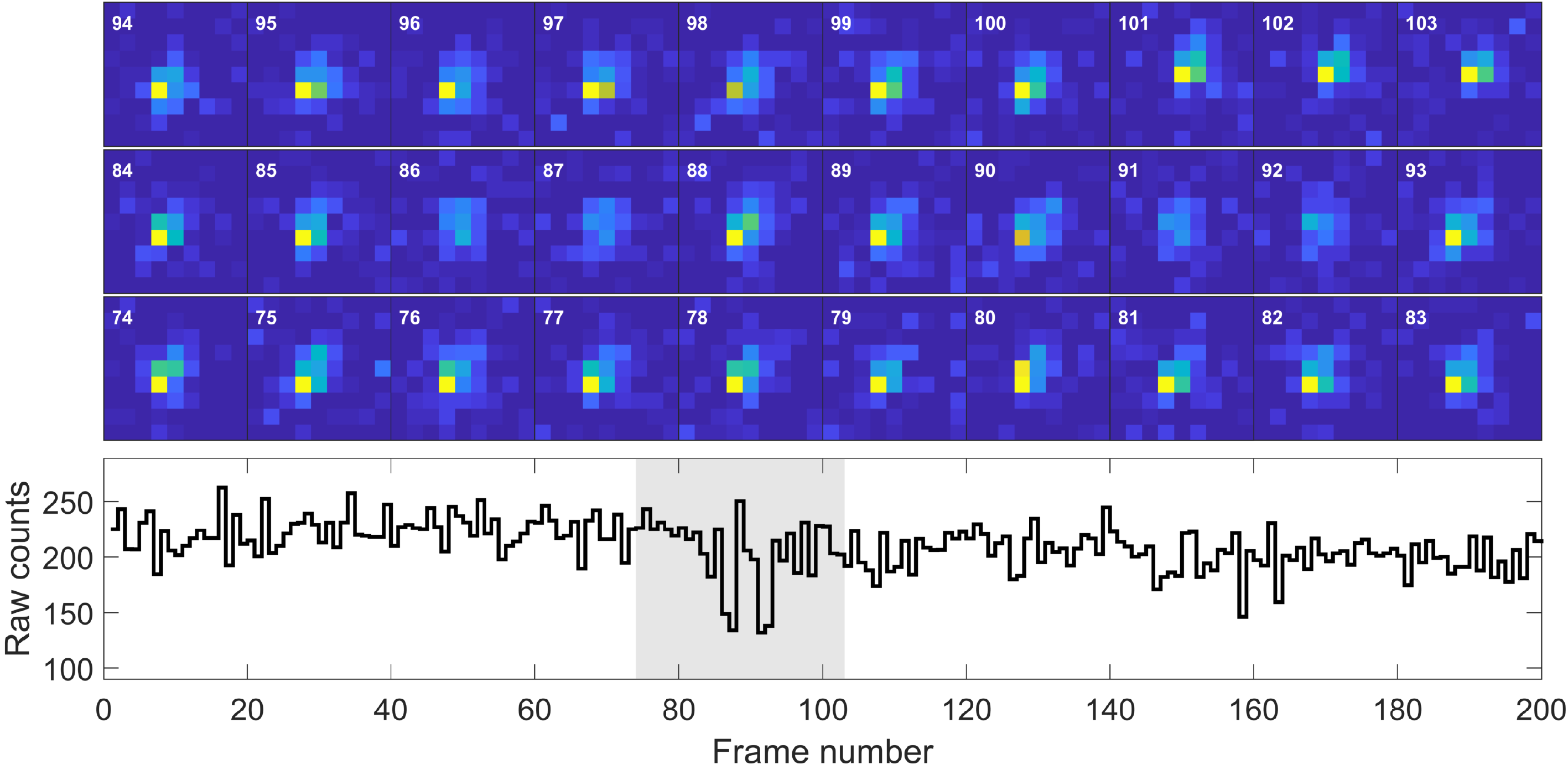}
    \caption{
         Raw counts and cutouts for the occultation candidate of 2021 April 11. 
         The cutouts are set to the same dynamic range, 
         showing that the dimming is not an artefact
         of the photometric reduction, but a feature of the underlying pixel measurements.
    }
    \label{fig:flux cutouts 2021-04-11}
\end{figure*}

This event is a multi-frame occultation candidate, 
composed of two dimming frames on either side 
of a relatively bright middle three frames.
The shape of the light-curve, as well as the cutout images around 
the time of the event are shown in Figure~\ref{fig:flux cutouts 2021-04-11}. 
The shape is reminiscent of occultations with 
strong diffraction effects, 
where in this case the middle brightening, 
also known as the Poisson peak, 
is brighter than the edges of the 
light-curve right before and after the dimming frames. 

\begin{figure*}
    \centering
    \pic[0.9]{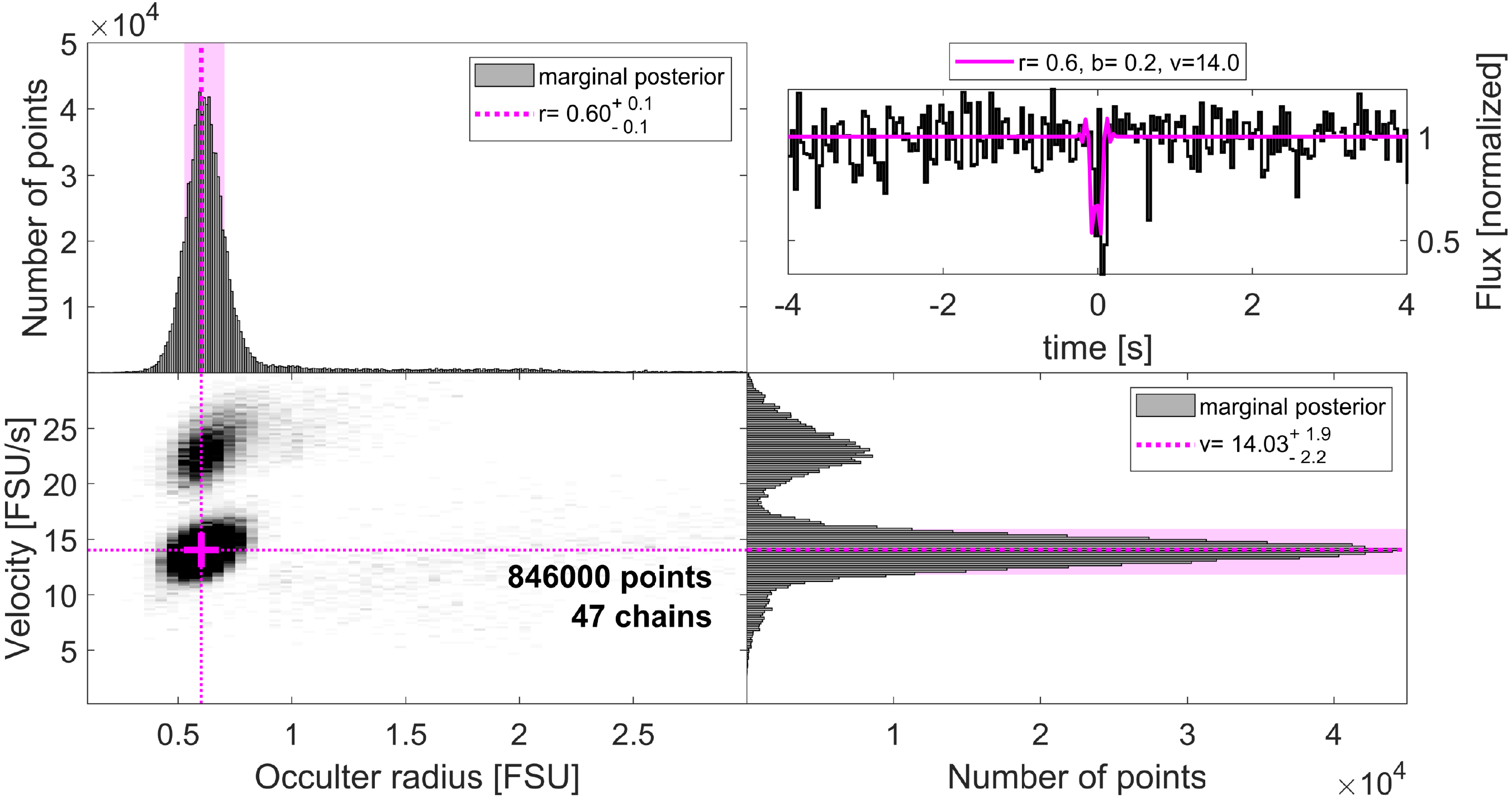}
    \caption{
        Results of an MCMC analysis (see \S\ref{sec:mcmc}) for the occultation candidate of 2021 April 11.
        The panels are the same as in Figure~\ref{fig:mcmc 2020-07-01}. 
        There are two clusters for the 2D posterior distribution. 
        The denser one, at a lower velocity of $\approx 18.2\pm3$\,km\,\persec, 
        is only marginally consistent with the Earth's transverse velocity of 9.5\,km\,\persec, 
        if we assume a retrograde occulter. 
        The strong Poisson peak in the event light-curve
        is not matched by the best-fit template, 
        which requires a small impact parameter which is 
        not in itself very likely. 
    }
    \label{fig:mcmc 2021-04-11}

\end{figure*}

The MCMC fit to this event follows the same methods as used for all events
(see \S\ref{sec:mcmc} and \S\ref{sec:candidate 2020-07-01}). 
The posterior distribution, shown in Figure~\ref{fig:mcmc 2021-04-11}, 
follows a bi-modal distribution, with two spots at different velocities. 
The larger spot, centered at $r=0.6$\,FSU and $v=14$\,FSU\,s$^{-1}$, 
strongly constrains the parameters of this event, mostly due to the 
presence of the Poisson peak and the many frames of the occultation. 
While this is marginally consistent with the observed Earth velocity of 7.3\,FSU\,s$^{-1}$, 
it once again requires a retrograde occulter to match the width of the light-curve. 
We also notice that the best-fit light-curve, 
shown in pink on the top-right panel of Figure~\ref{fig:mcmc 2021-04-11}, 
does not reproduce the strong Poisson peak of the event. 
Manually running some simulations, we conclude that 
a Poisson peak that is equally as bright as the 
ingress/egress fringes of the occultation is only 
possible when the impact parameter is close to zero, 
which is not a particularly likely value. 

\begin{figure*}
    \centering
    \pic[0.9]{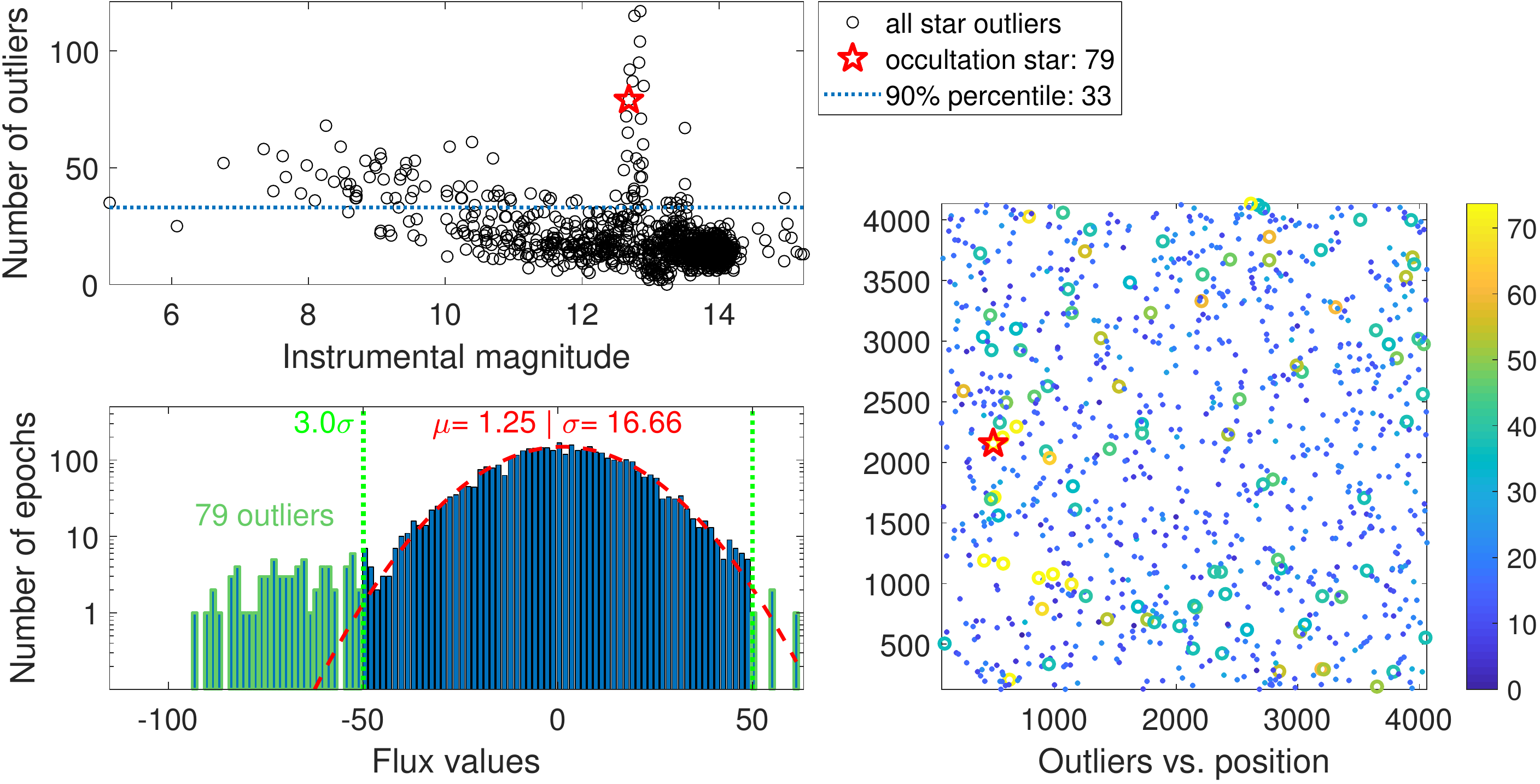}
    \caption{Outlier analysis (see \S\ref{sec:outliers}) for the occultation candidate of 2021 April 11.
             The panels are the same as in Figure~\ref{fig:outliers 2020-07-01}. 
             The occultation star has 79 outliers, which is more than double
             the 90th percentile (of 33 outliers). 
             On the bottom-right panel we see 
             the star appears to be in a region with many high-outlier stars, 
             which is indicative of an instrumental or atmospheric origin for the event. 
    }
    \label{fig:outliers 2021-04-11}
\end{figure*}

An outlier analysis of the star's long term light-curve, 
as described in \S\ref{sec:outliers} is shown in 
Figure~\ref{fig:outliers 2021-04-11}. 
This star's long term light-curve
(spanning 4000 frames, or 400\,s, around the event)
shows an unusually high number of dimming outlier events. 
A cluster of stars nearby on the sensor are seen to
have a high number of outliers (seen on the bottom-right panel)
which indicates some problem with that area of the detector, 
either due to defocus, vignetting, or a bad match of the 
photometric aperture to the star positions. 
The event star has 79 outlier frames in that time range, 
mostly dimming frames, 
while the rest of the stars' 90th percentile is only 33 outliers. 
This result makes it highly likely that the dimming
is caused by an instrumental effect, 
not a physical occultation. 

\begin{figure*}
    \centering
    \pic[0.9]{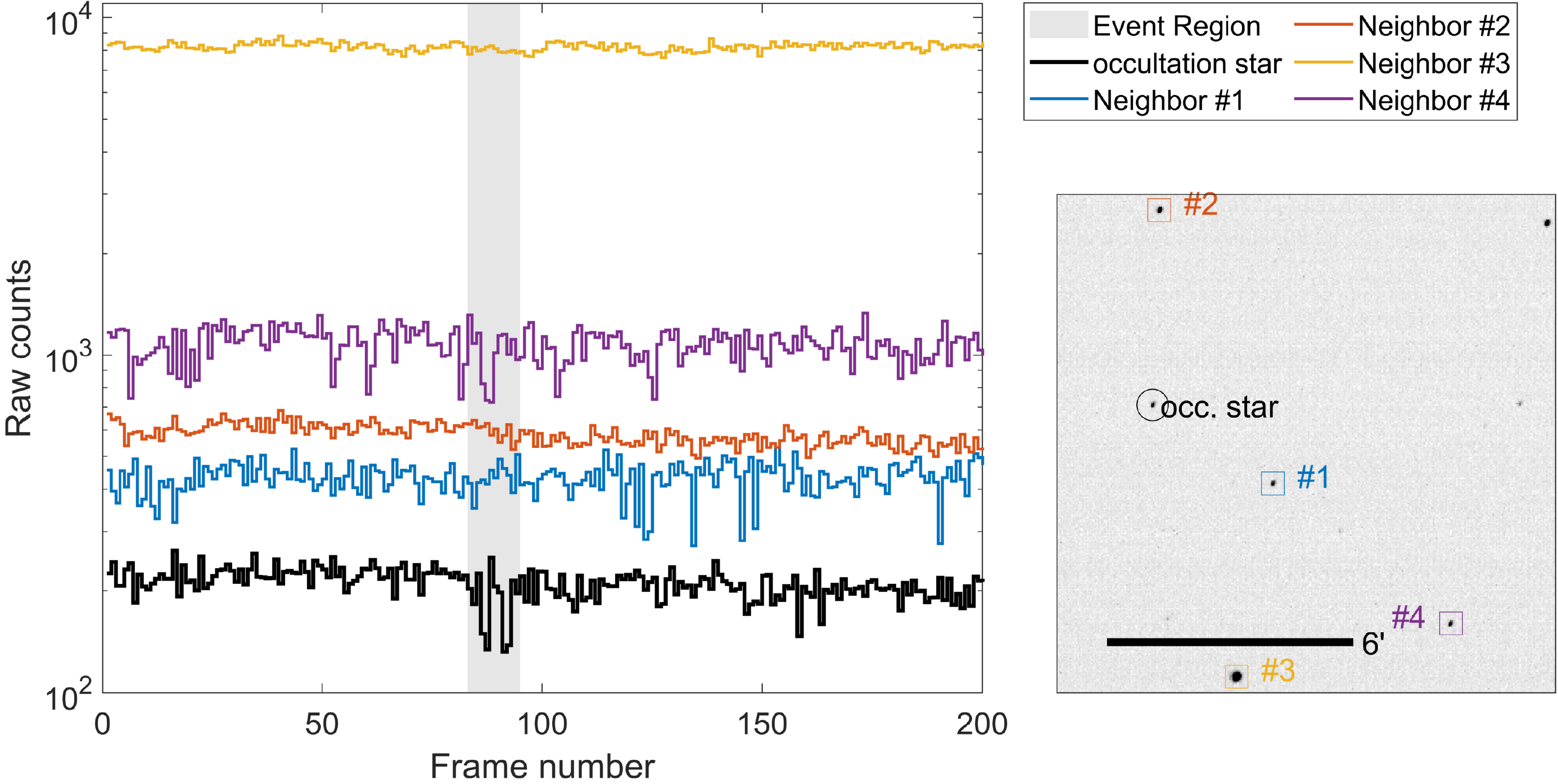}
    \caption{
        Flux values for nearby stars around the time of the event of 2021 April 11. 
        Stars  \#1 and \#4, shown as blue and purple lines, 
        show strong variability, 
        indicating an atmospheric or instrumental origin. 
    }
    \label{fig:neighbors 2021-04-11}
\end{figure*}

We also check if other stars nearby the event star 
are affected by any transient dimming. 
The results for four nearby stars with $S/N>5$ are shown in 
Figure~\ref{fig:neighbors 2021-04-11}. 
There is some evidence that the stars \#1 and \#4, 
shown as the blue and purple lines, respectively, 
also show strong variability on similar time-scales
as the event star. 
This is consistent with an instrumental effect 
but also with an atmospheric event that affected
some of the stars nearby. 

\subsection{Occultation candidate 2021-04-12}\label{sec:candidate 2021-04-12}

\begin{figure*}

    \centering
    \pic[0.9]{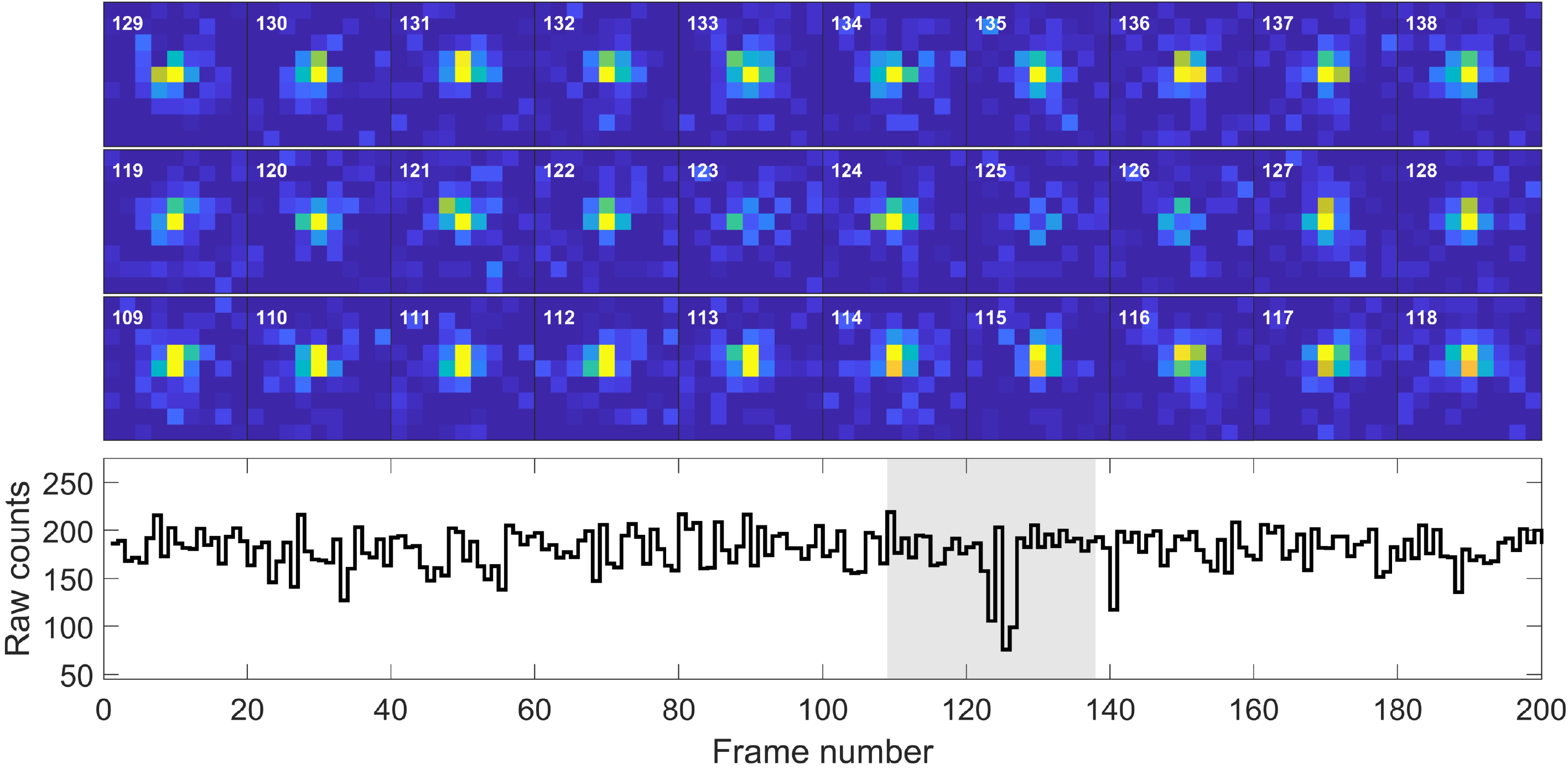}
    \caption{
         Raw counts and cutouts for the occultation candidate of 2021 April 12. 
         The cutouts are set to the same dynamic range, 
         showing that the dimming is not an artefact
         of the photometric reduction, but a feature of the underlying pixel measurements.
    }
    \label{fig:flux cutouts 2021-04-12}
\end{figure*}

This event is a multi-frame occultation candidate, 
composed of one and two dimming frames before and after
a normal brightness, single middle frame.
The shape of the light-curve, as well as the cutout images around 
the time of the event are shown in Figure~\ref{fig:flux cutouts 2021-04-12}. 
The shape is reminiscent of occultations with 
strong diffraction effects. 
It has fewer dimming frames, 
and a shorter and less prominent middle part, 
compared to the event detected only one day before it. 

\begin{figure*}
    \centering
    \pic[0.9]{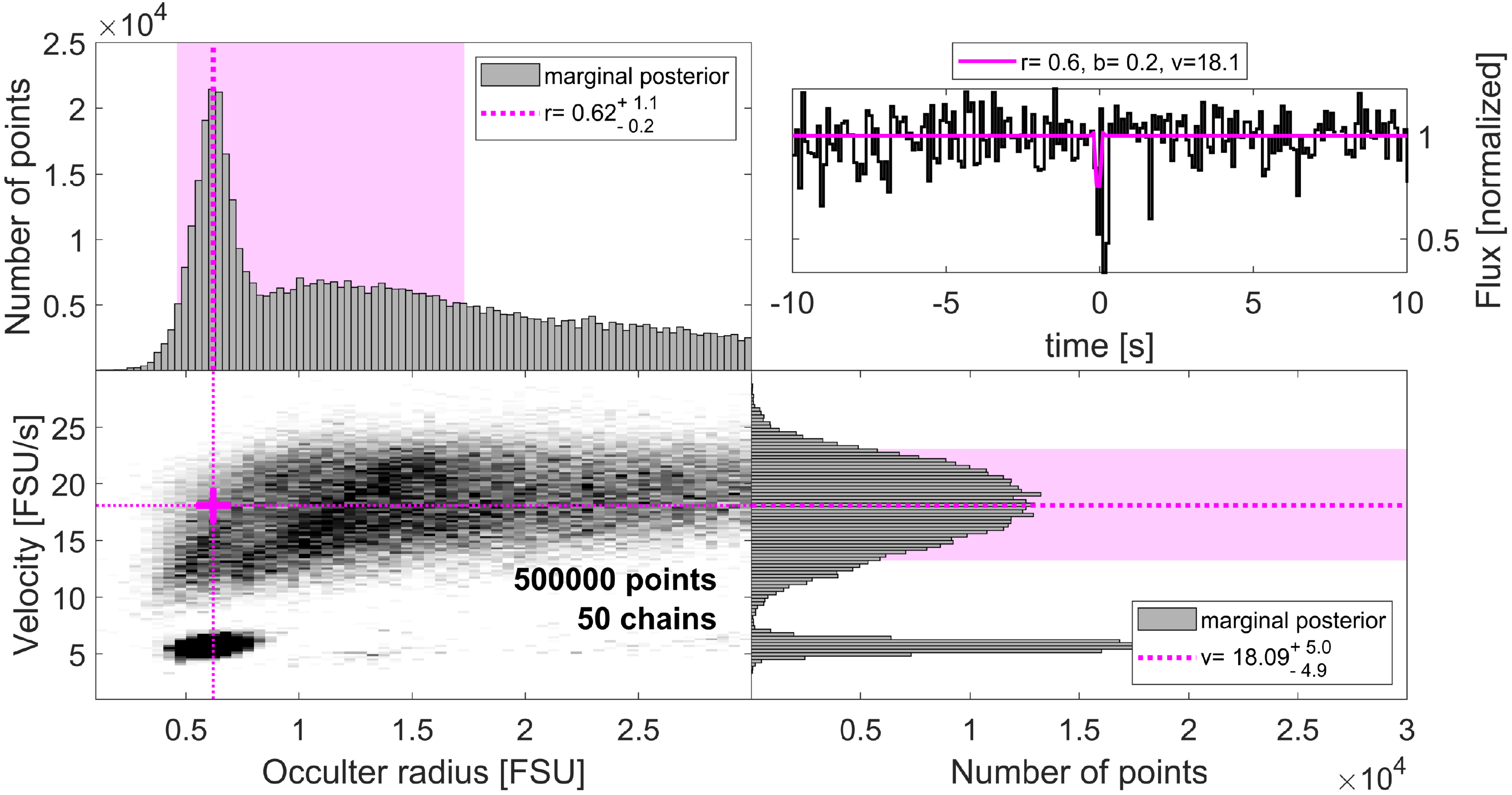}
    \caption{
        Results of an MCMC analysis (see \S\ref{sec:mcmc}) for the occultation candidate of 2021 April 12.
        The panels are the same as in Figure~\ref{fig:mcmc 2020-07-01}. 
        There are two clusters for the 2D posterior distribution. 
        The smaller one, at a lower velocity of $\approx 8\pm2$\,km\,\persec, 
        is consistent with the Earth's transverse velocity of 9.3\,km\,\persec, but the velocity of the upper cluster
        of $\approx 24\pm6.5$\,km\,\persec is too high to be consistent. 
        None of the points in either distribution 
        manages to reproduce the single-frame Poisson peak
        of this event, even with an impact parameter set to zero. 
    }
    \label{fig:mcmc 2021-04-12}

\end{figure*}

The MCMC fit to this event follows the same methods as those used for all events
(see Sections~\ref{sec:mcmc} and \ref{sec:candidate 2020-07-01}). 
The posterior distribution, shown in Figure~\ref{fig:mcmc 2021-04-12}, 
follows a bi-modal distribution, with one small spot at low velocity and radius, 
and another large distribution at higher velocities that spans most of the 
allowed range for the occulter radius. 
We also notice that the best fit light-curve, 
shown in pink on the top-right panel of Figure~\ref{fig:mcmc 2021-04-12}, 
does not reproduce the Poisson peak of the event. 

\begin{figure}
    \centering
    \pic[0.9]{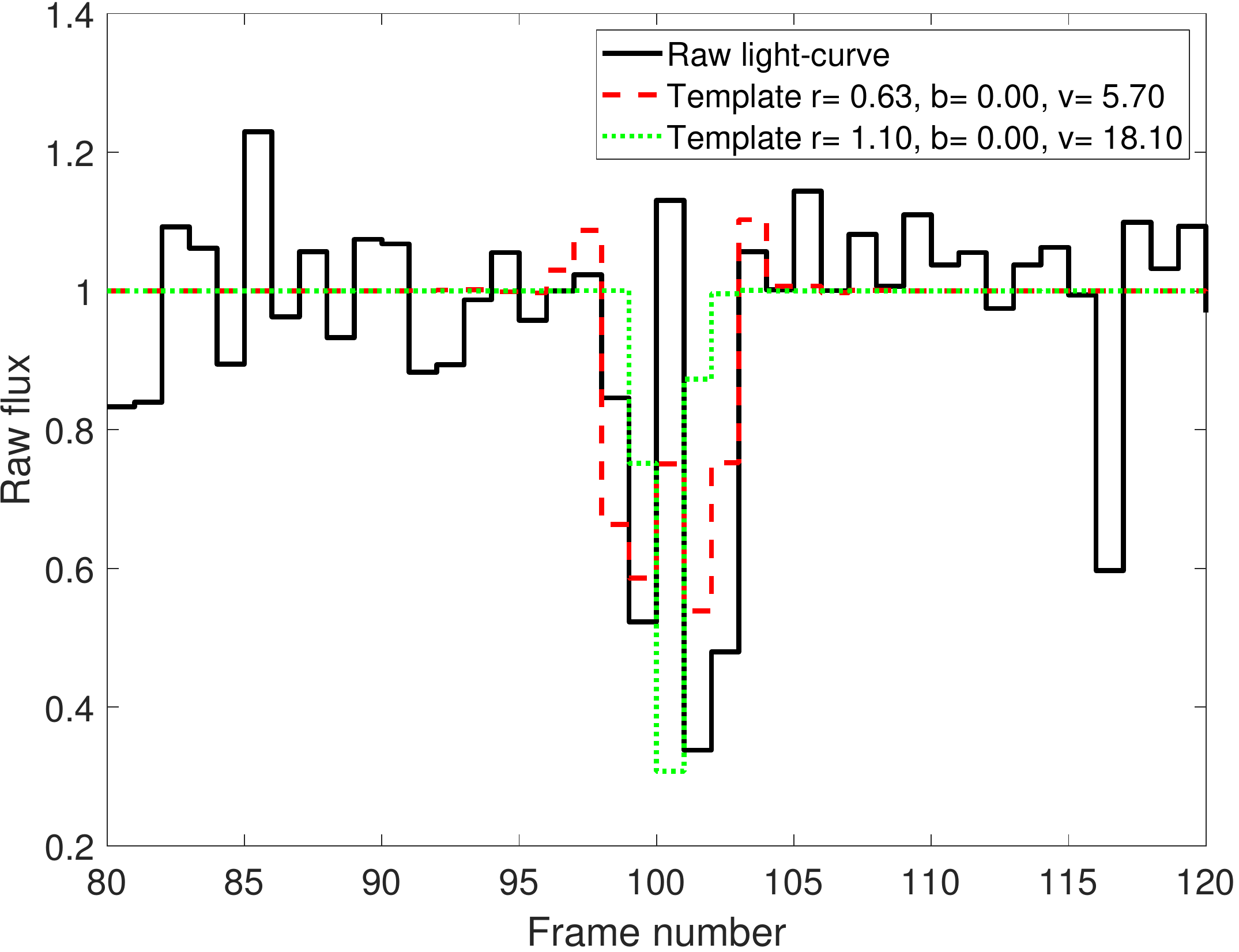}
    \caption{Comparison of occultation parameters to the light curve
             of the event at 2021-04-12. 
             We chose the central values from the bimodal posterior
             presented in Figure~\ref{fig:mcmc 2021-04-12}, 
             and show the resulting templates, 
             setting both to $R_\star=0.38$, the star's best fit value, 
             and $b=0$, the optimal impact parameter to recreate the 
             strong Poisson peak in this light curve. 
             The low-velocity template still cannot recreate
             the Poisson peak, 
             while the high-velocity template completely skips 
             that part and only produces a few dimming frames. 
    }
    \label{fig:bimodal 2021-04-12}
\end{figure}

We check whether a combination of parameters from 
one of these two distributions fits the light-curve
more successfully than using the ``best-fit'' parameters, 
which are calculated using the velocity 
of the larger distribution and the radius of the smaller one. 
We chose the central values of either distribution, 
along with $b=0$ and $R=0.38$\,FSU 
(the best fit to the star's Gaia magnitudes). 
We show the resulting light-curve templates, 
along with the real event counts in 
Figure~\ref{fig:bimodal 2021-04-12}. 
It is clear that neither of these templates
can reproduce the central peak in the raw light-curve.
Even at negligible impact parameter values, 
the low frame rate (10\,Hz) will wash out the 
brightening at the center of the illumination pattern,
such that it would not show a strong Poisson peak. 
It is much more likely that the event originates 
from an instrumental or atmospheric effect. 

\begin{figure*}
    \centering
    \pic[0.9]{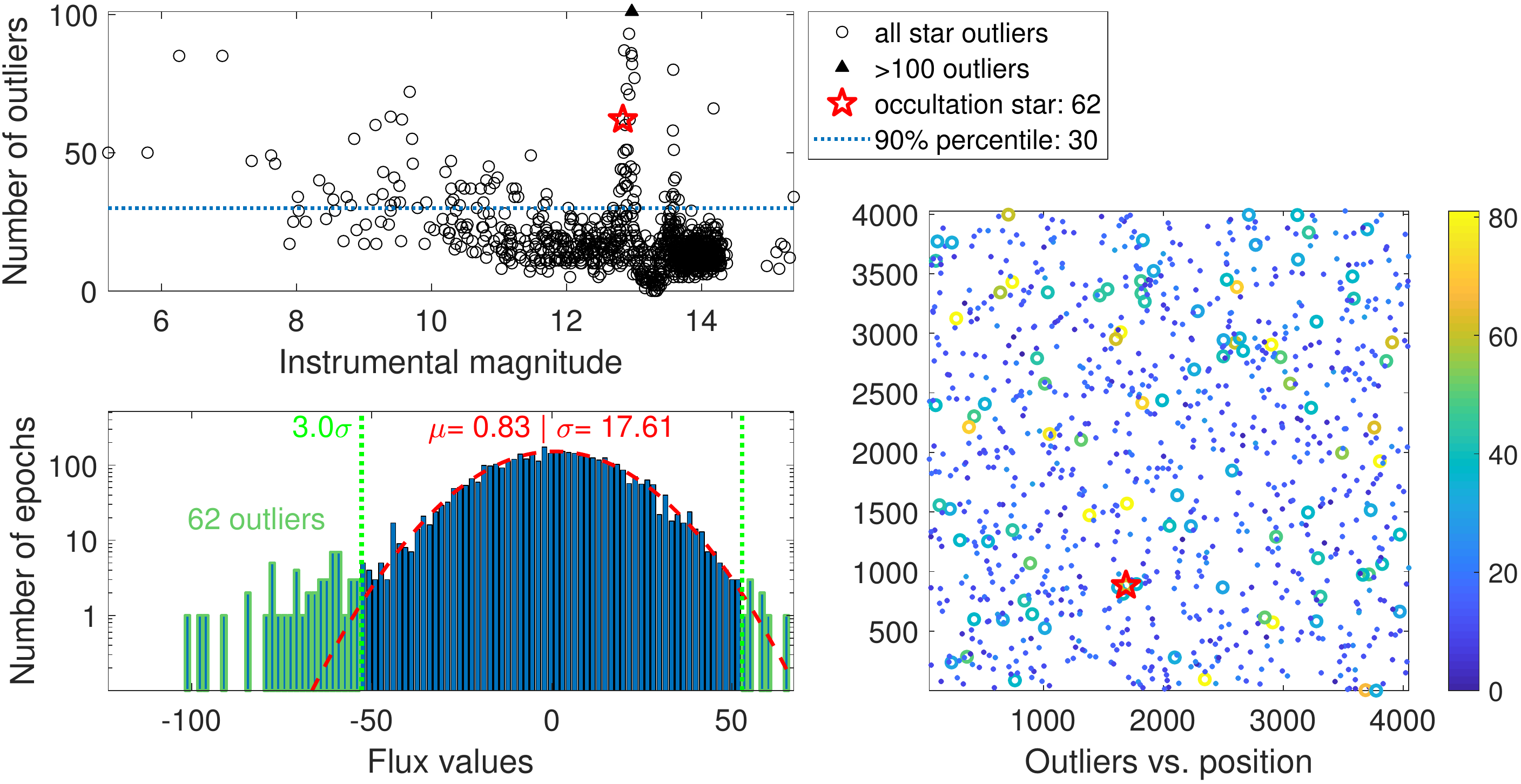}
    \caption{Outlier analysis (see \S\ref{sec:outliers}) for the occultation candidate of 2021 April 12.
             The panels are the same as in Figure~\ref{fig:outliers 2020-07-01}. 
             The occultation star has 62 outliers, which is more than double
             the 90th percentile (of 30 outliers). 
             On the bottom-right panel we see 
             the candidate star does not seem to be close to any over-abundance 
             in high-outlier stars. 
             However, just the number of outliers it has in the long-term light-curve
             indicates an instrumental effect is the likely cause of the event.          
    }
    \label{fig:outliers 2021-04-12}
\end{figure*}

An outlier analysis of the star's long term light-curve, 
as described in Section~\ref{sec:outliers} is shown in 
Figure~\ref{fig:outliers 2021-04-12}. 
As was the case for the event of 2021 April 11, 
this star's long term light-curve
(spanning 4000 frames, or 400\,s, around the event)
shows an unusually high number of dimming outlier events. 
The event star has 62 outlier frames in that time range, 
mostly dimming frames, 
while the rest of the stars' 90th percentile is only 30 outlier frames. 
This result makes it highly likely that the dimming
is caused by an instrumental effect, 
not a physical occultation. 

\begin{figure*}
    \centering
    \pic[0.9]{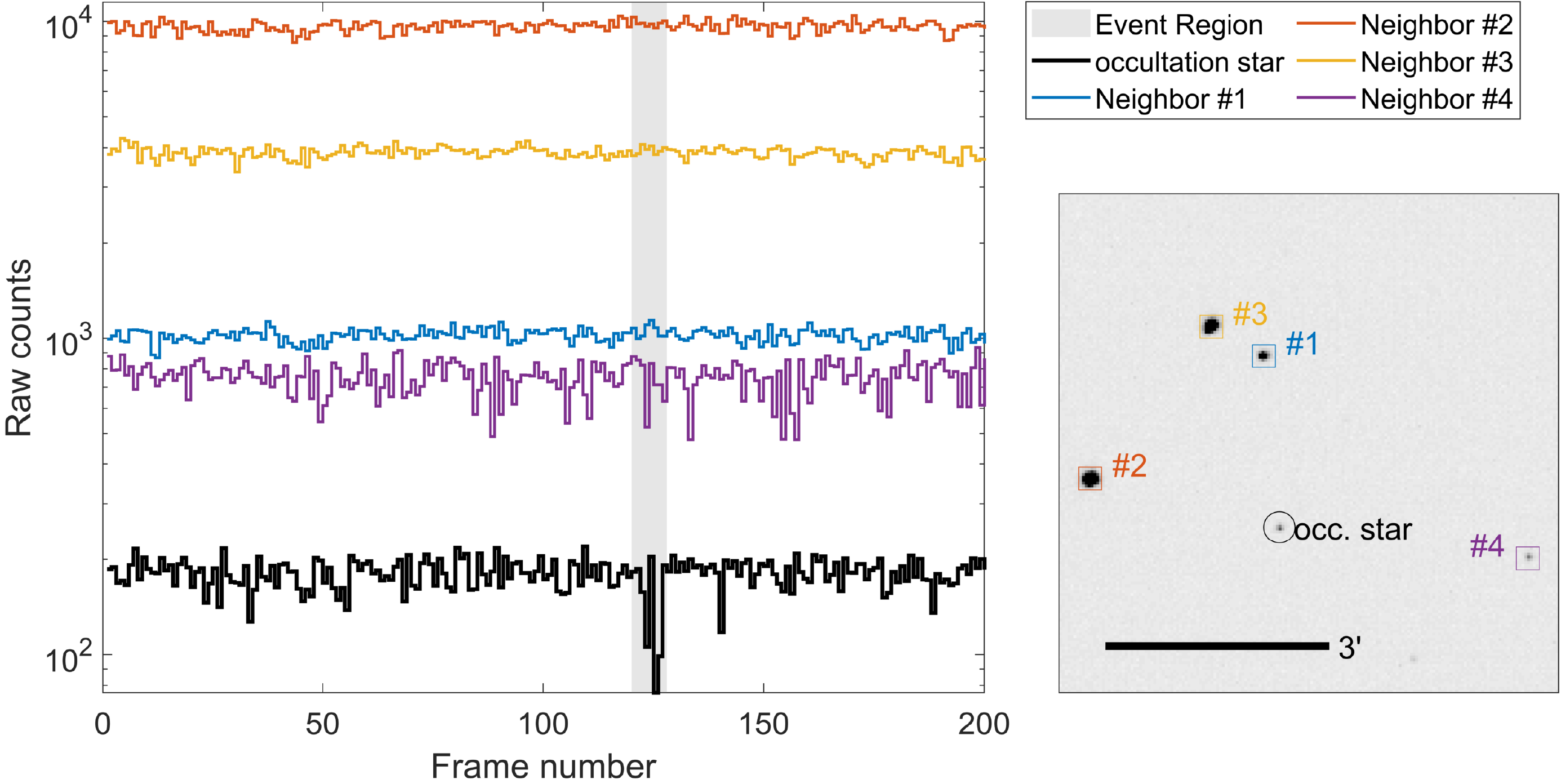}
    \caption{
        Flux values for nearby stars around the time of the event of 2021 April 12. 
        Star \#4, shown as a purple line, 
        shows strong variability throughout the event time. 
        Other stars do not show strong variability, 
        so it is hard to draw conclusions based on the nearby stars. 
    }
    \label{fig:neighbors 2021-04-12}
\end{figure*}

We also check if other stars nearby the event star 
are affected by any transient dimming. 
The results for four nearby stars with S/N>5 are shown in 
Figure~\ref{fig:neighbors 2021-04-12}. 
There is some evidence that star \#4, 
shown as the purple line,
also shows strong variability on similar time-scales
as the event star. 
This is consistent with an instrumental effect 
but also with an atmospheric event that affected
some of the stars nearby. 

\subsection{Occultation candidate 2021-04-16}\label{sec:candidate 2021-04-16}

\begin{figure*}

    \centering
    \pic[0.9]{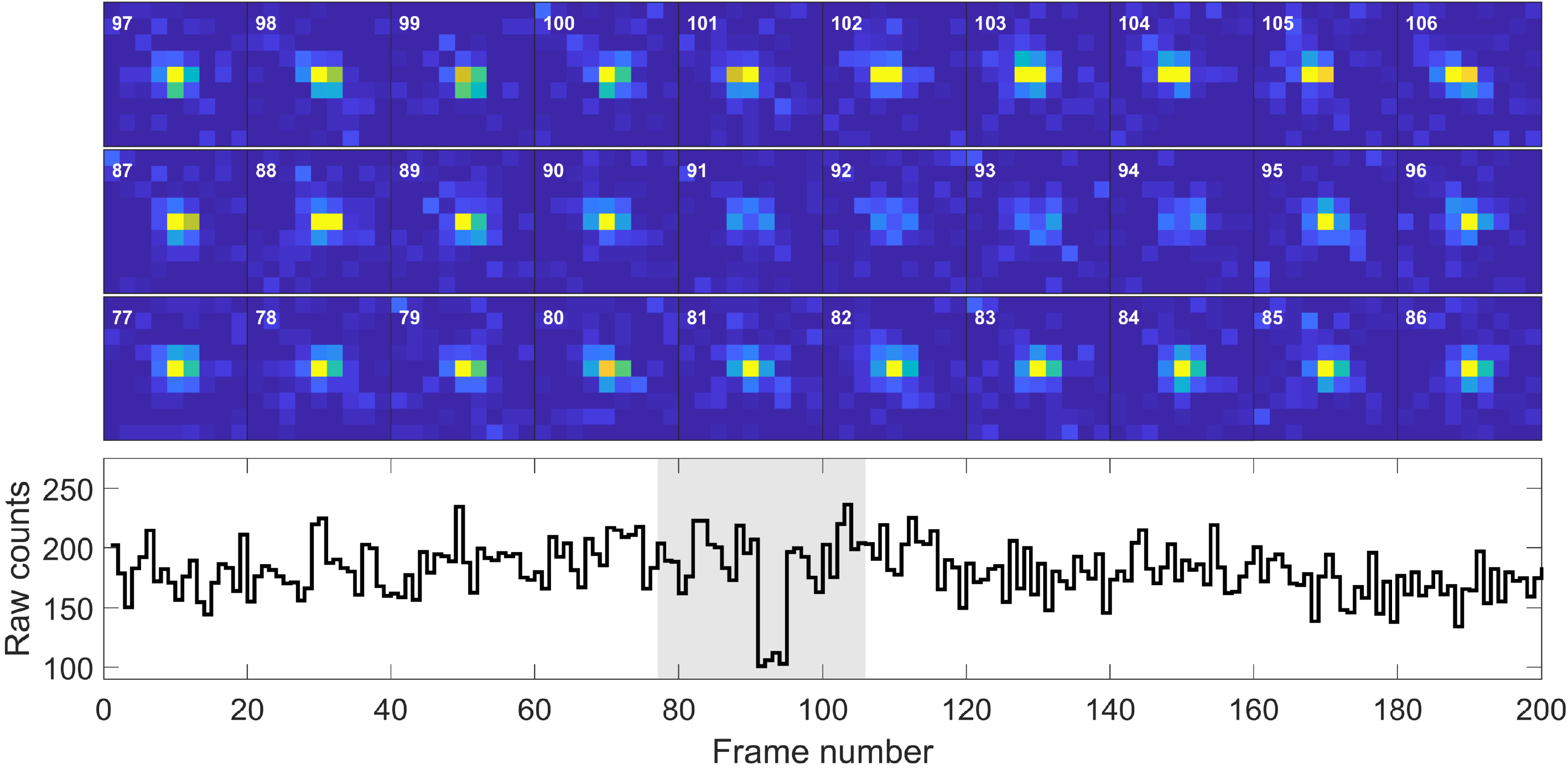}
    \caption{
         Raw counts and cutouts for the occultation candidate of 2021 April 16. 
         The cutouts are set to the same dynamic range, 
         showing that the dimming is not an artefact
         of the photometric reduction, but a feature of the underlying pixel measurements.
    }
    \label{fig:flux cutouts 2021-04-16}
\end{figure*}

This event is a multi-frame occultation candidate, 
composed of four consecutive frames of lower flux values. 
It is more promising as a true occultation candidate 
than the previous events, although still far from 
being a definite detection. 
The shape of the light-curve, as well as the cutout images around 
the time of the event are shown in Figure~\ref{fig:flux cutouts 2021-04-16}. 
The shape of the light-curve does not show any 
diffraction effects, and in fact the steep drop-off
of the intensity of light at the beginning and end of the dip
is surprising, given the high cadence of the observations. 

\begin{figure*}
    \centering
    \pic[0.9]{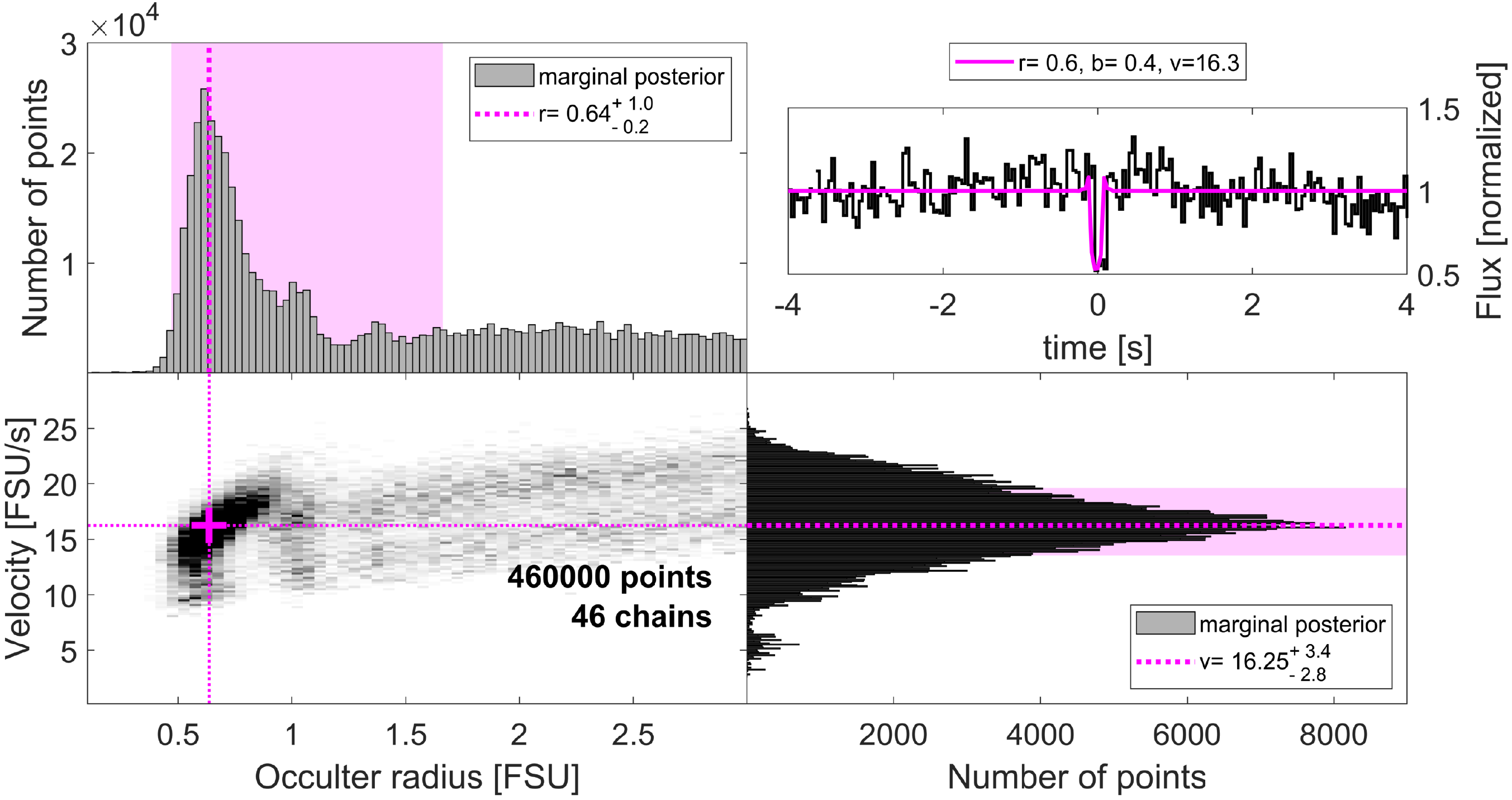}
    \caption{
        Results of an MCMC analysis (see \S\ref{sec:mcmc}) for the occultation candidate of 2021 April 16.
        The panels are the same as in Figure~\ref{fig:mcmc 2020-07-01}. 
        The majority of the posterior probability is focused on a small
        region of the parameter space which is consistent with the 
        transverse velocity of the observed field. 
        The size of the occulter is consistent with a small KBO
        ($r\approx 0.6$\,FSU$=0.9$\,km). 
    }
    \label{fig:mcmc 2021-04-16}

\end{figure*}

The MCMC fit to this event follows the same methods as used for all events
(see \S\ref{sec:mcmc} and \S\ref{sec:candidate 2020-07-01}). 
The posterior distribution, shown in Figure~\ref{fig:mcmc 2021-04-16}, 
shows a rather narrow peak centered around $r=0.6$\,FSU and $v=16.3$\,FSU\,\persec, 
and a wider range of points that likely arise from chains that are not fully converged. 
The majority of the probability, however, 
is densely concentrated, with radius and velocity
that are consistent with an occultation by a small KBO. 

\begin{figure*}
    \centering
    \pic[0.9]{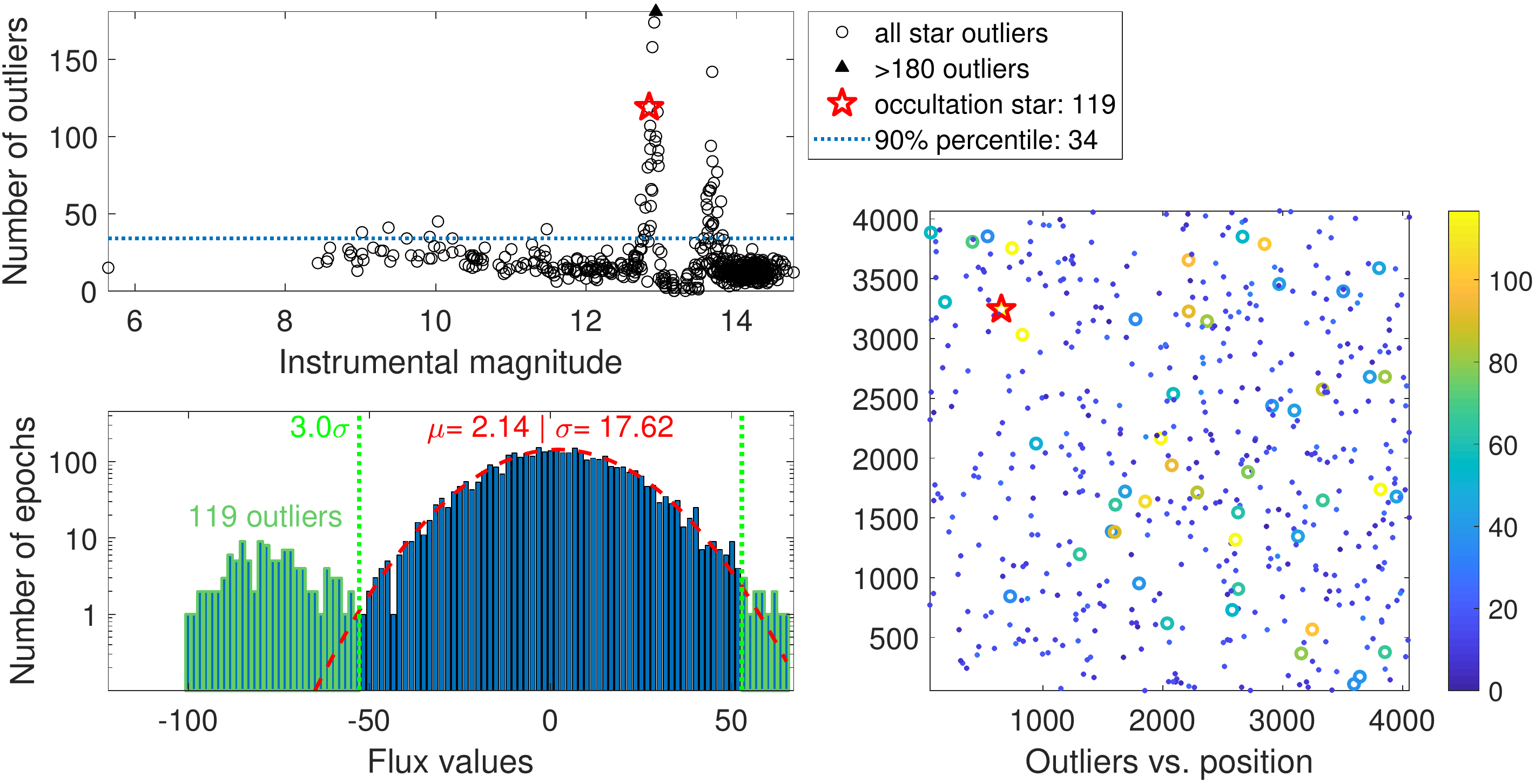}
    \caption{Outlier analysis (see \S\ref{sec:outliers}) for the occultation candidate of 2021 April 16.
             The panels are the same as in Figure~\ref{fig:outliers 2020-07-01}. 
             The occultation star has 119 outliers, which is almost three times
             the 90th percentile (of 34 outliers). 
             In the bottom-right panel we see that 
             the candidate star does not seem to be close to any over-abundance 
             in high-outlier stars. 
             However, the number of outliers it has in the long-term light-curve
             indicates an instrumental effect is the likely cause of the event. 
    }
    \label{fig:outliers 2021-04-16}
\end{figure*}

While the parameter estimation described above seem to be
consistent with a real occultation with the correct transverse velocity,  
an outlier analysis quickly shows that 
this is not likely to be the cause of the event. 
The results of such an analysis, as described in \S\ref{sec:outliers} 
is shown in Figure~\ref{fig:outliers 2021-04-16}. 
As was the case for the events of 2021 April 11 and 12, 
this star's long term light-curve
(spanning 4000 frames, or 400\,s, around the event)
shows an unusually high number of dimming outlier events. 
The event star has 119 outlier frames in that time range, 
mostly dimming frames, 
while the rest of the stars' 90th percentile is only 34 outlier frames. 
This result makes it most likely that the dimming
is caused by an instrumental effect, 
not a physical occultation. 

\begin{figure*}
    \centering
    \pic[0.9]{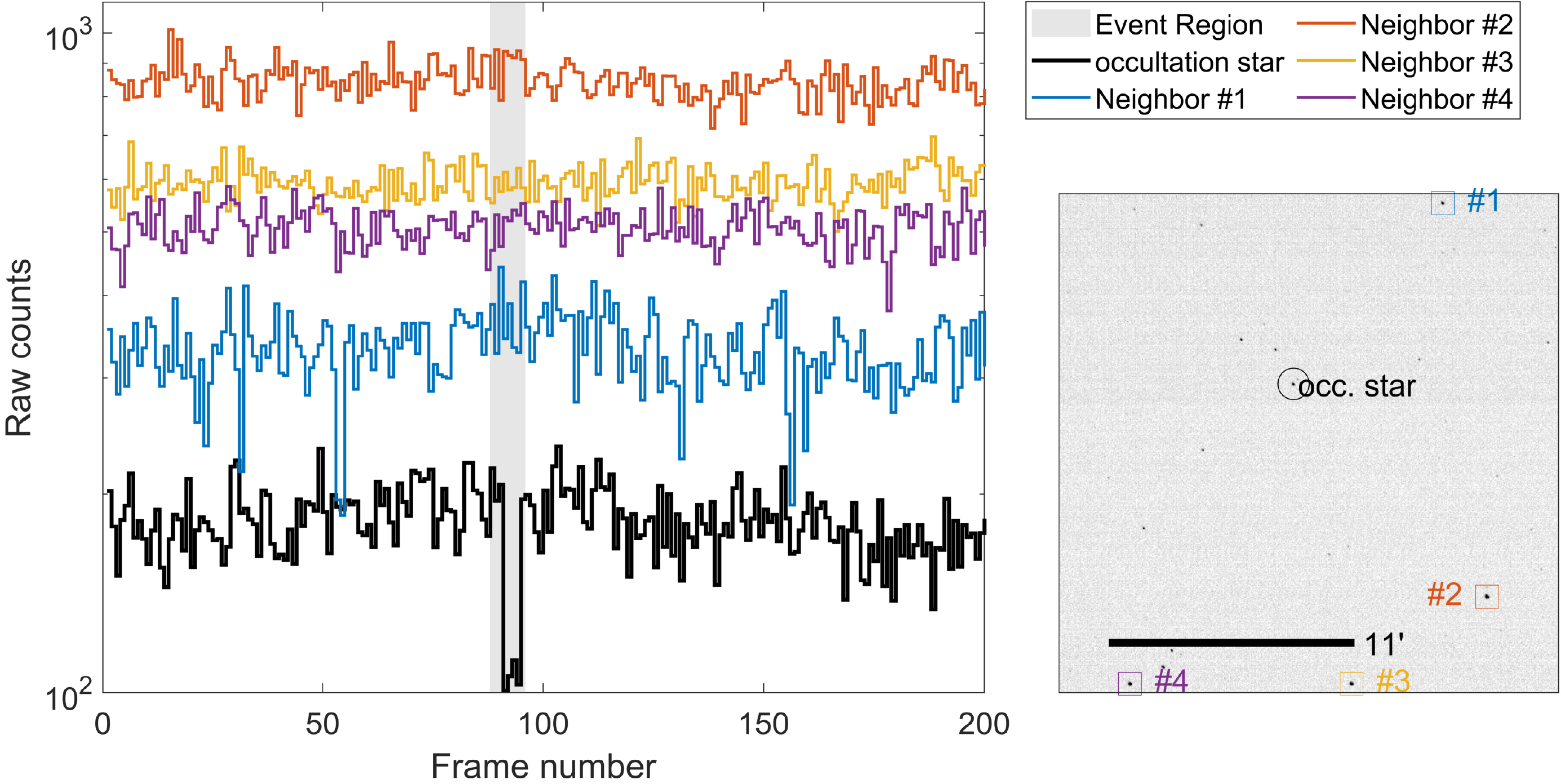}
    \caption{
        Flux values for nearby stars around the time of the event of 2021 April 16. 
        Star \#1, shown as a blue line, 
        shows strong variability throughout the frames surrounding the event. 
        Other stars do not show strong variability, 
        so it is hard to draw conclusions based on the nearby stars. 
    }
    \label{fig:neighbors 2021-04-16}
\end{figure*}

We also check if other stars nearby the event star 
are affected by any transient dimming. 
The results for four nearby stars with $S/N>5$ are shown in 
Figure~\ref{fig:neighbors 2021-04-16}. 
There is some evidence that star \#1 (the closest one, 
shown as the blue line), 
also shows strong variability on similar time-scales
as the event star. 
This is consistent with an instrumental effect 
but also with an atmospheric event that affected
some of the stars nearby. 
However, it should be noted
that the closest stars are more than 10 arc-minutes 
from the event star, so may not be good indicators
for variability for this event. 

\subsection{Occultation candidate 2021-09-14}\label{sec:candidate 2021-09-14}

\begin{figure*}

    \centering
    \pic[0.9]{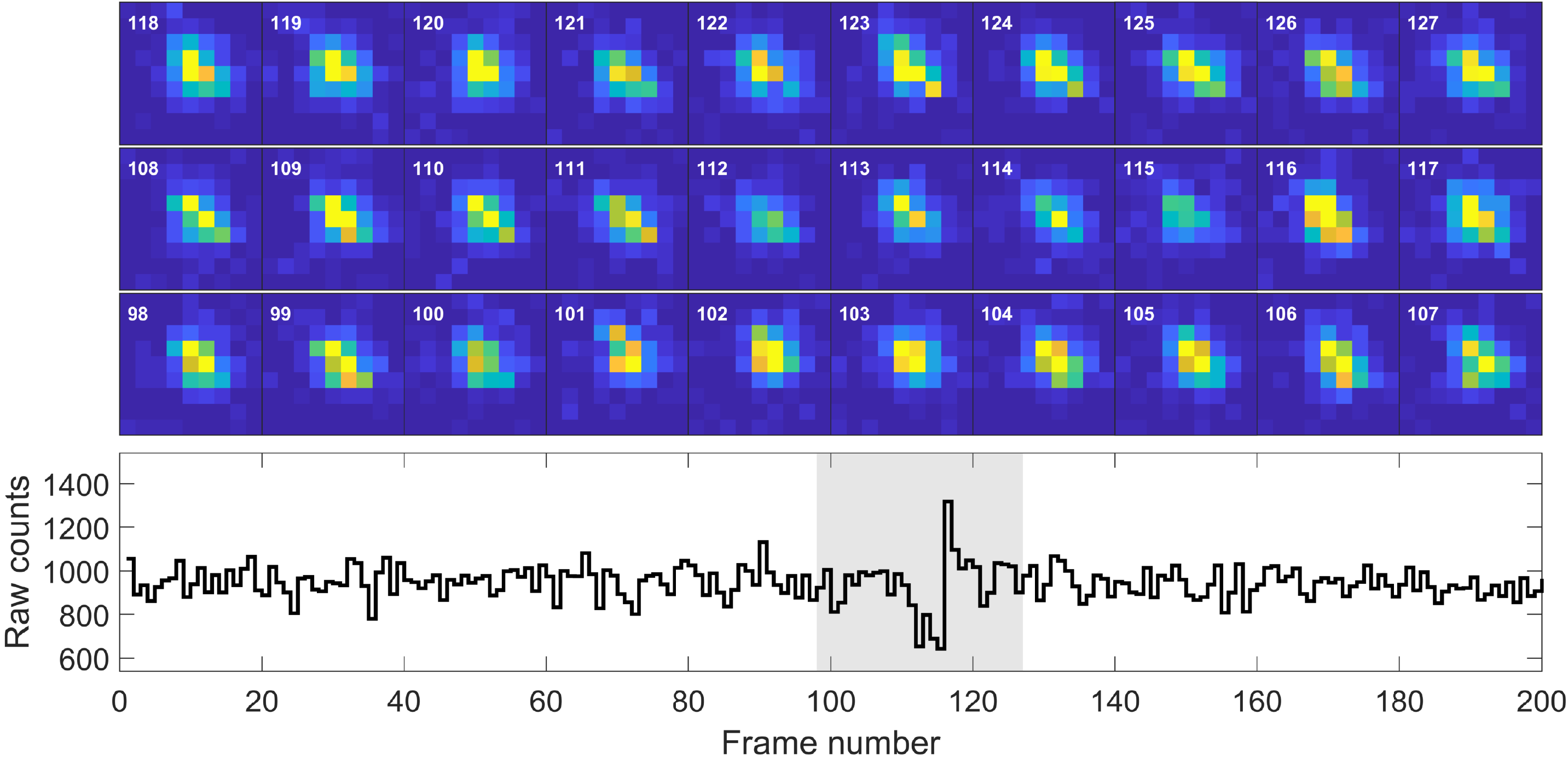}
    \caption{
         Raw counts and cutouts for the occultation candidate of 2021 September 14. 
         The cutouts are set to the same dynamic range, 
         showing that the dimming and brightening are not artefacts
         of the photometric reduction, but features of the underlying pixel measurements.
    }
    \label{fig:flux cutouts 2021-09-14}
\end{figure*}

This event is a multi-frame occultation candidate, 
composed of four consecutive frames of lower flux values, 
of which one is slightly brighter, 
and at least another frame where the flux exceeds the average.
It is the only event in 2021 that was not detected in 
the first two-and-a-half weeks of April, 
and it looks more consistent with a real 
occultation than all other events. 
It does not, however, appear to be a clear true
detection or a definite false-positive. 
The shape of the light-curve, as well as the cutout images around 
the time of the event are shown in Figure~\ref{fig:flux cutouts 2021-09-14}. 
The shape of the light-curve is asymmetric, 
with a clear brightening after an initial dip. 
Such light-curves often arise from binary or elongated 
occulters, according to simulations (see \cite{wfast_kbo_pipeline_Nir_2023}).

\begin{figure*}
    \centering
    \pic[0.9]{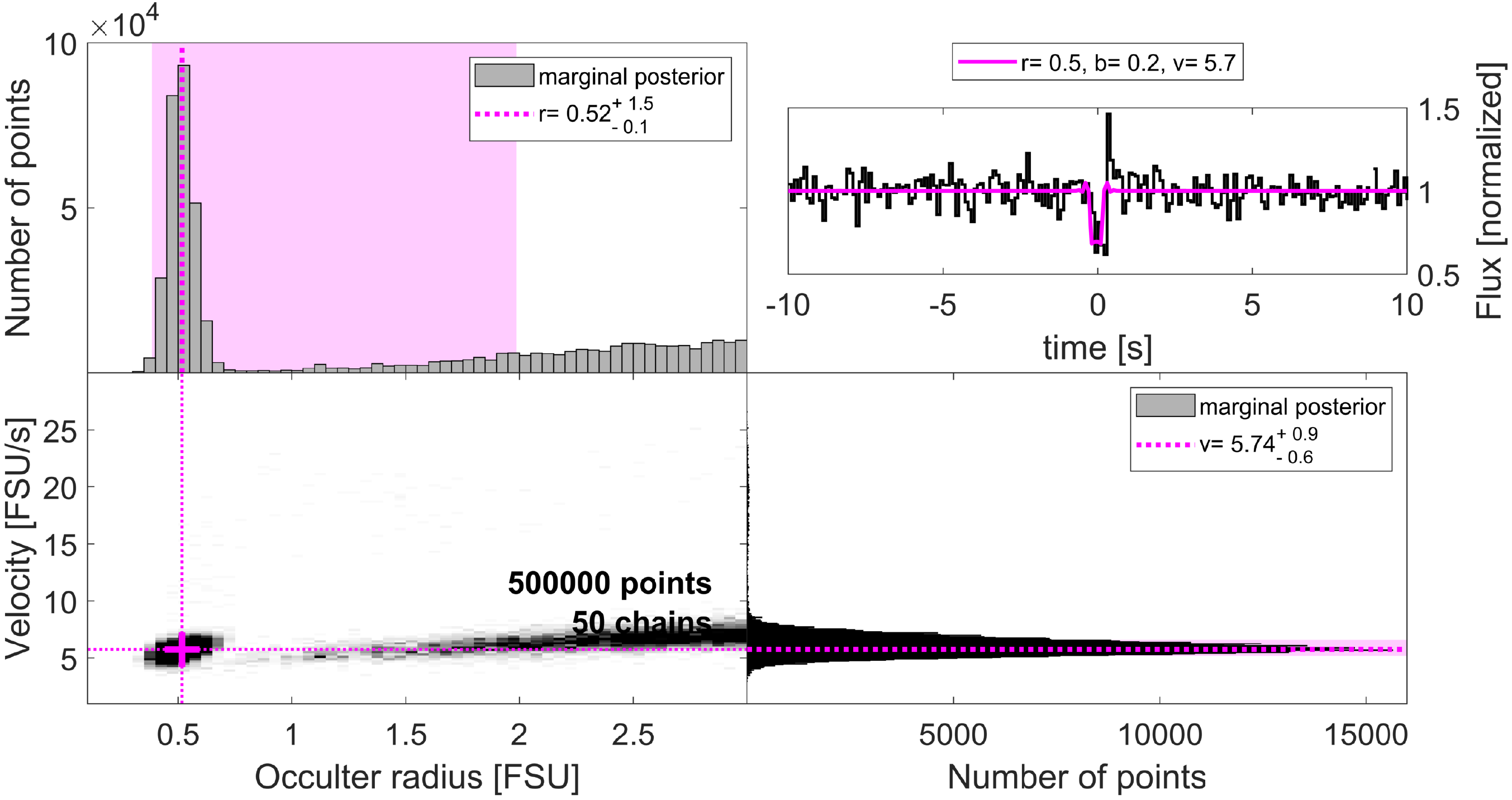}
    \caption{
        Results of an MCMC analysis (see \S\ref{sec:mcmc}) for the occultation candidate of 2021 September 14.
        The panels are the same as in Figure~\ref{fig:mcmc 2020-07-01}. 
        The majority of the posterior probability is focused on a small
        region of the parameter space with low occulter radii ($\approx 0.65$\,km)
        and velocities ($\approx 7.5\pm0.1$\,km\,\persec), 
        the latter being inconsistent with the field's transverse velocity of 20.3\,km\,\persec.
    }
    \label{fig:mcmc 2021-09-14}

\end{figure*}

The MCMC fit to this event follows the same methods as used for all events
(see Sections~\ref{sec:mcmc} and \ref{sec:candidate 2020-07-01}). 
The posterior distribution, shown in Figure~\ref{fig:mcmc 2021-09-14}, 
shows a rather narrow peak centered around $r=0.5$\,FSU and $v=5.7$\,FSU\,\persec, 
and another region with similar velocity and a wider occulter radius range. 
The majority of the probability, however, 
is densely concentrated, with radius and velocity
that are consistent with an occultation by a small KBO
at a low velocity. 
A light-curve template associated with such an occultation
is shown in the top-right panel of Figure~\ref{fig:mcmc 2021-09-14}, 
which fits the dip in flux fairly successfully, 
but does not account for the small brightening in the middle 
of the dip or the larger brightening after it. 
The inferred velocity is inconsistent with a KBO, 
which will require a transverse velocity of $\approx 10$\,FSU\,\persec$= 13$\,km\,\persec.

An outlier analysis, as described in Section~\ref{sec:outliers}
does not reveal an over-abundance of flux outliers in the 
4000 frame (400\,s) time span around the event. 
The star's light-curve contains 21 outliers in that time, 
less than the 90\% percentile of 29, and similar to the mean
of stars of that brightness. 

\begin{figure*}
    \centering
    \pic[0.9]{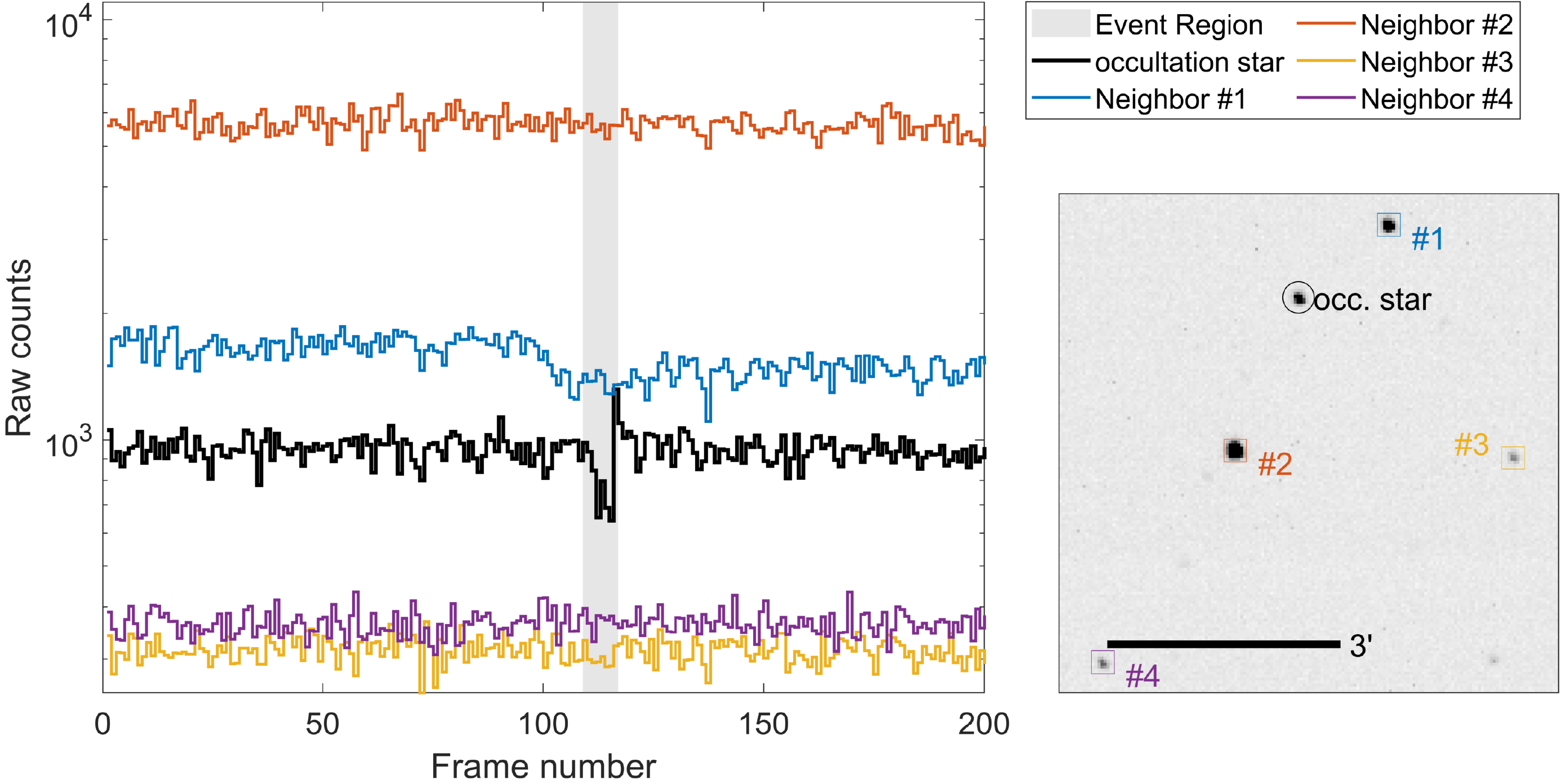}
    \caption{
        Flux values for nearby stars around the time of the event of 2021 September 14. 
        Star \#1, shown as a blue line, 
        seems to drop in flux right before the candidate occultation occurs, 
        but remains at a ``low state'' for the remaining frames in that batch. 
        The shape of the two changes in flux is not similar, 
        and it is unclear what sort of transient effect could cause both. 
        Other stars do not show strong variability, 
        so it is hard to draw conclusions based on the nearby stars. 
    }
    \label{fig:neighbors 2021-09-14}
\end{figure*}

We also check if other stars nearby the event star 
are affected by any transient dimming. 
The results for four nearby stars with S/N>5 are shown in 
Figure~\ref{fig:neighbors 2021-09-14}. 
There is some evidence that star \#1 (the closest one, 
shown as the blue line), 
changes brightness within about 10 frames (1\,s) 
from the time when the candidate event occurs. 
Other stars within a few arc-minutes
do not show any unusual variability. 
While the nearby star's behavior is suspicious, 
it does not appear similar in nature to the 
dimming and brightening of this event. 

\subsection{Candidate summary}\label{sec:candidate summary}

The seven candidate occultations presented here are all ruled out for 
one reason or another. Some are clearly false positives, 
such as events with many outliers in the long term light-curves. 
Others seem likely to be atmospheric, such as the events
of 2020-07-01 or 2021-09-14 that show both decrease and increase
of flux, appearing ``flare-like'' which makes them suspiciously similar
to the flare events presented in \S\ref{sec:flare events}. 
Two events are single-frame events where there is insufficient information 
to confirm or rule out the events, but the low 
transverse velocity makes them unlikely to be KBO occultations. 

We summarize the properties of all seven candidates
in Table~\ref{tab:candidate summary}. 
One event (2020 July 1) is flare-like and off-ecliptic. 
Three events have many outliers (2021 April 11, 12 and 16). 
Three events have inconsistent transverse velocities
(2020 July 1, 2021 April 1, 3 and September 14). 
It is clear that none of the candidates
is fully consistent with what we expect from a KBO occultation. 

We thus conclude that none of the candidate events 
are caused by KBO occultations. 
We calculate the limits on the number density based on 
this null result in \S\ref{sec:limits}. 
The variety of light-curve morphologies should be taken 
as an indication for the many types of false-positives that could arise in such a search for fast transients under atmospheric conditions.

\section{Flare events}\label{sec:flare events}

In addition to finding occultation events, 
our pipeline was also sensitive to ``flares'', 
i.e., events where the flux is mostly increasing, 
rather than dimming. 
Many such events were detected and flagged 
as satellite glints or tracks. 

\begin{figure}
    \centering
    \pic[1]{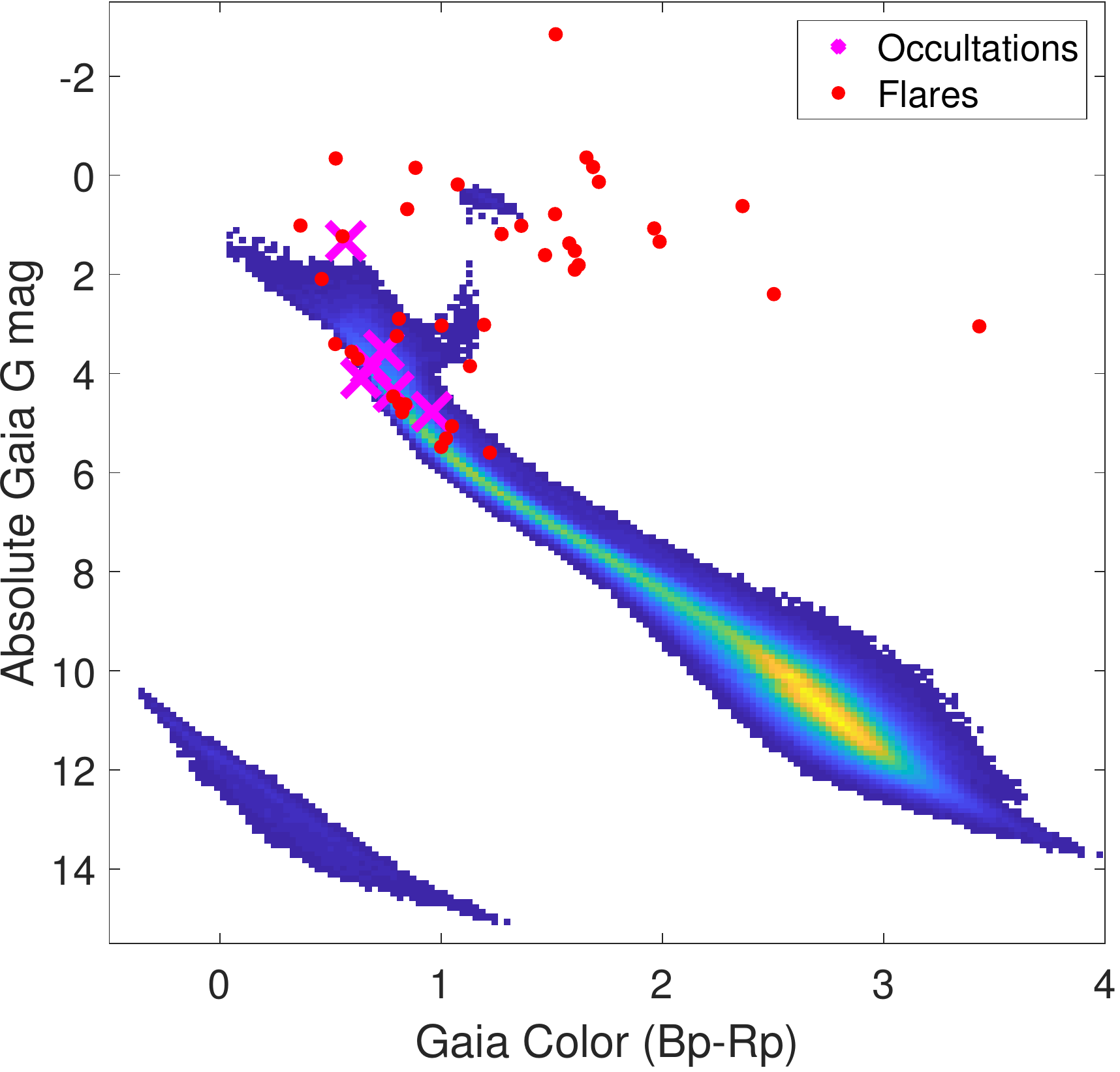}
    \caption{The color-magnitude of occultation-like events (magenta crosses)
             and flares (red circles) superimposed on the density map of stars 
             from Gaia DR2. 
             Events appear to occur on brighter stars, 
             which could be a detection bias
             or an indication that these are artefacts. 
             Neither type of event seems to be associated
             with a population of active stars.     
             Absorption was used for all stars with a known absorption from the Gaia catalog. 
             One occultation-like event and one flare are not shown as 
             the underlying star did not have positive parallax measured in the catalog.             
    }
    \label{fig:hr diagram}
\end{figure}

\begin{figure*}

    \centering
    \pic[1]{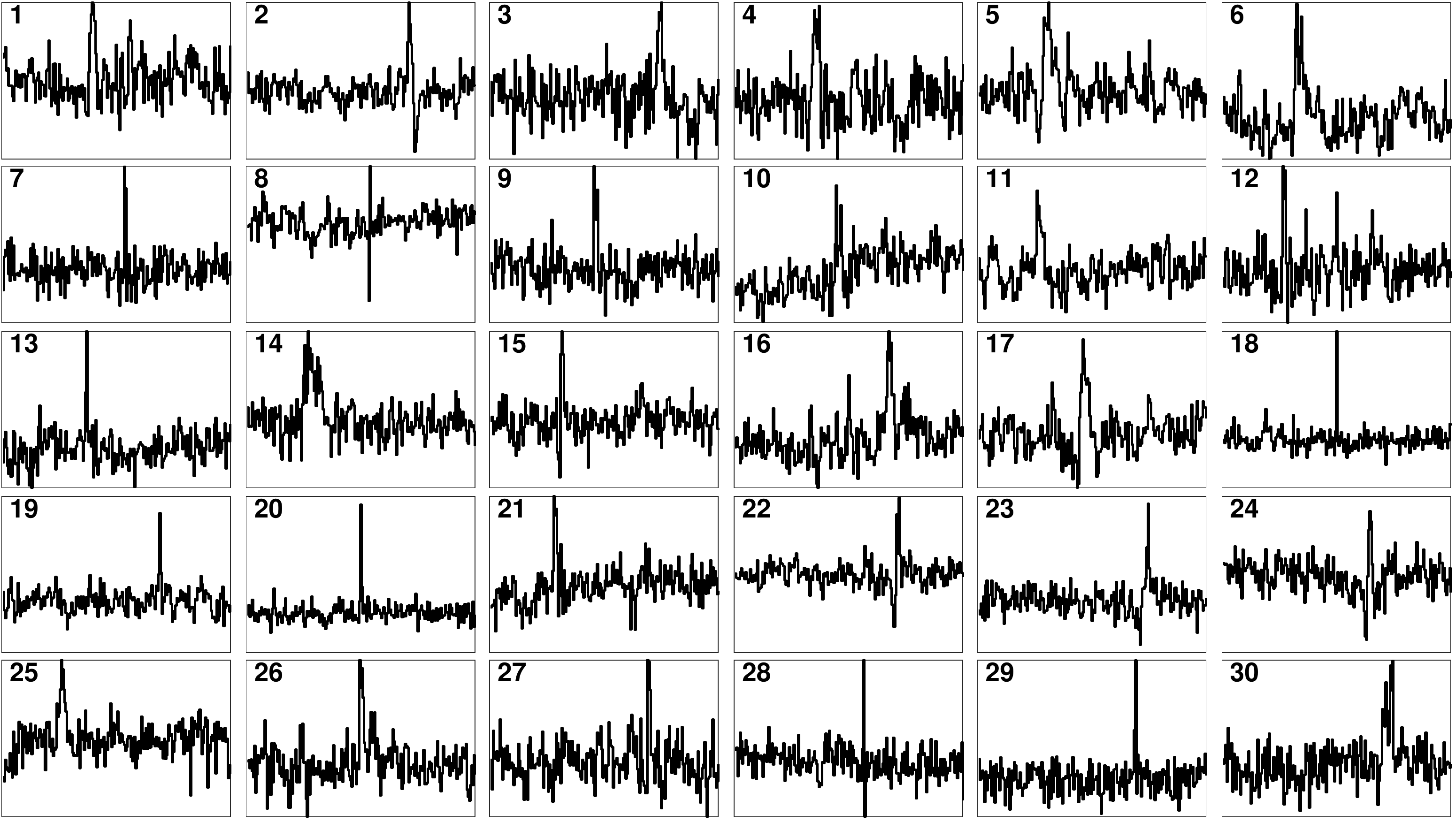}
    \caption{Light-curves for all flares detected in the 2020--2021 observing seasons. 
             These events remain after removing all visible indications of source motion, 
             caused by satellite glints. 
             The flux scale is normalized to the median of each light-curve, 
             to show the morphology of the flares. 
             Information about each individual flare is given in Table~\ref{tab:flares}. 
    }
    \label{fig:flare lightcurves}

\end{figure*}

After removing all obvious satellites we detected 41 
potential flares. 
Careful inspection of the cutout images has ruled out
a further 11 events where the satellite was seen 
glinting close to the edge of the cutout or 
just as a faint moving object relative to the star. 
The remaining 30 flares could be due to
(a) more satellites that somehow were visible only near the star;
(b) occultations where a bright diffraction fringe passes over the observer; 
(c) atmospheric fluctuations that momentarily focus the star's light;
(d) some unknown transient phenomena that increases the flux of stars for less than a second. 
The color-magnitude position of all 30 flares, and the occultation-like events, 
is shown in Figure~\ref{fig:hr diagram}, superimposed on the distribution
of colors and magnitudes of stars in Gaia DR3. 
The events are all on relatively bright stars, 
which could be interpreted as a detection bias
or as an indication that these are instrumental/atmospheric artefacts. 
Both kinds of events do not seem associated with 
a particular population of active stars. 
The lightcurves for all 30 flares is shown in 
Figure~\ref{fig:flare lightcurves} on arbitrary flux axes 
(normalized to the median of each light-curve). 

It is likely that some of the flares are due to satellite glints
that perfectly align with the star's position and do not
reflect any additional light before or after 
crossing close to the star's position. 
\cite{satellite_glints_Nir_2021} reported such glints, 
with a rate of $1.3 \times 10^{-7}$ glints per hour per arc-second squared, 
for glints brighter than 11th magnitude. 
With $\approx 10^6$ star-hours, and the images of stars
with a typical FWHM of $\approx 5''$, 
we can estimate there would be $\approx 3$ satellite glints in our data.  
The stars in our sample are commonly dimmer than 11th magnitude. 
If the magnitude distribution of glints 
does not drop off after the 11th magnitude, 
there could potentially be even more glints, 
with similar flux to the stars they are on. 

Bright diffraction fringes can also appear as flux increases, 
usually accompanied by dimming before or after the flare. 
This is indeed seen in some simulated light-curves (see \citealt{wfast_kbo_pipeline_Nir_2023}). 
However, this is relatively rare, compared to normal (dimming) occultations. 
Looking at simulated occultation templates within the range
of occulter radii $0.3<r<3$\,FSU and impact parameters $0<b<2$\,FSU, 
we find less than 30\% of events have more positive flux points
than dimming points. 
Therefore, even if a few events are due to this effect, 
there should be $\times 3$ more dimming than flaring 
occultations in our data. 
Binary or elongated occulters have more likelihood 
to produce flux increases, but even then most events
would also produce dimming, unless the major axis 
of the occulting body (or bodies) is perfectly 
aligned relative to the direction of motion, 
which is not particularly likely (see \citealt{wfast_kbo_pipeline_Nir_2023}). 

Atmospheric phenomena remains a viable explanation for
at least some of the flare events, and also most of the
occultation events discussed in the previous section. 
Inspecting histograms of outlier flux values does not 
immediately show a long tail of flux variations that 
would be expected from random atmospheric fluctuations. 
Observations with multiple telescopes could help 
conclude if these are extreme atmospheric events or 
caused by physical phenomena outside the Earth. 

Finally, although less likely, 
we cannot rule out that these are actual 
unknown transient phenomena occurring at the star or in 
the light's path through the Solar System. 
We find this explanation most unlikely. 

A summary of the properties of the flare events
is given in Table~\ref{tab:flares}. 
We do not apply the additional parameter estimation
or outlier analysis for these events, 
but we do note the flares that show some
dimming in the light-curve that makes them 
similar to occultation events. 

\begin{table*}
    \scriptsize
    \caption{Summary of flare events. 
             Most of the columns are the same as in Table~\ref{tab:candidate summary}. 
             The columns for parameter ranges, neighbors, and outlier numbers 
             have been removed as these events did not get the full processing
             as the occultation candidates. 
             The column for flare-like is replaced with ``occultation-like'', 
             where a flare would be occultation-like if it displays a
             pronounced dip in the flux, in addition to the brightening. 
             The first column is a serial number which refers to the numbers
             of the panels in Figure~\ref{fig:flare lightcurves}. 
    }
    \label{tab:flares}
    
    \begin{tabular}{cccccccccccccc}
    
        \hline
        number & Event time (UTC) & coordinates             & $\beta$ & event S/N & $M_G$ & $T_\text{eff}$ & $R_\star$ & phot.~S/N & $f$  & FWHM     & A.M. & $V_T$       & occultation-like  \\ 
               &                  & (hh:mm:dd$\pm$dd:mm:ss) & (deg)   &           &       & ($^\circ$K)    & (FSU)     &           & (Hz) & (arcsec) &      & (km\persec) &             \\ \hline
         1 & 2020-07-25 19:19:23 & 18:05:31.0-19:56:44.6 &   3.0 &  7.91 & 12.11 & 4995 &  0.4 &  6.63 &   25 &  4.79 &  1.59 &  27.1 & No \\ 
         2 & 2020-07-25 21:03:36 & 18:05:38.1-21:21:22.0 &   3.0 &  7.50 & 10.58 & 3977 &  2.5 &  8.68 &   25 &  5.72 &  1.76 &  27.1 & Yes \\ 
         3 & 2020-07-25 22:31:47 & 18:05:22.7-21:40:25.3 &   3.0 &  8.02 &  8.26 & 5867 &  2.0 & 10.62 &   25 &  5.41 &  2.51 &  27.1 & No \\ 
         4 & 2020-08-04 20:01:05 & 20:17:56.1-17:40:23.3 &   1.9 &  7.72 & 10.16 & 7762 &  0.5 & 13.86 &   25 &  5.28 &  1.58 &  25.7 & No \\ 
         5 & 2020-08-06 20:19:27 & 20:23:21.9-17:25:42.6 &   1.9 &  7.52 & 13.01 & 7656 &  0.1 &  5.72 &   25 &  5.89 &  1.52 &  25.6 & No \\ 
         6 & 2020-08-10 19:57:36 & 18:08:23.8-19:11:38.9 &   3.0 &  8.05 & 10.95 & 5802 &  0.8 & 12.20 &   25 &  4.76 &  1.75 &  24.9 & No \\ 
         7 & 2020-08-10 18:00:40 & 20:18:39.6-16:26:38.6 &   1.9 &  9.10 & 12.03 & 4134 &  0.9 &  5.31 &   25 &  6.01 &  2.13 &  25.4 & No \\ 
         8 & 2020-08-28 20:00:47 & 19:40:18.4+40:13:22.5 &  59.2 &  8.25 & 12.72 & 5643 &  0.3 &  4.14 &   25 &  7.21 &  1.05 &  16.8 & No \\ 
         9 & 2020-09-10 17:57:28 & 18:08:21.0-19:32:50.3 &   3.0 &  7.59 & 11.71 & 5838 &  0.9 &  7.29 &   25 &  9.20 &  1.75 &  21.0 & No \\ 
        10 & 2020-09-10 18:01:00 & 18:09:03.3-21:23:12.6 &   3.0 &  8.20 & 11.59 & 7117 &  0.3 &  7.85 &   25 &  5.76 &  1.77 &  21.0 & No \\ 
        11 & 2020-09-10 18:01:36 & 18:08:41.6-20:05:13.8 &   3.0 &  8.23 &  7.39 & 4332 &  5.3 & 10.31 &   25 &  8.46 &  1.77 &  21.0 & No \\ 
        12 & 2021-04-02 20:01:38 & 07:58:46.1+19:56:51.8 &  -1.5 &  7.51 & 11.30 & 4733 &  0.8 & 14.61 &   10 &  7.51 &  1.38 &   2.3 & No \\ 
        13 & 2021-04-03 17:14:36 & 06:33:20.5+20:28:52.9 &  -1.7 &  8.21 & 10.33 & 4613 &  1.3 & 28.12 &   10 &  5.27 &  1.11 &   5.0 & No \\ 
        14 & 2021-04-03 17:37:16 & 06:34:43.4+20:45:41.8 &  -1.7 &  7.89 & 11.11 & 6738 &  0.6 & 22.06 &   10 &  6.59 &  1.16 &   5.0 & No \\ 
        15 & 2021-04-21 00:08:59 & 18:09:13.2-20:27:20.7 &   3.0 & 13.53 & 12.32 & 3671 &  2.0 &  6.36 &   10 &  6.33 &  1.82 &  22.9 & Yes \\ 
        16 & 2021-06-01 01:20:54 & 18:05:14.0-19:35:31.1 &   3.0 &  8.16 & 10.92 & 4236 &  1.6 & 13.82 &   10 &  8.35 &  1.99 &  28.3 & No \\ 
        17 & 2021-06-04 20:02:52 & 15:43:03.3-19:41:04.7 &   0.2 &  7.61 & 10.45 & 7050 &  0.7 & 19.90 &   10 &  5.23 &  1.57 &  25.7 & No \\ 
        18 & 2021-06-06 21:49:23 & 18:04:02.3-21:09:07.5 &   3.0 &  7.61 & 10.63 & 7154 &  0.4 & 19.23 &   10 &  4.70 &  1.65 &  28.7 & No \\ 
        19 & 2021-06-11 22:09:13 & 18:05:02.3-20:32:03.4 &   3.0 &  7.52 & 10.79 & 5773 &  0.6 & 12.32 &   10 &  5.15 &  1.59 &  29.0 & No \\ 
        20 & 2021-06-15 22:21:43 & 16:57:45.0-25:51:04.0 &  -1.7 &  7.61 & 12.83 & 4391 &  0.5 & 10.57 &   10 &  5.83 &  1.89 &  28.6 & No \\ 
        21 & 2021-06-19 19:49:35 & 18:06:54.6-21:05:26.5 &   3.0 &  9.12 & 12.93 & 3670 &  1.9 &  6.55 &   10 &  4.85 &  1.96 &  29.3 & No \\ 
        22 & 2021-06-27 20:14:20 & 15:35:29.1-19:49:50.7 &   0.2 & 13.60 & 12.98 & 5086 &  0.4 & 12.16 &   10 &  5.85 &  1.67 &  23.4 & Yes \\ 
        23 & 2021-07-07 19:41:44 & 18:05:31.8-20:04:56.7 &   3.0 &  7.87 & 11.40 & 4360 &  1.2 & 18.42 &   10 &  4.90 &  1.67 &  28.8 & No \\ 
        24 & 2021-07-08 19:42:11 & 18:00:23.5-20:34:21.0 &   3.0 &  7.98 & 10.65 & 5314 &  0.8 & 25.22 &   10 &  6.50 &  1.66 &  28.8 & No \\ 
        25 & 2021-07-10 22:48:07 & 22:02:33.5-10:29:14.2 &   1.3 &  7.53 & 12.28 & 4973 &  0.4 & 13.22 &   10 &  6.59 &  1.48 &  17.5 & No \\ 
        26 & 2021-07-10 22:49:37 & 22:03:43.3-09:31:33.7 &   1.3 &  8.06 & 10.05 & 4127 &  2.3 & 16.47 &   10 &  5.11 &  1.48 &  17.5 & No \\ 
        27 & 2021-07-12 20:18:34 & 18:04:03.3-21:41:45.1 &   3.0 &  9.79 & 13.58 & 5350 &  0.3 &  8.17 &   10 &  8.10 &  1.59 &  28.5 & No \\ 
        28 & 2021-07-14 22:14:11 & 18:07:12.1-19:48:49.8 &   3.0 & 13.76 & 14.12 & 4882 &  0.3 &  7.11 &   10 &  6.72 &  1.89 &  28.3 & Yes \\ 
        29 & 2021-07-16 22:08:51 & 18:05:12.4-19:48:52.8 &   3.0 & 13.78 & 13.96 & 5993 &  0.2 &  7.36 &   10 &  5.91 &  1.91 &  28.1 & No \\ 
        30 & 2021-07-28 18:41:55 & 18:22:56.3-24:48:54.5 &  -1.6 &  7.69 & 12.70 & 4311 &  0.7 &  6.70 &   10 &  6.19 &  1.83 &  27.4 & No \\  \hline
        
    \end{tabular}
\end{table*}


\bsp	
\label{lastpage}
\end{document}